\documentclass{article}
\usepackage[11pt]{extsizes}
\usepackage[utf8]{inputenc}
\usepackage{ulem}
\usepackage[english]{babel}
\usepackage[T1]{fontenc}
\usepackage{listings}
\usepackage{amsmath}
\usepackage{mathrsfs}
\usepackage{hyperref}
\usepackage{ifthen}
\usepackage{blindtext}
\usepackage{mathtools}
\usepackage{amssymb}
\usepackage{url}
\usepackage{textcomp}
\usepackage{dsfont}
\usepackage{hyperref}
\usepackage{bbm}
\usepackage{verbatim}
\usepackage{glossaries}
\usepackage{appendix}

\usepackage[utf8]{inputenc}
\usepackage[T1]{fontenc}
\usepackage{array}
\PassOptionsToPackage{english}{babel}
\newcolumntype{R}[1]{>{\raggedleft\arraybackslash }b{#1}}
\newcolumntype{L}[1]{>{\raggedright\arraybackslash }b{#1}}
\newcolumntype{C}[1]{>{\centering\arraybackslash }b{#1}}
\usepackage{diagbox}

\usepackage{booktabs}
\usepackage{tabularx}
\usepackage{color}
\definecolor{mygreen}{RGB}{28,172,0} % color values Red, Green, Blue
\definecolor{mylilas}{RGB}{170,55,241}
\newcommand{\R}{\mathbb{R}}

\newcommand{\N}{\mathbb{N}}
\newcommand{\E}{\mathbb{E}}
\newcommand{\X}{\mathbb{X}}

\newcommand{\Prob}{\mathbb{P}}

\newcommand{\Tau}{\mathcal{T}}
\usepackage{mathtools}
\newcommand{\bH}{\overline{H}}

\DeclarePairedDelimiterX{\expectarg}[1]{(}{)}{%
  \ifnum\currentgrouptype=16 \else\begingroup\fi
  \activatebar#1
  \ifnum\currentgrouptype=16 \else\endgroup\fi
}

\newcommand{\innermid}{\nonscript\;\delimsize\vert\nonscript\;}
\newcommand{\activatebar}{%
  \begingroup\lccode`\~=`\|
  \lowercase{\endgroup\let~}\innermid 
  \mathcode`|=\string"8000
}
\usepackage[numbers]{natbib}   % numeric citations
\setcitestyle{square}

\newtheorem{theo}{Theorem}

\newtheorem{Corollary}{Corollary}
\newtheorem{prop}{Proposition}
\newtheorem{lem}{Lemma}
\newtheorem{remark}{Remark}
\newtheorem{proof}{Proof}
\newtheorem{mydeff}{Definition}
\usepackage{graphicx}
\usepackage[utf8]{inputenc}
\usepackage{mathenv}
\usepackage{gensymb}
\usepackage{subcaption}
\usepackage{tikz}
\usetikzlibrary{shapes}
\usetikzlibrary{trees}
\usetikzlibrary{calc}
\newcommand\tab[1][0.5cm]{\hspace*{#1}}
\usepackage[top=2cm, bottom=2cm, left=2cm, right=2cm]{geometry}
\usepackage{listings} % Para usar código fuente
\usepackage{relsize}
\usepackage{pgfplots}
\usepackage{multirow}
\usepackage{lipsum}
\usepackage[algo2e]{algorithm2e} 
\usepackage{hyperref}
\usepackage[section]{placeins}
\usepackage{caption}
\usepackage{float}
\usepackage[justification=centering]{caption}
\usepackage{bm}
\usepackage{algorithm}
\usepackage{algpseudocode}
\usepackage{mathtools, stmaryrd}
\usepackage{xparse} \DeclarePairedDelimiterX{\Iintv}[1]{\llbracket}{\rrbracket}{\iintvargs{#1}}
\NewDocumentCommand{\iintvargs}{>{\SplitArgument{1}{,}}m}
{\iintvargsaux#1} %
\usepackage{xparse}
\NewDocumentCommand{\iintvargsaux}{mm} {#1\mkern1.5mu..\mkern1.5mu#2}
\hypersetup{
   colorlinks=true, 
   breaklinks=true, 
   urlcolor= blue, 
   linkcolor= blue,
   pdfcreator  = {\LaTeX},
   pdfproducer = {Kile}
}

\DeclareMathOperator*{\argmin}{arg\,min} % Jan Hlavacek
\usepackage{natbib}
\usepackage{hyperref}

\usepackage{anysize}%para personalizar el ancho de los margenes
\marginsize{2cm}{2cm}{2cm}{2cm}%izquierda, derecha, arriba, abajo
\usepackage{anyfontsize}

\usepackage{appendix}

\definecolor{dkgreen}{rgb}{0,0.6,0} % Definimos colores para usar en el código
\definecolor{gray}{rgb}{0.5,0.5,0.5} 
\usepackage{fancyhdr} 
\usepackage{amssymb}
\usepackage{enumitem}   

\usepackage{hyperref}

\usepackage{authblk}
\title{From rough to multifractal multidimensional volatility: A multidimensional Log S-fBM model}
\author[1]{Othmane Zarhali}
\author[1]{Emmanuel Bacry}
\author[2]{Jean-François Muzy}
\affil[1]{Ceremade, CNRS-UMR 7534, Université Paris-Dauphine PSL, Place du Maréchal de Lattre de Tassigny, 75016 Paris, France}
\affil[2]{SPE CNRS-UMR 6134, Université de Corse, campus Grimaldi, BP 52, 20250 Corte, France}

\date{}
\hypersetup{
    colorlinks=False,
    linkcolor=black,
    citecolor=black,
    urlcolor=blue
}

\newcounter{subsubsubsection}[subsubsection]

\makeatletter
\renewcommand\paragraph{\@startsection{paragraph}{4}{\z@}%
  {3.25ex \@plus1ex \@minus.2ex}%
  {1em}%
  {\normalfont\normalsize\bfseries}}
\renewcommand\subparagraph{\@startsection{subparagraph}{5}{\parindent}%
  {3.25ex \@plus1ex \@minus .2ex}%
  {-1em}%
  {\normalfont\normalsize\bfseries}}
\makeatother

\begin{document}

\maketitle
\begin{abstract}

We introduce the multivariate Log S-fBM model (mLog S-fBM), extending the univariate framework proposed by Wu \textit{et al.} \cite{wu2022rough} to the multidimensional setting. We begin by defining the multidimensional Stationary fractional Brownian motion (mS-fBM). This process is characterized by marginals following S-fBM dynamics and a specific cross-covariance structure. It is parametrized by a correlation scale $T$, marginal-specific intermittency parameters $(\lambda_i^2)_{1\leq i \leq d}$ and Hurst exponents $(H_i)_{1\leq i \leq d}$, as well as their multidimensional counterparts: the co-intermittency matrix $(\xi_{i,j})_{1\leq i,j \leq d}$ and the co-Hurst matrix $(H_{i,j})_{1\leq i,j \leq d}$.
The mLog S-fBM is then constructed by modeling the volatility components as exponentials of the mS-fBM, thereby preserving the dependence structure encoded in the Gaussian core. We establish that the mLog S-fBM is well-defined for any co-Hurst matrix with entries in $[0, 1/2)$. Notably, the model supports vanishing co-Hurst parameters, allowing it to bridge the rough volatility and multifractal regimes. Furthermore, we generalize the small intermittency approximation technique to the multivariate setting to develop an efficient Generalized Method of Moments (GMM) calibration procedure. This method relies on estimating the cross-covariance parameters for pairs of marginals. We validate the procedure on synthetic data and subsequently apply it to S\&P 500 market data, modeling stock return fluctuations within this multivariate stochastic volatility framework. In particular, diagonal estimates of the stock Hurst matrix, corresponding to single-stock log-volatility Hurst exponents, are consistently found to be close to $H_{ii} \approx 0$, indicating a multifractal behavior in agreement with the findings of \cite{wu2022rough}. Conversely, we observe that the co-Hurst matrix off-diagonal entries are close to the Hurst exponent of the S\&P 500 index itself ($H \approx 0.15$), consistent with the rough volatility literature, while the off-diagonal entries of the co-intermittency matrix align closely with univariate intermittency estimates.
\end{abstract}


\section*{Introduction}

The Fractional Brownian Motion (fBm) $B_t^H$, introduced by Mandelbrot and Van Ness (1968), has emerged as the fundamental mathematical extension of classical Brownian motion for modeling systems with memory. It is a continuous-time, zero-mean Gaussian process with stationary increments, characterized by its auto-correlation function:
\begin{equation}\label{eq:fBM_def}\mathbb{E}\left(B_t^HB_s^H\right) = \frac{\sigma^2}{2}\left(t^{2H}+s^{2H}-|t-s|^{2H}\right),
\end{equation}
where $\sigma^2$ is the variance parameter and $H\in \left]0,1\right[$ is the Hurst parameter. This exponent $H$ is the critical parameter controlling the path regularity: for $H<\frac{1}{2}$, the process is anti-persistent and its paths are "rough" (less regular than Brownian motion), whereas for $H>\frac{1}{2}$ the paths are smoother and the process exhibits long-range persistence.

In quantitative finance, fBM and related fractional processes have become central tools for modeling volatility, originally motivated by strong empirical evidence of long memory and multiscale structure in realized volatility time series. The first study to introduce a stochastic volatility model driven by a persistent (i.e., $H> \frac{1}{2}$) fBm in order to account for volatility fluctuations is due to Comte and Renault~\cite{comte1998long}. In recent years, however, the paradigm has shifted dramatically from persistent to rough ($H<\frac{1}{2}$) fBm-based log-volatility models, following the seminal analysis of Gatheral, Jaisson and Rosenbaum~\cite{gatheral2018volatility}. These authors argued that log-volatility behaves essentially as a fractional Brownian motion with a very small Hurst exponent,
$H \approx 0.1$, implying anti-persistence and sample paths significantly rougher than those of standard Brownian motion. This insight led to the development of fractional Ornstein--Uhlenbeck (fOU) processes as building blocks for option pricing models and to the emergence of prominent frameworks such as the rough Bergomi model~\cite{bayer2016pricing}, which successfully reproduces the power-law explosion of the implied volatility skew at short maturities and the rough Heston model~\cite{ElEuchRosenbaum2019}. These non-Markovian models required the development of new mathematical tools, including fractional Riccati equations and stochastic Volterra equations.

In fact, all the core ideas underlying the "rough" log-volatility paradigm were already present within the earlier and also very popular class of models, namely the class of Multifractal models of volatility. More than a decade prior to the "Rough Volatility" literature, 
clustering and scaling properties of the realized volatility
were described in a framework initiated by Mandelbrot, Calvet and Fisher through their multifractal models of asset returns~\cite{mandelbrot1997multifractal}.
This discrete, non-stationary cascade construction was later generalized by Bacry, Delour and Muzy, who introduced the multifractal random walk (MRW)~\cite{muzy2000modelling,bacry2001multifractal}. From a theoretical standpoint, these multifractal models can be interpreted as exhibiting an ``extreme'' form of roughness. While rough volatility models typically assumes a power-law decay as $C-\tau^{2H}$ of the covariance behavior with $H \approx 0.1$, the MRW features a logarithmic decay of the covariance structure, which formally corresponds to the limit $H \to 0^+$, characteristic of the so-called $1/f$ processes \cite{muzy2000modelling}.  The log-normal MRW framework turns out to be intimately linked to the concept of ``Gaussian Multiplicative Chaos'' introduced by Kahane~\cite{kahane1985chaos} and was extended by two of us (E.B. and J.F.M.) to continuous cascades with arbitrary log-infinitely divisible laws~\cite{muzy2002multifractal,bacry2003log}.  A unifying mathematical framework that conciliates the rough volatility approach (where typically $H \approx 0.1$) with the asymptotic multifractal behavior ($H \to 0^+$) was introduced in 2022 by Wu, Bacry and Muzy: the Logarithmic Stationary fractional Brownian Motion (Log S-fBM)~\cite{wu2022rough}. Within this model, finite values of $H$ correspond to a variant of Gatheral et al. model while $H=0$ exactly recovers the MRW model.
 
Overall, rough volatility models provide a unified framework linking microstructural effects, statistical scaling properties and option-implied features of volatility, while remaining sufficiently tractable for calibration and numerical implementation. However, Wu \textit{et al.}~\cite{wu2022rough} showed, through calibration of the log S-fBM model on empirical equity data, that while stock indices, representing an ``average'' behavior of individual stocks, exhibit log-volatility roughness consistent with the findings of~\cite{gatheral2018volatility} ($H \approx 0.1$), individual stock log-volatilities display significantly lower regularity, bringing them closer to the multifractal limit ($H = 0$).
In a recent work~\cite{LogSfbmNFM}, we showed that this behavior can be explained within a ``Nested Factor'' framework, that is, a factor model for both asset returns and their log-volatilities~\cite{chicheportiche2015nested}, with log S-fBM components governing volatility dynamics. 
Because return as well realized volatility fluctuations are far from independent across various assets, it is important to develop models capable of fully capturing their joint dynamics.
The objective of the present paper is to extend the univariate log S-fBM model to a genuinely multivariate setting, thereby generalizing both multifractal and rough volatility models to multivariate contexts.

Let us notice that a simple multivariate extension of the MRW was already considered in~\cite{mMRW}, featuring a logarithmic decay of cross-autocovariances of log-volatilities and applications to optimal portfolio selection. Fractional volatility models have also been extended to multivariate and cross-asset frameworks, enabling the modeling of long-memory dependence in volatility co-movements and correlation dynamics~\cite{abijaber2019affine}. More recently, Bibinger, Yu and Zhang~\cite{bibinger2025} introduced a multivariate fBm-based model to capture cross-correlations and heterogeneous roughness levels across assets, leading to improved realized volatility forecasts in interconnected markets. Beyond applications to log-volatility in equity markets, the construction of multidimensional stationary fractional Brownian motions raises fundamental mathematical challenges that have been addressed in earlier studies. In particular, Lavancier \textit{et al.}~\cite{mfbm} introduced the multidimensional fractional Brownian motion (m-fBM), which preserves key properties of fBm—such as self-similarity and stationary increments—while allowing for flexible cross-covariance structures capable of capturing anisotropy and heterogeneous correlations across components. Their simulation scheme relies on a multidimensional extension of the circulant embedding method of Chan \textit{et al.} We may also cite the recent work of Dugo \textit{et al.}~\cite{Dugo2026}, who extended the framework of Lavancier \textit{et al.} to define and study a multivariate fractional Ornstein--Uhlenbeck process, that is, a stationary multivariate Gaussian process allowing for distinct Hurst exponents across components.\\

The remainder of the paper is organized as follows. We first briefly introduce the main models of interest: the one-dimensional S-fBM and the associated log S-fBM model, as extensively studied in~\cite{wu2022rough}. Section~\ref{sec:multidimsfbm} and \ref{sec:multidimlogsfbm} are devoted to the construction of a multidimensional version of the S-fBM and Log S-fBM processes and its logarithmic counterpart, including the derivation of appropriate cross-autocovariance functions. In Section~\ref{sec:SIAmultidimlogsfbm}, we leverage the small intermittency approximation developed in~\cite{wu2022rough} to derive a proper small intermittency approximation result related to the multidimensional model amenable to parameter estimation and we assess the accuracy of this calibration procedure using a generalized method of moments on synthetic data in Section~\ref{sec:numerics}. Finally, Section~\ref{sec:empirics} presents financial applications. We first discuss the connection between the mLog S-fBM model and the Nested stationary fractional factor model (N-SfFM) of Zarhali et al.~\cite{LogSfbmNFM} and then explore the interpretation of the mLog S-fBM as a multivariate stochastic volatility model, supported by empirical results on S\&P~500 stock data.

\section{The stationary fractional Brownian Motion (S-fBM) and the Log stationary fractional brownian motion model (Log S-fBM)}
\label{sec:sectionSfbm}
    In this section,
    the aim is to review the Log S-fBM model of Wu \textit{et al.} in \cite{wu2022rough} defined for $H\in \left[0,\frac{1}{2}\right[$. For that sake, we start by recalling the definition of the Stationary fractional Brownian Motion (S-fBM) process introduced by Wu \textit{et al.} in \cite{wu2022rough} which is defined for $H\in \left] 0,\frac{1}{2}\right[$ alongside its main properties: its construction as an integrated inhomogenious gaussian white noise, its derived autocovariance function and the Log S-fBM random measure evoking its limit when the Hurst exponent vanishes enabling to define the Log S-fBM (as well as the S-fBM) model for $H=0$. Here, the focus is made on the relevant features of the S-fBM process necessary to introduce its multidimensional analogue in the following section. A comprehensive study of the S-fBM process is available in \cite{wu2022rough}.
\subsection{The S-fBM process}
\label{subsec:Sfbm}
As mentioned in the introduction earlier, we consider $H\in \left] 0,\frac{1}{2}\right[$ and introduce the necessary ingredients to construct the S-fBM process. Later on, we will extend its definition for $H=0$ in the sense that is explained in Section \ref{subsec:MRMSfBM}.
   \\\\ 
First and foremost, we denote for any $t\geq 0$ the domain $C_T(t)$ defined as:
\begin{eqnarray}
        \forall T\in \R_{+},\forall t^{*}\in \R,\tab 
        C_{T}(t):=\left\{(t,h)/h>0 ,|t-t^{*}|<\frac{1}{2}\min(h,T)\right\}
\end{eqnarray}
which is particularly useful in the logarithmically correlated process involved in the construction of the MRW process (see \cite{bacry2001multifractal}).\\
 \begin{figure}
    \centering
    \includegraphics[width=0.5\linewidth]{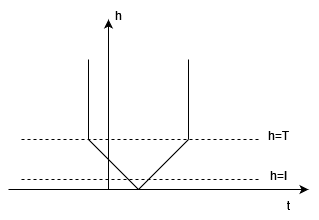}
    \caption{Representation of $C_{T}(.)$.}
    \label{fig:takenakadomain}
\end{figure}
The S-fBM process is preliminarly defined for $H\in \left]0,\frac{1}{2}\right[$ as an integral of a gaussian noise on $C_T(.)$ as follows:
\begin{eqnarray}
   \forall t \geq 0,\tab  \omega_{H,T}(t)=\mu_{H}+\int_{C_T(t)} {\rm d}G_{H}
\end{eqnarray}
\\
Above, $\mu_{H}$ is a normalizing constant defined such that:
    \begin{eqnarray}
    \label{eq:normalizationSfbm}
        \forall t\geq 0, \tab \E\left(e^{\omega_{H,T}(t)}\right)=1
    \end{eqnarray}
and ${\rm d}G_H$ is a non homogeneous gaussian white noise of variance measure given by:
     \begin{eqnarray}
     \label{eq:variancenonhomogeniousgaussiannoise}
         \E\left( {\rm d}G_{H}(t,h)^2 \right)=\lambda^2\frac{h^{2H-2}}{T^{2H}} {\rm d}t {\rm d}h
    \end{eqnarray}
where $\lambda$ is the intermittency coefficient and $T>0$.\\
As ${\rm d}G_H$ is centered, $\mu_{H}$ also plays the role of the mean of $\omega_{H,T}$.\\\\
We define $\nu^2$ as:
\begin{eqnarray}
    \nu^2=\frac{\lambda^2}{H(1 - 2H)}
\end{eqnarray}
The detailed computations of Appendix A.1 in \cite{wu2022rough} demonstrate that the autocovariance function of the S-fBM process is given by:
\begin{equation}
\label{eq:covsfbm}
C_{\omega_{H,T}}(\tau) := \mathrm{cov}(\omega_{H,T}(t), \omega_{H,T}(t+\tau)) =
\begin{cases}
\displaystyle \frac{\nu^2}{2} \left(1 - \left(\dfrac{|\tau|}{T}\right)^{2H} \right) \quad \text{if } |\tau| < T \\[1.2ex]
0 \quad\hspace{3.2cm}\text{otherwise}
\end{cases}
\end{equation}
For a given Hurst exponent $H\in \left]0,\frac{1}{2}\right[$, the S-fBM process $(\omega_{H,T}(t))_t$ is defined as a stationary Gaussian process where $\nu$, a dimensionless parameter, controls the amplitude of the process and $T$ represents the correlation limit. Thus, for any $t\geq 0$, one has  $\text{var}\left(\omega_{H,T}(t)\right)=\frac{\nu^2}{2}$. Straightforward computations using the latter variance expression in Eq.~\eqref{eq:normalizationSfbm} lead to:
    \begin{eqnarray}
    \label{eq:defmu_H}
        \mu_{H}=-\frac{\nu^2}{4}
    \end{eqnarray}
\subsection{The Log S-fBM random measure and the Log S-fBM model}
\label{subsec:MRMSfBM}
Wu \textit{et al.} defined analoguously to Bacry \textit{et al.} in \cite{bacry2001modelling} for $H\in \left]0,\frac{1}{2}\right[$ the Log S-fBM random measure $(M_{H,T}(t))_t$ as:
\begin{eqnarray}
\label{eq:mrm}
    M_{H,T}(\mathrm{d}t) = \exp(\omega_{H,T}(t)) \mathrm{d}t
\end{eqnarray}
with $(\omega_{H,T}(t))_{t}$ being a S-fBM process with parameters $H$, $\lambda$ and $T$.\\
 Notably, Wu \textit{et al.} demonstrated that as $H$ approaches $0$, $(M_{H,T}(t))_t$ converges to the multifractal random measure of the MRW process, as introduced in \cite{bacry2001multifractal, bacry2001modelling, muzy2000modelling, muzy2002multifractal}. In the following, we present the construction of such limit.\\\\
We start by recalling the logarithmically correlated Gaussian process $\omega_{\ell,T}$ studied in \cite{bacry2001multifractal} for any arbitrary $(\ell,T)\in \R_+^2$ with autocovariance:
\begin{eqnarray}
\label{eq:logmrm}
    \begin{aligned}
    C_{\tilde \omega_{\ell,T}}(\tau) =\mathrm{cov}(\tilde \omega_{\ell,T}(t),\tilde \omega_{\ell,T}(t+\tau)) =  \begin{cases}
        -\lambda^2 (\ln({\ell}/{T}) - 1 + \tau/\ell),\quad \text{ when } |\tau|<\ell\\
        -\lambda^2 \ln({\tau}/{T}),\quad \text{ when } \ell< |\tau|<T\\
        0 \quad \text{otherwise}
    \end{cases}.
\end{aligned}
\end{eqnarray}
Its mean value is chosen such that the identity:
\begin{eqnarray}
\label{eq:normalizationlogmrm}
\mathbb{E}(e^{\tilde \omega_{\ell,T}(t)}) = 1
\end{eqnarray}
is satisfied.\\
By introducing the random measure:
\begin{eqnarray}
    \widetilde{M}_{\ell,T}(dt)=\exp\left(\tilde \omega_{\ell,T}(t) \right) dt,
\end{eqnarray}
one retrieves the multifractal random measure (MRM) through the limit:
\begin{eqnarray}
\label{eq:weakcvgmrm1D}
    \Tilde{M}_{\ell,T}(\mathrm{d}t) \xrightarrow[\ell \to 0]{\mathrm{w}}
\tilde{M}_T(\mathrm{d}t)
\end{eqnarray}
where $\xrightarrow{\mathrm{w}}$ stands for the weak convergence.\\
Notably, for $H=0$, $M_{H,T}$ is given as the multifractal random measure $\tilde{M}_T$. Proposition 2 in \cite{wu2022rough} states that:
\begin{eqnarray}
\label{eq:limitHzerologSfbmmeasure}
    M_{H,T}(\mathrm{d}t) \xrightarrow[H \to 0]{\mathrm{w}}
\tilde{M}_T(\mathrm{d}t)
\end{eqnarray}
This framework integrates cases where $H > 0$ and where $H \to 0$ that can be interpreted respectively as the rough regime and multifractal regime, both observed in empirical financial data. Wu \textit{et al.} further propose the Log S-fBM process $X$ which has the following dynamic:
\begin{eqnarray}
    \mathrm{d}X_t = \sqrt{\frac{M_{H,T}(\mathrm{d}t)}{\mathrm{d}t}} \mathrm{d}B_t
\end{eqnarray}
 where $(B_t)_t$ is a standard Brownian motion.\\
 
 From a financial modeling perspective, this was proposed as a stochastic volatility model able to account for both rough and multifractal log volatility sample paths. As an example, we cite the numerical experiments of Figure 10 in \cite{wu2022rough} where indices shows a low roughness $H\simeq 0.1$ compared to single stocks having a high roughness characterized by $H\simeq 0.01$. This stylized fact was also reproduced by the N-SfFM in \cite{LogSfbmNFM} (see Section \ref{sec:linkwithLogSfbmNFM}). 
 
 The calibration of the Log S-fBM parameters to observed data is performed using the so called small intermittency approximation as presented in \cite{wu2022rough} (also used in the context of multifractal models, see for instance \cite{bacry2008log}).  It allows to derive closed form approximations of the log volatility measure generalized moments on arbitrary periods when the intermittency coefficient goes to $0$ as advocated in Proposition 4 in \cite{wu2022rough}.

\section{The multidimensional S-fBM (mS-fBM)}
\label{sec:multidimsfbm}
In this section, we introduce the mS-fBM process, a $\R^d$ valued process with marginals being correlated S-fBM processes for which we derive proper cross covariation properties. Herein, we start by the construction of the mS-fBM process from a $\R^d$ valued non homogeneous white noise with a proper correlation structure between its marginals. Then, we derive a closed form formula of the mS-fBM cross covariances. In the following, we proceed in a similar way as in Section \ref{sec:sectionSfbm}: we start by defining the mS-fBM for a co-Hurst matrix (analogue of the Hurst exponent in the multidimensional case) denoted  $\bm{H}:=\left(H_{i,j}\right)_{1\leq i,j \leq d} \in \left]0,\frac{1}{2}\right[^{d \times d }$ and a co-intermittency matrix (analogue of the intermittency coefficient in the multidimensional case) $\bm{\xi}:=\left(\xi_{i,j}\right)_{1\leq i,j \leq d} \in \R^{d\times d}$.
\begin{comment}
     later in Section \ref{sec:multidimlogsfbm}, we will extend the definition of the mLog S-fBM random measure (in the weak sense) and model  for a vanishing and non vanishing co-Hurst matrix.\\
\end{comment}
In the sequel, we consider $T>0$ being the correlation limit and  a co-intermittency matrix having diagonals of the form $\xi_{i,i}=\lambda_i^2$ for all $i\in \llbracket 1,d \rrbracket$ such that $\left(\lambda_i\right)_{1\leq i\leq d}\in \left]0,\frac{1}{2}\right[^d$.
The diagonals of the co-Hust and co-intermittency matrices refer respectively refer to the Hurst exponents and intermittency coefficients such that the autocovariance structures of the marginal S-fBM processes are well defined, as we are going to see in the upcoming definition.\\

We denote the mean value Hurst exponent for any $(i,j)\in \llbracket 1,d \rrbracket^2$,
\begin{eqnarray}
   \bH_{i,j} = \frac{H_{i,i}+H_{j,j}}{2}
\end{eqnarray}
\begin{comment}
     Analoguously to Sections \ref{sec:sectionSfbm} (Sections \ref{subsec:Sfbm} and \ref{subsec:MRMSfBM}), we suppose that for any $(i,j)\in \llbracket 1,d \rrbracket^2$, $H_{i,j}$ is in $\left]0,\frac{1}{2}\right[$ and will prove a limit theorem (Theorem \ref{theo:weakcontinuitylinkmLogSfBMMMRW} and Proposition \ref{prop:linkmLogSfBMMMRW} in Section \ref{sec:multidimlogsfbmmrm}) when the co-Hurst exponents goes to $0$ enabling to defined our model for non vanishing and vanishing co-Hurst matrices.\\\\
\end{comment}
We introduce:
 
\begin{eqnarray}
\label{eq:posdef-intermittency}
\bm{\bm{\mathcal{H}_{\mathrm{1}}}}:\quad 
\bm{\xi}\text{ is positive definite}
\end{eqnarray}

\begin{equation}
\label{eq:negdef-coHurst}
\bm{\mathcal{H}_{2}}:\qquad
\bm{H}\ \text{is conditionally negative definite, i.e:}
\\[0.2cm]
\sum_{i,j=1}^d c_i c_j H_{ij}\le 0,
\text{ for any } c\in\mathbb R^d
\ \text{such that}\
\sum_{i=1}^d c_i=0.
\end{equation}
The following results concern a necessary condition useful for the mS-fBM's construction.
\begin{lem}
\label{lemma:covdGi}
    If $\bm{\xi}$ and $\bm{H}$ satisfy respectively \hyperref[eq:posdef-intermittency]{\(\bm{\mathcal{H}_{\mathrm{1}}}\)} and \hyperref[eq:negdef-coHurst]{\(\bm{\mathcal{H}_{\mathrm{2}}}\)},
then, for any fixed $(t,h)\in C_T(.)$, the centered gaussian vector denoted $(dG_{i}(t,h))_{i\in \llbracket 1,d \rrbracket} $ with the following covariance matrix:
 \begin{equation}
        \forall(i,j)\in \llbracket 1,d \rrbracket^2, \tab \mathrm{cov}\left({\rm d}G_{i}(t,h),{\rm d}G_{j}(t,h)\right) =\begin{cases}
        \xi_{i,j} \frac{h^{2H_{i,j}-2}}{T^{2H_{i,j}}} \; {\rm d}t \; {\rm d}h, 0<h\leq T\\
        \xi_{i,j} \frac{h^{2\overline{H}_{i,j}-2}}{T^{2\overline{H}_{i,j}}} \; {\rm d}t \; {\rm d}h,h>T
        \end{cases}
        \label{eq:covmultidimsfbm_dG}
    \end{equation}
is positive definite.
\end{lem}

The latter Lemma, which proof is in Appendix \ref{app:proofexistencedGi},  states the positive definiteness of $(dG_{i}(t,h))_{i\in \llbracket 1,d \rrbracket} $'s covariance matrix.
The diagonal of the matrix whose entries are the terms of Eq.~\eqref{eq:covmultidimsfbm_dG} correspond to the variance measures of the one dimensional case in Eq.~\eqref{eq:variancenonhomogeniousgaussiannoise}.
\begin{comment}
     The autocovariance structure of Eq.~\eqref{eq:covmultidimsfbm_dG} can also be interpreted for large scales ($h>T$) so as the following identity holds $H_{i,j}=\bH_{i,j}$.\\
\end{comment}

Similarly, one can deduce the correlation structure using the variance measure of Eq.~\eqref{eq:variancenonhomogeniousgaussiannoise}:
 \begin{equation}
\label{eq:corrmultidimsfbm_dG}
    \forall (i,j)\in \llbracket 1,d \rrbracket^2,\tab \rho_{G}^{i,j}\left(h\right):=\frac{ \mathrm{cov}\left({\rm d}G_{i}(t,h),{\rm d}G_{j}(t,h)\right)}{\sqrt{\text{var}\left({\rm d}G_{i}(t,h)\right)}\sqrt{\text{var}\left({\rm d}G_{j}(t,h)\right)}}= \begin{cases}
        g_{i,j}\left(\frac{h}{T}\right)^{2(H_{i,j}-\bH_{i,j})}, h\leq T\\
        g_{i,j}, h>T
    \end{cases},
\end{equation}
where $g_{i,j}$ represents the co-intermittency correlation coefficient defined by:
\begin{eqnarray}
\label{eq:co-intermittencycorrelationdef}
    \forall (i,j)\in \llbracket 1,d \rrbracket^2, \tab g_{i,j}=\frac{\xi_{i,j}}{\lambda_i\lambda_j}.
\end{eqnarray}\hyperref[eq:posdef-intermittency]{\(\bm{\mathcal{H}_{\mathrm{2}}}\)} leads to the following condition:
\begin{eqnarray}
\label{eq:lowerboundco-Hurst}
\forall (i,j)\in \llbracket 1,d \rrbracket^2,\tab
    H_{i,j}\geq \bH_{i,j}
\end{eqnarray}
enabling the correlation structure of Eq.~\eqref{eq:corrmultidimsfbm_dG} to be well defined. Besides,
Eq.~\eqref{eq:corrmultidimsfbm_dG} infers that in small scales $(h \leq  T)$, the correlation of the gaussian noises ${\rm d}G_{i}$ is a power of the scale. In high scales ($h > T$), this correlation is constant and equal to the co-intermittency correlation. Thanks to \hyperref[eq:posdef-intermittency]{\(\bm{\mathcal{H}_{\mathrm{2}}}\)}, one has:
\begin{eqnarray}
\label{eq:upperboundco-intermittencycorr}
    \forall (i,j)\in \llbracket 1,d \rrbracket^2, \quad \left|g_{i,j}\right|\leq 1
\end{eqnarray}
\begin{remark}
The correlation structure of Eq.~\eqref{eq:corrmultidimsfbm_dG} displays a scale invariance behaviour similar to the one of the $p^{th}$ order moment of the log volatility increments (see Gatheral \textit{ et
al.} in\cite{gatheral2018volatility}). This property enables to estimate the Hurst exponent by regressing the log $p^{th}$ order moment of the log volatility increments on the log scale and identifying the Hurst exponent from the slope value.
\end{remark}
The m-SfBM process is inspired from the S-fBM construction from Section \ref{subsec:Sfbm} as the integrated multivariate non homogenious white noise of covariance the one in Lemma \ref{lemma:covdGi} over the domain $C_T(.)$, which is the object of the following definition.

\begin{mydeff}[\textbf{mS-fBM process}]
\label{def:multidimsfbm}
The mS-fBM process $((\omega_{i}(t))_t,{i\in \llbracket 1,d \rrbracket})$ is a $\R^d$ valued Gaussian process whose marginals are defined as:
\begin{equation}
    \omega_{i}(t)=  \mu_{i}+\int_{C_T(t)}{\rm d}G_{i},i\in \llbracket 1,d \rrbracket
    \label{eq:sfbmdefinition}
\end{equation}
Where:
\begin{enumerate}[label=(\roman*)]
   \item $(dG_{i}(t,h))_{i\in \llbracket 1,d \rrbracket} $ is the non homogeneous white noise of Lemma \ref{lemma:covdGi} with the autocovariance structure defined in Eq.~\eqref{eq:covmultidimsfbm_dG}.
    \item The mean value $\mu_i$ associated to the $i^{th}$ component is defined the same way as in Eq.~\eqref{eq:defmu_H} through the normalization condition of Eq.~\eqref{eq:normalizationSfbm}:
\begin{eqnarray}
    \forall i\in \llbracket 1,d \rrbracket,\tab \mu_i=-\frac{\nu_i^2}{4}=-\frac{\lambda_i^2}{4H_{i,i}(1-2H_{i,i})}
\end{eqnarray}
\end{enumerate}
\end{mydeff}
\begin{remark}
For any $i\in \llbracket 1,d \rrbracket$, $\omega_{i}$ can be seen as a set indexed process whose covariance function writes:
     \begin{equation}
         \mathrm{cov}\left(\omega_{i}(t),\omega_{j}(s)\right) = \int_{C_T(t)\cap C_T(s)} \mathrm{cov}\left({\rm d}G_{i}(t^{'},h),{\rm d}G_{j}(t^{'},h)\right) .
         \label{eq:setindexcov}
    \end{equation} 
\end{remark}

The mS-fBM process via the autocovariance structure of  $(dG_{i}(t,h))_{i\in \llbracket 1,d \rrbracket} $ in Eq.~\eqref{eq:covmultidimsfbm_dG} is characterized by a double coupling: a coupling via the Hurst exponents and a coupling via the co-intermittency coefficient. 
\begin{remark}
The previous formulation leads that the marginals of the mS-fBM process are S-fBM processes. The formulation of Dugo \textit{et al.} in \cite{Dugo2026} introduced a co-Hurst exponent in the m-fOU process that corresponds to the lower bound in Eq.~\eqref{eq:lowerboundco-Hurst}.Consequently, the mS-fBM lies in allowing more flexibility in terms of the coroughness parameter $H_{i,j}$. In that sense, the cross covariance structure of Eq.~\eqref{eq:covmultidimsfbm_dG} is more general.
\end{remark}

\begin{comment}
   The construction of the covariance structure of the gaussian noise  $(dG_{H_i}(t,h))_{i\in \llbracket 1,d \rrbracket} $ on the Takenaka domain through Eq.~\eqref{eq:covmultidimsfbm_dG} enables to see that the correlation of such a noise scales to the power $H_{i,j}$, that is to say that:
\begin{eqnarray}
\label{eq:scaleco-HurstdG}
    \forall (i,j)\in \llbracket 1,d \rrbracket^2,\tab \rho_{G}^{i,j}\left(h\right):=\frac{ \mathrm{cov}\left({\rm d}G_{i}(t,h),{\rm d}G_{j}(t,h)\right)}{\sqrt{\text{var}\left({\rm d}G_{i}(t,h)\right)}\sqrt{\text{var}\left({\rm d}G_{j}(t,h)\right)}}=g_{i,j}\left(\frac{h}{T}\right)^{2H_{i,j}-H_i-H_j}
\end{eqnarray}
where $g_{i,j}$ is the co-intermittency correlation defined as:
\begin{eqnarray}
    g_{i,j}=\frac{\xi_{i,j}}{\lambda_i\lambda_j}
\end{eqnarray} 
\end{comment}

In the light Definition \ref{def:multidimsfbm} and using Eq.~\eqref{eq:covmultidimsfbm_dG}, one can derive the cross covariance function of the mS-fBM marginals.

\begin{prop}[\textbf{cross covariance of the mS-fBM}]
\label{prop:autocovariancemultidimsfbm}
$((\omega_{i}(t))_t,{i\in \llbracket 1,d \rrbracket})$ is a stationary gaussian field with the following covariance structure:
    \begin{equation}
         \forall(t,s)\in \R_{+}^2,\forall(i,j)\in \llbracket 1,d \rrbracket^2,\tab \mathrm{cov} \left( \omega_i(t), \omega_j(s) \right) =
\begin{cases}
\begin{aligned}
\xi_{i,j} \Bigg[
&\frac{1+2 H_{i,j}-2\bH_{i,j}}{2H_{i,j}(1-2 \bH_{i,j})} 
- \left(\frac{\tau}{T}\right)^{2H_{i,j}} \frac{1}{2H_{i,     j}(1-2H_{i,j})} \\
& - \frac{\tau}{T}  \left(\frac{2H_{i,j}-2 \bH_{i,j}}{(2H_{i,j}-1)(1-2 \bH_{i,j})}
\right)
\Bigg], \quad \text{if} \quad \tau \leq T
\end{aligned} \\
0, \quad \text{otherwise}
\end{cases}
         \label{eq:autocovmultidimsfbm}
    \end{equation}
    where $\tau=|t-s|$ 
  
\end{prop}
The proof can be found in the first section of Appendix \ref{app:proofautocovariancemultidimsfbm}.\\\\
In the sequel, we denote the latter cross covariance function $C_{i,j}^{\omega}(.)$ as a function of $\tau$.\\

\begin{remark}
We notice that in Eq.~\eqref{eq:autocovmultidimsfbm}, if $i=j$, $H_{i,j}=H$ and $\xi_{i,j}=\lambda^2$, $C_{i,i}^{\omega}(.)$ corresponds to the autocovariance function of the monovariate case of Eq.~\eqref{eq:covsfbm}. Moreover, for any fixed $t\in \R_+$, $\left(\omega_{i}(t)\right)_{i\in \llbracket 1,d \rrbracket}$ is a gaussian vector with the covariance:
\begin{eqnarray}
    C_{i,j}^{\omega}(0)=\frac{\xi_{i,j}}{2H_{i,j}(1-2H_{i,j})}
\end{eqnarray}
\end{remark}

\section{The multidimensional Log S-fBM model (mLog S-fBM)}
\label{sec:multidimlogsfbm}

This section is devoted to the mLog S-fBM random measure's construction using the mS-fBM process of Section \ref{sec:multidimsfbm}, its main properties as well as the mLog S-fBM model.\\
Analoguously to the one dimensional case tackled in \cite{wu2022rough}, our aim here is to demonstrate that the mLog S-fBM framework is defined for both non vanishing and vanishing co-Hurst exponents. For this sake, we start by defining the mLog S-fBM random measure for non vanishing co-Hurst exponents. Then, we introduce a random measure namely a multidimensional multifractal random measure as the limiting measure in the case of vanishing co-Hursts, its multifractal property as well as a reference to the small intermittency approximation in this context. Finally, we present the definition of the mLog S-fBM random measure for any co-Hurst matrix satisfying the constraints of Section \ref{sec:multidimsfbm}. finally we present the associated mLog S-fBM model.\\

In the following, we denote:
\begin{equation}
\begin{cases}
\displaystyle
\mathcal{S}_d
:=
\left\{
\mathbf H \in \left[0,\frac12\right[^{d\times d}
\;\middle|\;
\mathbf H^\top=\mathbf H,
\quad
\mathbf H \text{ satisfies \hyperref[eq:negdef-coHurst]{\(\bm{\mathcal H}_{2}\)}}
,
\quad
\forall\, i\neq j \in \llbracket 1,d\rrbracket,
\quad
H_{ij}\neq 0
\right\},
\\[2ex]
\displaystyle
\mathcal{S}_d^{*}
:=
\left\{
\mathbf H \in \left]0,\frac12\right[^{d\times d}
\;\middle|\;
\mathbf H^\top=\mathbf H,
\quad
\mathbf H \text{ satisfies \hyperref[eq:negdef-coHurst]{\(\bm{\mathcal H}_{2}\)}}
,
\quad
\forall\, i\neq j \in \llbracket 1,d\rrbracket,
\quad
H_{ij}\neq 0
\right\}.
\end{cases}
\end{equation}
being respectively the set of symmetric matrices satisfying \hyperref[eq:lowerboundco-Hurst]{\(\bm{\mathcal{H}_{\mathrm{1}}}\)} with non-vanishing off-diagonal entries and the set of symmetric matrices satisfying \hyperref[eq:lowerboundco-Hurst]{\(\bm{\mathcal{H}_{\mathrm{1}}}\)} with non-vanishing entries.
\subsection{The mLog S-fBM random measure for 
$\mathbf{H}\in \mathcal{S}^{*}_{d}$}

\label{sec:multidimlogsfbmmrm}
\begin{mydeff}[\textbf{mLog S-fBM random measure}]
   The mLog S-fBM random measure is the $\R^d$ valued random measure denoted as:
\begin{eqnarray}
    \bm{M}^{*}_{\bm{H}}(dt):= \begin{pmatrix}
M_{1} \left(dt\right) \\
M_{2} \left(dt\right) \\
\vdots \\
M_{d} \left(dt\right)
\end{pmatrix},
\end{eqnarray}
each marginal is defined similarly as in Eq.~\eqref{eq:mrm}:
\begin{eqnarray}
\label{eq:logsfbmmrmnonvanishH}
     \forall i\in \llbracket 1,d \rrbracket,\tab M_{i}(dt)=\exp\left(\omega_{i}(t)\right)dt
\end{eqnarray} 
where
$\left(\omega_{i}\right)_{i\in \llbracket 1,d \rrbracket}$ is a mS-fBM process (see Definition \ref{def:multidimsfbm}).
\end{mydeff}

Its counterpart on an arbitrary period $[0,t]$ for $t>0$ is given by:
\begin{eqnarray}
   \forall i\in \llbracket 1,d \rrbracket,\tab M_{i}(t):=M_{i}([0,t])
\end{eqnarray}
as well as over an interval of and arbitrary length $\Delta>0$:
\begin{eqnarray}
   \forall i\in \llbracket 1,d \rrbracket,\tab M_{i,\Delta}(t):=M_{i}([t,t+\Delta]) ,\tab t\geq 0
    \label{eq:mrmoverint}
\end{eqnarray}
In Appendix \ref{app:proofautocovMRM}, a semi-closed form formula of the cross covariance function of $\left(\left(M_{i,\Delta}(t)\right)_t\right)_{i\in \llbracket 1,d \llbracket}$ has been derived in the form of a series expansion. Leveraging the small intermittency approximation, one can derive a small intermittency approximation of the latter cross covariance function expressed in Eq.~\eqref{eq:crossautocovmrmSIA} where one can see an additional term that vanishes in the one dimensional case ($i=j$) retrieving the results of Wu \textit{et al.} reported in Section 3.1 in \cite{wu2022rough}.
\subsection{The multidimensional multifractal random measure (mMRM)}
\label{sec:multidimmmrm}
Herein, we introduce the $\R^d$ valued stationary Gaussian process $\left(\tilde \omega^i_{\ell}(.),i\in \llbracket 1,d \rrbracket \right)$ as the multivariate version of $\tilde\omega_{\ell,T}$ introduced in Eq.~\eqref{eq:logmrm}, also known as the multivariate multifractal random walk model proposed in \cite{mMRW} and \cite{mMRW1}. We choose Eq.~\eqref{eq:logmrm} to become in the multidimensional setting of the form :
\begin{eqnarray}
\label{eq:mlogmrm}
    \begin{aligned}
    \widetilde{C}_{\ell,T}^{i,j}(\tau) =\mathrm{cov}\left(\tilde \omega^i_{\ell}(t),\tilde \omega^j_{\ell}(t+\tau)\right) =  \begin{cases}
        -\xi_{i,j} (\ln({\ell}/{T}) - 1 + \tau/\ell),\quad \text{ when } |\tau|<\ell\\
        -\xi_{i,j} \ln({\tau}/{T}),\quad \text{ when } \ell< |\tau|<T\\
        0 \quad \text{otherwise}
    \end{cases}.
\end{aligned}
\end{eqnarray}
The construction of such process is recalled in Appendix \ref{app:linkMMRW}, it ensures that each marginal of the multivariate multifractal random walk is a multifractal random walk process. The following is the definition of the mMRM.
\begin{mydeff}
The $\R^d$ valued random measure associated with the process $\left(\omega^i_{\ell,T}\right)_{i\in \llbracket 1,d \rrbracket}$ reads:
\begin{eqnarray}
\widetilde{\bm{M}}_{\ell}(dt):= \begin{pmatrix}
 \widetilde{M}^1_{\ell} \left(dt\right) \\
 \widetilde{M}^2_{\ell} \left(dt\right) \\
\vdots \\
 \widetilde{M}^d_{\ell} \left(dt\right)
\end{pmatrix}
\end{eqnarray}
where:
\begin{eqnarray}
\label{eq:Mtildel}
    \widetilde{M}^i_{\ell}(dt):=\exp\left(\omega^i_{\ell,T}(t) \right) dt
\end{eqnarray}

The mMRM $\widetilde{\bm{M}}$ is the $\R^d$ valued random measure defined as the weak limit:
\begin{eqnarray}
\label{eq:mmrmMRW}
\widetilde{\bm{M}}_{\ell}(dt)\xrightarrow[\ell \to 0]{\mathrm{w}} \widetilde{\bm{M}}(dt)
\end{eqnarray}
\end{mydeff}
Along the same line as in the proof Proposition \ref{prop:linkmLogSfBMMMRW} in Appendix \ref{app:linkMMRW} (see Appendix \ref{app:prooflinkmLogSfBMMMRW}), one can prove the existence of this weak limit by using a tightness argument that simply extents to multidimensional framework the former one established in the 1D case (see \cite{wu2022rough}). For the rest, we denote the associated $\R^d$ valued process as:
\begin{eqnarray}
\forall t\geq 0, \tab 
\widetilde{\bm{M}}(t):= \widetilde{\bm{M}}([0,t])
\end{eqnarray}
The process $\widetilde{\bm{M}}$ is a multidimensional multifractal process as shown in the following proposition proved in Appendix \ref{app:proofmutidimensionalmultifractalMRW}.
\begin{prop}[\textbf{multifractality of $\widetilde{\bm{M}}$}\\]
\label{prop:mutidimensionalmultifractalMRW}
    For any $t>0$, $(p,q)\in\R^2$ and $(i,j)\in \llbracket 1,d \rrbracket^2$, there exist $K^{i,j}_{p,q}>0$ and $\zeta_{i,j}(p,q)$ such that:
    \begin{eqnarray}
    \E\left[ \widetilde{M}^i(t)^p\widetilde{M}^j(t)^q\right]=K^{i,j}_{p,q}\left(\frac{t}{T}\right)^{\zeta_{i,j}(p,q)}
\end{eqnarray}
where $$\zeta_{i,j}(p,q)=\left(p+q\right)\left(1+\frac{\xi_{i,j}}{2}(p+q-1)\right) \; .$$
\end{prop}

We define in the next section the mLog S-fBM random measure which is the analogue of the mMRM associated with the mS-fBM process. A particular focus will be made on the mLog S-fBM random measure when the co-Hurst matrix (or some of its entries) goes to $0$, enabling to obtain, as a limit, the mMRM presented earlier.

\subsection{The mLog S-fBM random measure and process}
Here, we formulate the definition of the mLog S-fBM random measure for any arbitrary co-Hurst matrix in $\mathcal{S}_d$. To this end, we first identify the multifractal random measure $\widetilde{\bm{M}}$ introduced in Section~\ref{sec:multidimmmrm} as the weak limit of $\bm{M}_{\bm{H}}$ when the co-Hurst matrix vanishes. Building on this result, we introduce a component-wise definition of the mLog S-fBM random measure allowing an arbitrary number of multifractal components. This formulation makes it possible to define the mLog S-fBM random measure for vanishing and non vanishing co-Hurst entries and leads to introducing the mLog S-fBM process.  

One can extend the definition of mLog S-fBM random measure to the case of $\bm{H}\in \mathcal{S}_{d}$ allowing to have pure multifractal components as well as rough components. For that sake, we define the random measure $\bm{M}_{\bm{H},\ell}(dt)$ for any $\ell >0$ and $\bm{H} \in \mathcal{S}_{d}$ whose component are such that:
\begin{eqnarray}
\label{eq:mLogSfbmRMell}
    \left(\bm{M}_{\bm{H},\ell}\right)_j(dt)=
    \begin{cases}
        M_j(dt), \tab \text{if $H_{j,j} \neq 0$}\\
        \widetilde{M}^j_{\ell}(dt), \tab \text{if $H_{j,j} =0$}
    \end{cases}
\end{eqnarray}
where the $j^{th}$ marginals $M_j$ and $\widetilde{M}^j_{\ell}$ are respectively given in Eq.~\eqref{eq:logsfbmmrmnonvanishH} and Eq.~\eqref{eq:Mtildel}.\\
The following result claims the weak convergence when the scale parameter $\ell$ is close to $0$.
\begin{prop}
\label{prop:mixedmlogsfbmMRM}
For  $\bm{H} \in \mathcal{S}_{d}$, the following weak convergence holds:
\begin{eqnarray}
\label{eq:mixedmlogsfbmmeasure}
\bm{M}_{\bm{H},\ell}(dt)\xrightarrow[\ell \to 0]{\mathrm{w}} \bm{M}_{\bm{H}}(dt)
\end{eqnarray}

where $\bm{M}_{\bm{H}}$ is an $\R^d$ valued random measure defined component-wisely as:
\begin{eqnarray}
\label{eq:MLogSfbmweaklimit}
    \left(\bm{M}_{\bm{H}}(dt)\right)_j=
    \begin{cases}
        M_j(dt), \tab \text{if $H_{j,j} \neq 0$}\\
        \widetilde{M}^j(dt), \tab \text{if $H_{j,j} = 0$}
    \end{cases}
\end{eqnarray}
\end{prop}
The proof is in Appendix \ref{app:prooflemmaexistencemixedmlogsfmbmrm}.\\

Notice that in Proposition \ref{prop:mixedmlogsfbmMRM}, the rough components of Eq.~\eqref{eq:mLogSfbmRMell}  remain unchanged when the cutoff scale $\ell$ goes to $0$ as the weak limit operates only on the pure multifractal components.\\

Here, we denote $\|. \|$ and arbitrary matrix norm on $\R^{d\times d}$.
When $d=1$,  the weak convergence toward the multifractal random measure of the Log S-fBM has been proved in Proposition 2 of \cite{wu2022rough} as mentioned earlier.
Before, we state this continuity theorem related to the measure $\bm{M}_{\bm{H}}$.

\begin{theo}[\textbf{Weak continuity}\\]
\label{theo:weakcontinuitylinkmLogSfBMMMRW} 
Considering a sequence of co-Hurst matrices $\left(\bm{H}_n\right)_n\in \mathcal{S}_d^{\N}$ and $\bm{H}\in \mathcal{S}_d$, the sequence of measures $\left(\bm{M}_{\bm{H}_n}\right)_n$ converges weakly to $\bm{M}_{\bm{H}}$ as $\|\bm{H}_n-\bm{H} \|\rightarrow 0$.
\end{theo}
The proof is in Appendix \ref{app:proofweakcontinuitylinkmLogSfBMMMRW}.\\

Following Theorem \ref{theo:weakcontinuitylinkmLogSfBMMMRW}, similar arguments as in Proposition 2 in \cite{wu2022rough} have been employed to derive the limit of $\bm{M}_{\bm{H}}$ as the co-Hurst matrix vanishes.
\begin{prop}[\textbf{The limit $\|\bm{H} \|\rightarrow 0$}\\]
\label{prop:linkmLogSfBMMMRW} 
    The measure $\bm{M}_{\bm{H}}$ converges weakly to $\widetilde{\bm{M}}$ as $\|\bm{H} \|\rightarrow 0$.
\end{prop}
The proof is in Appendix \ref{app:prooflinkmLogSfBMMMRW}.

Furthermore, using the same arguments as in Proposition \ref{prop:linkmLogSfBMMMRW}, one can state the existence of the weak limit of $\bm{M}_{\bm{H}}$ as:
\begin{eqnarray}
\label{eq:mixedmlogsfbmmeasurefinallimit}
\bm{M}_{\bm{H}}(dt)\xrightarrow[\|\bm{H} \|^{\operatorname{diag}}_\infty\rightarrow 0]{\mathrm{w}} \widetilde{\bm{M}}(dt)
\end{eqnarray}
where for any $A\in \R^{d\times d}$, $\|.\|^{\operatorname{diag}}_\infty$ is the semi-norm:
\begin{eqnarray}
\|\bm{A} \|^{\operatorname{diag}}_\infty = \sup_{1 \leq i \leq d} \left|A_{i,i}\right|
\end{eqnarray}
This means that the diagonal entries of the co-Hurst matrix determine the transition from the rough to the multifractal setting.\\

Consequently, one may defined the mLog S-fBM random measure (the multidimensional analogue of the Log S-fBM random measure, see Section \ref{subsec:MRMSfBM}) as $\bm{M}_{\bm{H}}$.
\begin{mydeff}[\textbf{mLog S-fBM random measure}]
    \label{def:extensionMlogsfbmRM} For any $\bm{H}\in \mathcal{S}_d$, the mLog S-fBM random measure $\bm{M}_{\bm{H}}$ is the $\R^d$ valued random measure defined component-wisely as follows:
    \begin{comment}
         \begin{eqnarray}
    \left(\bm{M}_{\bm{H} ,T}(dt)\right)_j=
    \begin{cases}
        \begin{cases}
        M_j(dt), \tab \text{if $\bm{H}_{j,j} \neq 0$}\\
        \widetilde{M}^j_{T}(dt), \tab \text{if $\bm{H}_{j,j} = 0$}
    \end{cases} \text{ if } $ \bm{H}\neq 0_{d\times d}$\\
        \widetilde{M}^j_{T}(dt), \tab \text{for $j \in \llbracket 1,d \rrbracket$} \text{ if } $\bm{H}=0_{d\times d}$
    \end{cases}
\end{eqnarray}
    \end{comment}
    \begin{eqnarray}
    \left(\bm{M}_{\bm{H}}(dt)\right)_j=
    \begin{cases}
        M_j(dt), \tab \text{if $H_{j,j} \neq 0$}\\
        \widetilde{M}^j(dt), \tab \text{if $H_{j,j} = 0$}
    \end{cases}
\end{eqnarray}
\end{mydeff}
By the mean of Definition \ref{def:extensionMlogsfbmRM}, the mLog S-fBM process encompasses both the multidimensional rough and multifractal settings. Similarly and without loss of generality, one may extend the the mS-fBM process for $\bm{H}$ with null diagonal entries as the log correlated gaussian process $\left(\omega^i_{\ell,T}(.),i\in \llbracket 1,d \rrbracket \right)$ for $\ell$ arbitrary small. \\

Thus, the mLog S-fBM process can be the defined using the mLog S-fBM random measure as follows.
 \begin{mydeff}[\textbf{mLog S-fBM process}\\]
    For any $\bm{H}\in \mathcal{S}_{d} $, a mLog S-fBM process is the $\R^d$ valued process $(\X_t)_t:=((X^{i}_t)_t,i\in \llbracket 1,d \rrbracket)$ whose dynamics are defined as:
    \begin{eqnarray}
    \label{eq:mlogsfbm}
        \begin{cases*}
      dX^{i}_t = \sqrt{\frac{\left(\bm{M}_{\bm{H} }(dt)\right)_i}{dt}} dB^{i}_t\\
      X^i_0=x^i_0 \in \R
    \end{cases*}
    \end{eqnarray}
    Here,  $\bm{M}_{\bm{H}}(dt)$ is a mLog S-fBM random measure as defined in Definition  \ref{def:extensionMlogsfbmRM} and $((B^{i}_t)_t,i\in \llbracket 1,d \rrbracket)$ is a $d$-dimensional brownian motion (independent components) independent of $\bm{M}_{\bm{H}}$.
    
\end{mydeff}

In the next section, we are interested in an extension of the so called small intermittency approximation (see Bacry \textit{et al.} in \cite{bacry2008log}, \cite{bacry2001multifractal} and \cite{bacry2001modelling}, as well as \cite{wu2022rough} concerning the one dimensional Log S-fBM model) to the multidimensional case with a particular focus on dimension two that is useful in the estimation method of Section \ref{sec:estimation}.

\section{The small intermittency approximation}
\label{sec:SIAmultidimlogsfbm}
 The small intermittency approximation is of particular interest in calibrating the model parameters from observed data which will be the purpose of Section \ref{sec:estimation}. In fact, we will proceed similarly and see that within the small intermittency approximation in the mLog S-fBM model the generalized moments depend on the cointermitency and the co-Hurst matrices entries.

In the following, we start by deriving the cross covariance structure of the normalized integrated mS-fBM:
\begin{equation}
    \forall i\in \llbracket1,d\rrbracket,\forall I\subset K,\tab \Omega_{i}(I):=\frac{1}{\lambda_i}\int_I \left(\omega_{i}(u)-\mu_{i} \right) du
    \label{eq:Omega}
\end{equation}
useful in the small intermittency results. We present the small intermittency approximation result in dimension $d$. A special focus will be made for the case $d=2$ as it allows to calibrate the corresponding co-Hurst exponent and co-intermittency parameter, more details will be discussed in Section \ref{sec:estimation} concerning the estimation procedure.

\subsection{The cross covariance function of $\left(\Omega_{i}\right)_{i \in \llbracket 1,d \rrbracket}$}
For any $t\geq0$, we denote the integrated mS-fBM process over an interval of size $\Delta>0$ as:
\begin{equation}
\label{eq:Omega_bis}
\Omega_{i,\Delta}(t):=\Omega_{i}([t,t+\Delta]) \; .
\end{equation}
One can then derive the following result using the properties of $((\omega_{i}(t))_t,{i\in \llbracket 1,d \rrbracket})$ in Definition \ref{def:multidimsfbm} and Proposition \ref{prop:autocovariancemultidimsfbm}.
\begin{prop}[\textbf{cross covariance of $\Omega$}]
\label{prop:autocovarianceOmegamultidim}
   For any $\Delta>0$,
  $((\Omega_{i,\Delta}(t))_t,{i\in \llbracket 1,d \rrbracket})$ is a stationary gaussian field with the following covariance structure:
  
   \begin{eqnarray}
   \label{eq:covOmega}
    \forall (t,s)\in \R_{+}^2,\tab \mathrm{cov}\left(\Omega_{i,\Delta}(t),\Omega_{j,\Delta}(s) \right)=g_{i,j}\Delta^2 
    \left[\frac{1+2 H_{i,j}-2\bH_{i,j}}{2H_{i,j}(1-2 \bH_{i,j})}-\phi_{i,j}\left(\frac{\Delta}{\tau}\right) \right]
    \end{eqnarray}
     
   where 
   \begin{eqnarray}
       \begin{cases}
           \tau=|t-s| \\
           \phi_{i,j}\left(z\right)=f\left(z,1\right)\frac{2\bH_{i,j}-2H_{i,j}}{(2H_{i,j} - 1)(2\bH_{i,j} - 1)}-f\left(z,2H_{i,j}  \right)
    \frac{1}{2H_{i,j}(2H_{i,j} - 1)} 
 \\
           f(z,\alpha)=\left(\frac{\tau}{T}\right)^{\alpha}\frac{|1+z|^{\alpha +2}+|1-z|^{\alpha +2}-2}{z^2 (1+\alpha)(\alpha +2)},\alpha \in \left]0,1\right]
       \end{cases}
   \end{eqnarray}
\end{prop}
The proof is in the second section of Appendix \ref{app:multidimautocovs}.\\\\
We point out that in the high frequency regime $\left(\Delta \ll \tau \right)$, one has thanks to the second order Taylor expansion around 0 of $\phi_{i,j}$:
\begin{multline}
\mathrm{cov}\!\left(\Omega_{i,\Delta}(t),\Omega_{j,\Delta}(s) \right)
= g_{i,j}\Delta^2\Bigg(
 \frac{1+2 H_{i,j} - 2 \bar{H}_{i,j}}
                 {2 H_{i,j} \left(1 - 2 \bar{H}_{i,j}\right)}
 - \frac{ \tau^{2 H_{i,j}} (2H_{i,j}+2)(2H_{i,j}+1)}
        { 2 H_{i,j} T^{2H_{i,j}} \left[ 1 - (2H_{i,j})^2 \right] (2H_{i,j}+2)}
\\
+ \frac{\tau}{T} \cdot 
\frac{ 2 \bar{H}_{i,j} - 2 H_{i,j} }
     { (2 H_{i,j} - 1) \left( 2 \bar{H}_{i,j} - 1 \right) }
\Bigg)
+ o\!\left( \left( \frac{\Delta}{\tau} \right)^2 \right).
\end{multline}

By identifying with Eq.~\eqref{eq:autocovmultidimsfbm}, we obtain for any $(t,s)\in\R_{+}^2$: 
 \begin{eqnarray}
    \label{eq:covOmeganormalized}
        \mathrm{cov}\left(\frac{\Omega_{i,\Delta}(t)}{\Delta},\frac{\Omega_{j,\Delta}(s)}{\Delta} \right)=\frac{1}{\lambda_i\lambda_j}
      \mathrm{cov}\left(\omega_{i}(t),\omega_{j}(s)\right)+ o\left(\left(\frac{\Delta}{\tau}\right)^2\right)
    \end{eqnarray}

\subsection{General result for arbitrary $d\in \N^*$}
Herein, we denote the intermittency vector:
\begin{equation}
\label{def:Lambda_vec}
\Lambda = \begin{pmatrix}
\lambda_1 \\
\lambda_2 \\
\vdots \\
\lambda_d
\end{pmatrix} \in \R_{+}^d
\end{equation}
For any arbitrary subintervals $\left(I_i\right)_{i\in \llbracket 1,d \rrbracket}\subset K^d$ where the interval $K\subset \R_{+}$ with $|K|\leq T$, we introduce the generalized moments:
\begin{eqnarray}
\forall k\in \llbracket 1,d \rrbracket,\tab 
\begin{cases}
    C^{\ln}(I_1,...,I_k):= \E\left(\prod_{i=1}^{k} \ln\left(\frac{M_{i}(I_i)}{|I_{i}|}\right) \right)\\
    C^{\Omega}(I_1,...,I_k):= \E\left(\prod_{i=1}^{k} \frac{\Omega_{i}(I_i)}{|I_{i}|} \right)
\end{cases}
\end{eqnarray}
where $\Omega_{i}(I_i)$ are defined in Eq. \eqref{eq:Omega}. In the light of Proposition 4 in \cite{wu2022rough}, we can leverage the small intermittency approximation of the generalized moments of normalized Log S-fBM random measures over arbitrary intervals $I_1,...,I_d$ of length lower than $T$ to extend it for the generalized moments of $\left(\ln\left(\frac{M_{i}(I_i)}{|I_{i}|}\right)\right)_{i\in \llbracket 1,d \rrbracket}$, which is claimed in the following result.

\begin{theo}[\textbf{small intermittency approximation}]
\label{thm:smallintermlogmrmtheo}

 For any $\left(I_i\right)_{i\in \llbracket 1,d \rrbracket}\subset K^d$ and any $I\subset K$, one has:
     \begin{multline}
    C^{\ln}(I_1,...,I_d)-C^{\ln}(I,...,I)= 
    \sum_{\substack{m=0 }}^d\left(-1\right)^{d-m} 
    \sum_{\substack{\mathcal{I}_m=\left(i_k\right)_{k\in \llbracket 1,m \rrbracket} \\ \mathcal{I}_m \subset \mathcal{I}_d}}
    \Bigg[
    \sum_{s=1}^{m}
    \Bigg(
    \prod_{\substack{\left\{k,l\right\} \in P_s}} \lambda_{i_k}\lambda_{i_{l}}
    \mathbb{E} \left(
    \frac{\Omega_{i_k}(I_{i_k})}{|I_{i_k}|} 
    \frac{\Omega_{i_{l}}(I_{i_{l}})}{|I_{i_{l}}|}     
    \right)\\
    - 
    \prod_{\substack{\left\{k,l\right\} \in P_s}} \lambda_{i_k}\lambda_{i_{l}}
    \mathbb{E} \left(
    \frac{\Omega_{i_k}(I)}{|I|} 
    \frac{\Omega_{i_{l}}(I)}{|I|}     
    \right)
    \Bigg)
    \Bigg] 
    + o\left(\|\Lambda\|^d\right) \nonumber
\end{multline}

where $\mathcal{I}_d:=\llbracket 1,d \rrbracket$ and $P_s$ is the partition of pairs of successive integers of $\llbracket 1,s \rrbracket $ for $s \in \llbracket 1,d \rrbracket$.
         
\end{theo}
The proof is in Appendix \ref{app:thm1proof}.\\\\

\subsection{The case $d=2$}
\label{secsec:SIA2D}
The case of $d=2$ corresponds to considering two distinct marginals. It is key to estimate the model parameters from observed data. In practice, one can derive from Theorem \ref{thm:smallintermlogmrmtheo} a small intermittency approximation of the cross covariance function of the log normalized multifractal random measures. In fact, when $d=1$ one can straightforwardly see that the Log S-fBM random measures on any arbitrary $I\subset K$ are centered in small intermittencies which is due to the fact that  $\left(\Omega_{i}(I)\right)_{i\in \llbracket 1,d \rrbracket}$ are all centered for any arbitrary subinterval $I\subset \R_+$.\\
In the light of the computations of Appendix \ref{app:thm1proof} (see Remark \ref{rmk:SIAdim2simplification}), the small intermittency approximation in dimension $2$ lead to the following formulation:
     \begin{eqnarray}
     \label{eq:smallintermitformulamomentdim2}
           \E\left(\ln\left(\frac{M_{i}(I)}{|I|}\right)\ln\left(\frac{M_{j}(J)}{|J|}\right)\right)=\lambda_i\lambda_j\E\left(\frac{\Omega_{i}(I)}{|I|}\frac{\Omega_{j}(J)}{|J|}\right) +o\left(\|\Lambda_{i,j}\|^2\right) 
       \end{eqnarray}
       where: 
       \begin{eqnarray}
\Lambda_{i,j} = \begin{pmatrix}
\lambda_i \\
\lambda_j 
\end{pmatrix} \nonumber
\end{eqnarray}
Written differently, we have that:
\begin{eqnarray}
     \label{eq:smallintermitformulamomentdim2qv}
           \E\left(\ln\left(\frac{\left<X^i\right>_{I}}{|I|}\right)\ln\left(\frac{\left<X^j\right>_{J}}{|J|}\right)\right)=\lambda_i\lambda_j\E\left(\frac{\Omega_{i}(I)}{|I|}\frac{\Omega_{j}(J)}{|J|}\right) +o\left(\|\Lambda_{i,j}\|^2\right) 
       \end{eqnarray}
where $\left<.\right>_{I}$ stands for the quadratic variation over the interval $I$.
\\\\
Furthermore, the scale property of Eq.~\eqref{eq:corrmultidimsfbm_dG} can be retrieved in the small intermittency limit. We denote the normalized multifractal random measures  over a period of length $\Delta$: $\left(M^{i}_{\Delta}(t):=\frac{M_{i,\Delta}(t)}{\Delta}\right)_{t\geq 0}$ and $\left(M^{j}_{\Delta}(t):=\frac{M_{j,\Delta}(t)}{\Delta}\right)_{t\geq 0}$.\\
 We consider here the log volatility increments for two arbitrary marginals defined as:
\begin{eqnarray}
\forall k\in \{i,j\},\forall t\geq0\tab 
    \begin{cases}
        \delta_{\tau}\ln\left(M_{\Delta}^{k}\left(t\right)\right) := \ln\left( M_{\Delta}^{k}(t+\tau )\right) - \ln\left( M^{k}_{\Delta}(t)\right) \\
        \delta_{\tau}\Omega_{k,\Delta}\left(t\right):= \frac{1}{\Delta}\left(\Omega_{k,\Delta}\left(t+\tau\right) - \Omega_{k\Delta}\left(t\right)\right)
    \end{cases}
\end{eqnarray}
where $\Omega_{i,\Delta}(t)$ is defined in Eq. \eqref{eq:Omega_bis}. Using Eq.~\eqref{eq:smallintermitformulamomentdim2qv}, one has for any $t,\tau\geq 0$:
\begin{eqnarray}
    \mathrm{cov}\left(\delta_{\tau}\ln\left(M_{\Delta}^{i}\left(t\right)\right),\delta_{\tau}\ln\left(M_{\Delta}^{j}\left(t\right)\right)\right)=\lambda_i\lambda_j\mathrm{cov}\left(\delta_{\tau}\Omega_{i,\Delta}\left(t\right),\delta_{\tau}\Omega_{j,\Delta}\left(t\right)\right)+o\left(\|\Lambda_{i,j}\|^2\right) 
\end{eqnarray}
After some algebra, we have:
\begin{eqnarray}
     \mathrm{cov}\left(\delta_{\tau}\Omega_{i,\Delta}\left(t\right),\delta_{\tau}\Omega_{j,\Delta}\left(t\right)\right)=2g_{i,j}\Tilde{\phi}_{i,j}\left(\frac{\Delta}{\tau}\right)
 \end{eqnarray}
where:
\begin{eqnarray}
\label{eq:phiijtilde}
\Tilde{\phi}_{i,j}\left(z\right) = 
    \left(\frac{f\left(z,2H_{i,j}  \right)}{2H_{i,j}(2H_{i,j} - 1)} -\left(\frac{\Delta}{T}\right)^{2H_{i,j}}\frac{1}{H_{i,j}(1+H_{i,j})(4H_{i,j}^2-1)}\right)-\frac{2\bH_{i,j}-2H_{i,j}}{(2H_{i,j} - 1)(2\bH_{i,j} - 1)}\left(f\left(z,1\right)-\frac{\Delta}{6T}\right)\nonumber\\
 \end{eqnarray}
Which means that:
\begin{eqnarray}
    \mathrm{cov}\left(\delta_{\tau}\ln\left(M_{\Delta}^{i}\left(t\right)\right),\delta_{\tau}\ln\left(M_{\Delta}^{j}\left(t\right)\right)\right)=2g_{i,j}\Tilde{\phi}_{i,j}\left(\frac{\Delta}{\tau}\right)+o\left(\|\Lambda_{i,j}\|^2\right) .
\end{eqnarray}
Consequently, one can express the correlation coefficient:
\begin{eqnarray}
    \rho_{log}^{i,j}\left(\tau\right):=\frac{\mathrm{cov}\left(\delta_{\tau}\ln\left(M_{\Delta}^{i}\left(t\right)\right),\delta_{\tau}\ln\left(M_{\Delta}^{j}\left(t\right)\right)\right)}{\sqrt{\text{var}\left(\delta_{\tau}\ln\left(M_{\Delta}^{i}\left(t\right)\right)\right)}\sqrt{\text{var}\left(\delta_{\tau}\ln\left(M_{\Delta}^{j}\left(t\right)\right)\right)}}
\end{eqnarray}
and we obtain:
\begin{eqnarray}
\label{eq:scaleco-HurstSI}
    \rho_{log}^{i,j}\left(\tau\right)\underset{\|\Lambda_{i,j}\|\rightarrow 0}\sim g_{i,j}\frac{\Tilde{\phi}_{i,j}\left(\frac{\Delta}{\tau}\right)}{\sqrt{\Tilde{\phi}_{i,i}\left(\frac{\Delta}{\tau}\right)}\sqrt{\Tilde{\phi}_{j,j}\left(\frac{\Delta}{\tau}\right)}}.
\end{eqnarray}

\section{Estimation method}
\label{sec:estimation}
The estimation of the multidimensional Log S-fBM parameters leverages the generalized methods of moments (GMM) and relies on the same approach of the calibration method in the monovariate
Log S-fBM proposed by Wu \textit{et al.} in \cite{wu2022rough}.  This method consists in considering the logarithm of the measure over an interval of size $\Delta$,   $\ln(M_{H,T,\Delta})$ and identifying its (empirical) autocovariance function $\tau\mapsto \mathrm{Cov}\left(\ln\left(M_{H,T,\Delta}(t)\right),\ln\left(M_{H,T}(t+\tau)\right)\right)$ across a set of time lags derived with the one of the integrated S-fBM process $\Omega_{H,T,\Delta}$ given in closed form. We aim through the following section to generalize such estimation procedure to the multidimensional setting in order to estimate  the parameters of interest: the co-intermittency parameters and the co-Hurst matrix based on the small intermittency approximation results of subsection \ref{secsec:SIA2D}. The correlation scale $T$ is not relevant to us since, as in \cite{wu2022rough} we assume that we are in the ``high frequency regime'' where the maximum accessible time lag is smaller than $T$. As advocated in \cite{MuzyBaileBacry2013_PRE} (Figs 8 and 9) where the estimated value of 
$T$ is shown to systematically increase with the length of the available sample (see also \cite{wu2022rough}), this is consistent with observations when one consider applications to volatility fluctuations in financial market.

Since the diagonal values of parameter matrices (namely Hurst exponents and intermittency coefficients) can be directly estimated using the one dimensional GMM method proposed in \cite{wu2022rough} for each marginal component of mLog S-fBm, our concern is to develop a calibration method for the covariate parameters, namely the non-diagonal co-Hurst and co-intermittency matrix entries.  Consequently, it is sufficient to focus on the upper diagonals of $\left(g_{i,j}\right)_{1\leq i<j \leq d}$ and  $\left(H_{i,j}\right)_{1\leq i<j \leq d}$ as the matrices are symmetric and the co-intermittency coefficient can be deduced from the co-intermittency correlation via  Eq.~\eqref{eq:co-intermittencycorrelationdef}. In the limit of small intermittencies, the left hand side of Eq.~\eqref{eq:smallintermitformulamomentdim2qv} can be measured from observed data and the right hand side has the closed form formula of Eq.~\eqref{eq:covOmega}. The latter closed formula depends on the co-intermittency coefficient $g_{i,j}$ as well as the co-Hurst exponent $H_{i,j}$. As a result, this can be used to estimate  them using observed data  by minimizing the discrepancy between the left hand side term (derived from observed data) and the right had side term given via the closed formula of Eq.~\eqref{eq:covOmega}.

The following part is devoted to the theoretical guarantees of the estimation method. We start by introducing the notations of the respective covariances needed, then we state a first theorem on the convergence of the empirical covariances towards the theoretical ones and a second theorem on the weak consistency of the estimated parameters. In the sequel, we fix a scale $\Delta >0$ and consider $Q \in \N$ and $\Tau_Q:=\left(\tau_l\right)_{l\in \llbracket 1,Q \rrbracket}$ the time lag scale (such that $\Delta\Tau_Q$ is the sequence of time lags).\\
For any $k\in \Tau_Q$, we consider for any $\theta_{i,j} = \left(g_{i,j},H_{i,j},T\right) \in [-1,1]\times\left]0,\frac{1}{2}\right[\times\R_{+}$ the following autocovariance functions:
\begin{eqnarray}
\forall (i,j)\in \llbracket 1,d \rrbracket,\tab 
    \begin{cases}
        C_{i,j}^{\ln}\left(k,\theta_{i,j}\right):= \mathrm{cov}\left(\ln\left(\frac{M_{i,\Delta}(0)}{\Delta}\right),\ln\left(\frac{M_{j,\Delta}(k\Delta)}{\Delta}\right) \right)\\
        C_{i,j}^{\Omega}\left(k,\theta_{i,j}\right):= \mathrm{cov}\left(\frac{\Omega_{i,\Delta}(0)}{\Delta},\frac{\Omega_{j,\Delta}(k\Delta)}{\Delta} \right)
    \end{cases}\nonumber
\end{eqnarray}
where $\Omega_{i,\Delta}(t)$ is defined in Eq. \eqref{eq:Omega_bis}.
We introduce as well their empirical counterparts:
\begin{eqnarray}
    \begin{cases}
        \hat{C}_{i,j}^{\ln}\left(\Delta,k\Delta\right):= \frac{1}{N}\sum_{l=1}^{N-k}\left(\ln\left(\frac{M_{i,\Delta}(l\Delta)}{\Delta}\right)-m_{i,\Delta}\right)\left(\ln\left(\frac{M_{j,\Delta}((l+k)\Delta)}{\Delta}\right)-m_{j,\Delta}\right) \\
        \hat{C}_{i,j}^{\Omega}\left(\Delta,k\Delta\right):=\frac{1}{N}\sum_{l=1}^{N-k}\frac{\Omega_{i,\Delta}(l\Delta)}{\Delta}\frac{\Omega_{j,\Delta}((l+k)\Delta)}{\Delta}
    \end{cases}\nonumber
\end{eqnarray}
Where $m_{i,\Delta}=\frac{1}{N}\sum_{l=1}^{N}\ln\left(\frac{M_{i,\Delta}(l\Delta)}{\Delta}\right)$ is the empirical mean.
We also denote for any $k\in \Tau_Q$ and $\Delta>0$ the moments: 
    \begin{itemize}
        \item $\Hat{D}_{i,j}^{\Omega}\left(\Delta,k\Delta\right):=\hat{C}_{i,j}^{\Omega}\left(\Delta,k\Delta\right)-\hat{C}_{i,j}^{\Omega}\left(\Delta,0\right)$,   $D^{\Omega}\left(k,\theta_{i,j}\right) :=C^{\Omega}\left(k,\theta_{i,j}\right)-C^{\Omega}(0,\theta_{i,j})$
        \item $\Hat{D}_{i,j}^{\ln}\left(\Delta,k\Delta\right) :=\hat{C}_{i,j}^{\ln}\left(\Delta,k\Delta\right)-\hat{C}_{i,j}^{\ln}\left(\Delta,0\right)$,   $D^{\ln}\left(k,\theta_{i,j}\right) :=C^{\ln}\left(k,\theta_{i,j}\right)-C^{\ln}(0,\theta_{i,j})$
    \end{itemize}
Moreover, we consider the discrete autocovariance curves on the sequence of time lags:

\begin{equation}
\begin{array}{c@{\hspace{2cm}}c}
\begin{cases}
     \bm{\hat{C}}_{i,j}^{\ln}=\left( \hat{C}_{i,j}^{\ln}\left(\Delta,k\Delta\right),k\in\Tau_Q\right) \\
     \bm{\hat{C}}_{i,j}^{\Omega}=\left(\hat{C}_{i,j}^{\Omega}\left(\Delta,k\Delta\right),k\in\Tau_Q\right)\\
     \bm{C}^{\Omega}\left(\theta_{i,j}\right)=\left(C^{\Omega}\left(k,\theta_{i,j}\right),k\in \Tau_Q\right)
\end{cases}
& 
\begin{cases}
     \bm{\hat{D}}_{i,j}^{\ln}=\left( \hat{D}_{i,j}^{\ln}\left(\Delta,k\Delta\right),k\in\Tau_Q\right) \\
     \bm{\hat{D}}_{i,j}^{\Omega}=\left(\hat{D}_{i,j}^{\Omega}\left(\Delta,k\Delta\right),k\in\Tau_Q\right)\\
     \bm{D}^{\Omega}\left(\theta_{i,j}\right)=\left(D^{\Omega}\left(k,\theta_{i,j}\right),k\in \Tau_Q\right)
\end{cases}
\end{array}
\label{eq:autocovariancesandvectors}
\end{equation}

valued in $\mathbb{R}^Q$.\\
As announced on the beginning of the section, the following theorem concerns the convergence in probability of the empirical cross covariate quantities $\Hat{D}_{i,j}^{\Omega}$ and the small intermittency counterpart of $\Hat{D}_{i,j}^{\ln}$ defined as the almost sure small intermittency limit:
\begin{eqnarray}
       \Tilde{D}^{\ln}_{i,j}\left(\Delta,k\Delta\right):=\lim_{\|\Lambda_{i,j}\|^2 \to 0} \frac{\Hat{D}_{i,j}^{\ln}\left(\Delta,k\Delta\right)}{\lambda_i\lambda_j}  , \tab a.s
       \label{eq:aslimit}
    \end{eqnarray}
toward the theoretical counterpart $D^{\Omega}$.
\begin{theo}
\label{thm:converD}
For any $(i,j)\in \llbracket 1,d \rrbracket$, $\Delta\geq0$, $\theta_{i,j}\in[-1,1]\times\left]0,\frac{1}{4}\right[\times\R_{+} $ and $k\in \Tau_Q$, the following holds:
    \begin{enumerate}
        \item 
        \begin{eqnarray}
        \frac{N}{N-k}\Hat{D}_{i,j}^{\Omega}\left(\Delta,k\Delta\right) \xrightarrow[N \rightarrow \infty]{\Prob} D^{\Omega}\left(k,\theta_{i,j}\right)
    \end{eqnarray}
    \item 
    \begin{eqnarray}
        \frac{N}{N-k}\Tilde{D}^{\ln}_{i,j}\left(\Delta,k\Delta\right)  \xrightarrow[\substack{N \rightarrow \infty }]{\Prob} D^{\Omega}\left(k,\theta_{i,j}\right)
    \end{eqnarray}
    \end{enumerate}
    
\end{theo}
The proof is in Appendix \ref{app:proofconverD}.\\

As the co-Hurst and the co-intermittency correlation matrices are symmetric, so the calibration can be done through identifying the co-intermittency correlation and the co-Hurst exponent associated to each pair of the mLog S-fBM model. Particularly, the estimation method used here is the general method of moment (GMM) in the same spirit of \cite{bolko2020roughness} and also \cite{wu2022rough}.
Without loss of generality, the calibration based on the general method of moments can be formulated as follows. If $\left(S^i_t,i\in \llbracket 1,d \rrbracket\right)_t$ is a random field observed $N$ times at a frequency $\Delta>0$ with the corresponding theoretical cross covariance vector $\bm{C}\left(x\right)=\left(C_{i,j}\left(k,x\right),k\in \Tau_Q\right)$ where $x$ is the vector of the underlying parameters of the bidimensional model $\left(S^i_t,S^j_t\right)_t$ as well as the observed cross covariance function:
\begin{eqnarray}
     \forall k\in\Tau_Q,\tab \hat{C}_{i,j}(k)=\frac{1}{N}\sum_{l=1}^{N-k}\left(S^i_{l\Delta}-m_{i}\right)\left(S^j_{\left(l+k\right)\Delta}-m_{j}\right). \nonumber
 \end{eqnarray}
where $m_{i}=\frac{1}{N}\sum_{l=1}^{N}S^i_{l\Delta}$ is the empirical mean of the $i^{th}$ component. We denote similarly the observed cross covariance vector $ \bm{\hat{C}}_{i,j}=\left( \hat{C}_{i,j}(k),k\in\Tau_Q\right)$.\\
The calibration method broadly known as the minimum distance estimator method (MDE) (tackled in \cite{MDE} and  \cite{bolko2020roughness}) can be formulated as follows:

  \begin{eqnarray}
  \label{eq:calibrationproblemOmega}
       \eta_{i,j}^{N}=\argmin_{x\in\Xi_{i,j}} \bm{h}_{i,j}\left(x\right)^{T}W\bm{h}_{i,j}\left(x\right)
  \end{eqnarray}
where for any $(i,j)\in \llbracket 1,d \rrbracket^2$, $\Xi_{i,j}$ stands for the feasible domain of the vector of the true model parameters, $W\in \R^{Q\times Q}$ is a symmetric positive definite matrix and the map $\bm{h}_{i,j}:\Xi_{i,j} \mapsto \R^{Q}$ that represents the discrepancy between the empirical cross covariations and the theoretical ones.

If the observed process is the mS-fBM process (respectively the integrated Gaussian process of  Eq.~\eqref{eq:Omega}), $\bm{\hat{C}}$ will be the cross covariance vector whose components are of the form of Eq.~\eqref{eq:autocovmultidimsfbm} (respectively Eq.~\eqref{eq:covOmega}). Otherwise, if the observed process is the log volatility, given $\Delta$, $Q$ and $N$ one can compute the empirical autocovariance function for any $(i,j)\in \llbracket 1,d \rrbracket^2$:
\begin{eqnarray}
     \forall k\in\Tau_Q,\tab \hat{C}_{i,j}^{\ln}\left(\Delta,k\Delta\right)=\frac{1}{N}\sum_{l=1}^{N-k}\left(\ln\left(\frac{M_{i,\Delta}(l\Delta)}{\Delta}\right)-m_{i,\Delta}\right)\left(\ln\left(\frac{M_{j,\Delta}((l+k)\Delta)}{\Delta}\right)-m_{j,\Delta}\right). \nonumber
 \end{eqnarray}
 where $m_{q,\Delta}=\frac{1}{N}\sum_{l=1}^{N}\ln\left(\frac{M_{H_q,T,\Delta}(l\Delta)}{\Delta}\right),q\in \{i,j\}$.\\
In that case, the estimation of the mLog S-fBM $i^{th}$ and $j^{th}$ covariate parameters  is performed by solving this minimization problem:
 \begin{eqnarray}
 \label{eq:calibrationproblemlogvol}
       \theta_{i,j}^N\left(\Lambda_{i,j}\right)=\argmin_{x\in \Xi}\bm{h}_{i,j}^{\ln}\left(x,\Lambda_{i,j}\right)^{T}W\bm{h}_{i,j}^{\ln}\left(x,\Lambda_{i,j}\right)
  \end{eqnarray}
  Where $\Xi=[-1,1]\times\left]0,\frac{1}{2}\right[\times\R_{+}$ the support of all the covariate parameters (which does not depend on the pair $(i,j)$) and the map $\bm{h}_{i,j}^{\ln}:\Xi\times \R_{+}^2 \mapsto \R^{Q}$. This map depends on the intermittency vector in the perspective of the results that follows where the small intermittency approximation is involved.\\

  We introduce the following first order small intermittency approximations for any $\Lambda_{i,j} \in \R_{+}^2$:
  \begin{eqnarray}
      \theta_{i,j}^N\left(\Lambda_{i,j}\right)=\Hat{\theta}_{i,j}^N\left(\Lambda_{i,j}\right)+o\left(\|\Lambda_{i,j} \| ^2\right)
  \end{eqnarray}
One can derive the theoretical consistency guarantee (weak sense) of the the estimators $\eta_{i,j}^N$ and $\Hat{\theta}_{i,j}^N\left(\Lambda_{i,j}\right)$ as the number of observations $N$ is arbitrarily high.

  \begin{theo}[\textbf{weak consistency}]
\label{thm:convergenceprobcalibration}
   For two arbitrary marginals $\left(i,j\right)\in\llbracket 1,d \rrbracket^2$ and for $\bm{h}_{i,j}$ and $\bm{h}_{i,j}^{\ln}$ defined as:
\begin{eqnarray}
    \forall (x,\Lambda_{i,j})\in \Xi_{i,j}\times \left]0,\frac{1}{2}\right[^2,\tab \begin{cases}
         \bm{h}_{i,j}\left(x\right)=\bm{\hat{D}}^{\Omega}_{i,j}-\bm{D}^{\Omega}\left(x\right)\\
         \bm{h}_{i,j}^{\ln}\left(x,\Lambda_{i,j}\right)=\bm{\hat{D}}_{i,j}^{\ln}-\lambda_i\lambda_j\bm{D}^{\Omega}\left(x\right)
    \end{cases}
\end{eqnarray}
   if the true vector is $\theta_{i,j}^{*}$, then:
      \begin{enumerate}
          \item \begin{eqnarray}
    \Hat{\theta}_{i,j}^N\left(\Lambda_{i,j}\right) \xrightarrow[N\rightarrow \infty]{\Prob}\theta_{i,j}^{*}
\end{eqnarray}
\item 
\begin{eqnarray}
    \eta_{i,j}^N \xrightarrow[N\rightarrow \infty]{\Prob}\theta_{i,j}^{*}
\end{eqnarray}
      \end{enumerate}
  \end{theo}
  The proof can be found in Appendix \ref{app:proofconvergenceprobcalibration}.
\begin{remark}
Using similar arguments as in Theorem \ref{thm:convergenceprobcalibration}, one can see that the following holds:
\begin{eqnarray}
    \theta_{i,j}^N\left(\Lambda_{i,j}\right) \xrightarrow[N\rightarrow \infty]{\Prob}\theta_{i,j}
\end{eqnarray}
where $\theta_{i,j}$ is the first order approximation of the true vector of parameters $\theta_{i,j}^{*}$ in the small intermittency limit:
\begin{eqnarray}
      \theta_{i,j}^{*}=\theta_{i,j}+o\left(\|\Lambda_{i,j} \| ^2\right)
  \end{eqnarray}
\end{remark}
The consistency of the estimated parameters was also evoked in \cite{bacry2008log} in the context of the MRW model. As long as the moments are identifiable only in the small intermittency regime, the consistency guarantee of the GMM estimator is also obtained in the small intermittency regime.\\

For the numerical experiments of the next section, we choose the function $\bm{h}_{i,j}^{\ln}$ closer to the small intermittency approximation formulated in Eq.~\eqref{eq:smallintermitformulamomentdim2}: $\bm{h}_{i,j}^{\ln}\left(.,\Lambda_{i,j}\right)=\bm{\hat{C}}_{i,j}^{\ln}-\lambda_i\lambda_j\bm{C}^{\Omega}\left(.\right)$.\\
In the spirit of \cite{bolko2020roughness}, $W$ is considered a symmetric random weight matrix $W_N$ admitting a Cholesky decomposition of the form $W_N=A_N^TA_N$ where $A_N$ converges almost surely to a deterministic matrix $A$. In our experiments, we consider $W$ being the empirical covariance matrix of the errors meaning that the calibration algorithm penalizes highly correlated discrepancies between empirical cross covariations and theoretical ones.

\begin{comment}
    \\
\par\vspace{5cm}
https://projecteuclid.org/journals/annals-of-probability/volume-22/issue-4/Limit-Theorems-for-Nonlinear-Functionals-of-a-Stationary-Gaussian-Sequence/10.1214/aop/1176988503.full

https://arxiv.org/pdf/1708.03313

https://www.bauer.uh.edu/rsusmel/phd/ec1-7.pdf

https://perso.telecom-paristech.fr/sabourin/mdi720/main.pdf
\end{comment}

\section{Numerical experiments}
\label{sec:numerics}

\begin{figure}[H]
    \centering
    % ------- First Row -------
    \begin{minipage}[b]{\linewidth}
        \centering
        \includegraphics[width=13cm, height=5cm]{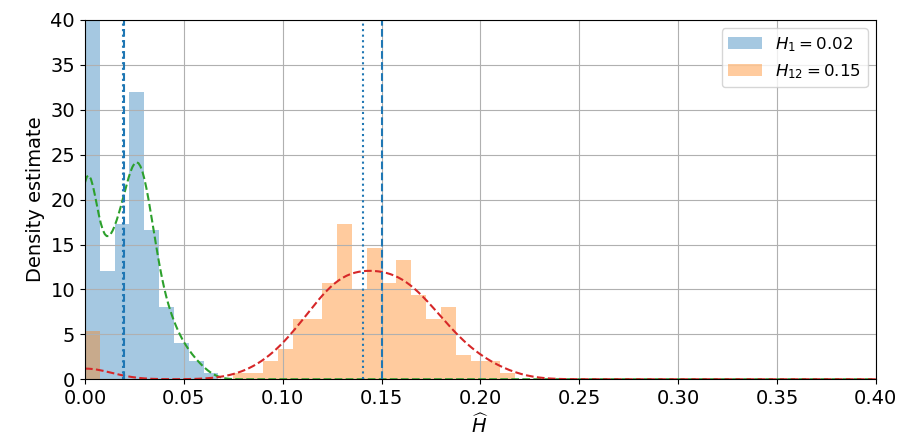}
    \begin{minipage}[b]{\linewidth}
        \centering
        \includegraphics[width=13cm, height=5cm]{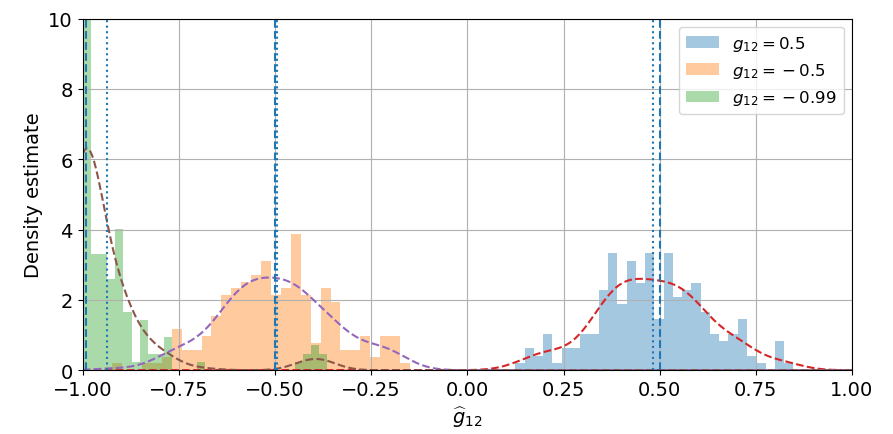}
    \end{minipage}
    \end{minipage}

    \caption{Empirical distributions of respectively $H_{i,j}$(\textbf{top}) and $g_{i,j}$(\textbf{bottom}) from a sample of $200$ copies of each obtained from synthetically generated data as well as the corresponding theoretical values. 
    Simulations with $H_1=H_2=0.02$, $\lambda_1^2=\lambda_2^2=0.05$, $T=N=2^{14}$ and subsampling length $2^4$.}
    \label{fig:histogramsg_ijHij}
\end{figure}

The objective of this section is to validate the proposed GMM estimation approach using synthetic data. Our primary focus is on the accurate retrieval of the co-intermittency and co-Hurst matrices (and therefore the co-intermittency coefficient) rather than the correlation limit $T$ that we exclude from this analysis; as discussed in Section \ref{sec:estimation} and \cite{wu2022rough}, this parameter is effectively intractable in the so-called high-frequency regime (i.e. when the observed time series length $L$ satisfies $L \ll T$) which corresponds to the practical setting of interest. For that purpose, we simulate a couple of distincts marginals as explained in Appendix \ref{app:simulationmarginalsmlogsfbm}, we calibrate first each marginal using the GMM method related to the Log S-fBM model described in Wu \textit{et al.} in \cite{wu2022rough} in order to estimate the corresponding Hurst exponents and intermittency coefficients. Then, we apply the GMM associated to the multidimensional version described in Section \ref{sec:estimation} to retrieve the calibrated co-Hurst exponent and co-intermittency correlation. This latter 
is constrained to be in $[-1,1]$ by the mean of a hyperbolic tangent parametrization. 
Finally, using  Eq.~\eqref{eq:co-intermittencycorrelationdef} we retrieve the co-intermittency coefficient. We implement the estimation procedure described in Section \ref{sec:estimation}  with $ Q = 19 $ and $ \{\tau_k\}_{k=0,\ldots,19} = \lfloor \sqrt{2^k} \rfloor $. Our implementation is similar to the approach employed for the one-dimensional case in \cite{wu2022rough} and thoroughly detailed in \cite{bolko2020roughness}. The matrix error covariance matrix $W$ of Eq.~\eqref{eq:calibrationproblemlogvol} (and also Eq.~\eqref{eq:calibrationproblemOmega} )  is estimated using the Newey-West HAC type estimator with a lag $ N^{1/3} $, while the initial values are selected arbitrarily but aligned closely with the standard values reported in the empirical findings of \cite{wu2022rough}. The calibration employs the L-BFGS-B  minimization algorithm, as implemented in the \texttt{scipy.optimize} library in Python. Similar results are observed when using alternative optimization algorithms such as SL-SQP or Nelder-Meald.

Herein, we evaluate the estimation robustness of the model parameters for various examples  of a two dimensional  mS-fBM process with different values of the co-intermittency coefficient and the co-Hurst exponent respecting the constraints of both Eq.~\eqref{eq:upperboundco-intermittencycorr} and  Eq.~\eqref{eq:lowerboundco-Hurst} and re-estimate them using the synthetic sample paths. For that purpose, we conduct two numerical experiments: the first one consists in simulating a two dimensional mS-fBM process with $g_{1,2}=0.5$, $H_1=H_2=0.02$, $\lambda_1^2=\lambda_2^2=0.05$, $T=N=2^{14}$ considering respectively $H_{1,2}=0.02$ and $H_{1,2}=0.15$ and re-estimate the co-Hurst exponents. On top of Figure \ref{fig:histogramsg_ijHij}, we plot the corresponding empirical distributions of the estimated co-Hurst exponents from a sample of $200$ estimations. The second experiment consists in simulating a two dimensional mS-fBM process with $H_{1,2}=0.15$, $H_1=H_2=0.02$, $\lambda_1^2=\lambda_2^2=0.05$, $T=N=2^{14}$ considering respectively this time $g_{1,2}=0.5$, $g_{1,2}=-0.5$ and $g_{1,2}=-0.99$ and re-estimate the co-intermittency correlations. On the bottom of Figure \ref{fig:histogramsg_ijHij}, we plot the corresponding empirical distributions of the estimated co-intermittency correlations from a sample of $200$ estimations. We observe that in each case, the retrieval of the estimated parameter is unbiased. Concerning the empirical distribution of the estimators, one can see that it is centered around the theoretical value of $H_{i,j}$ and $g_{i,j}$ considered at each case. 

\begin{figure}[H]
    \centering

    \begin{minipage}[b]{\textwidth}
        \centering
        \includegraphics[width=\linewidth]{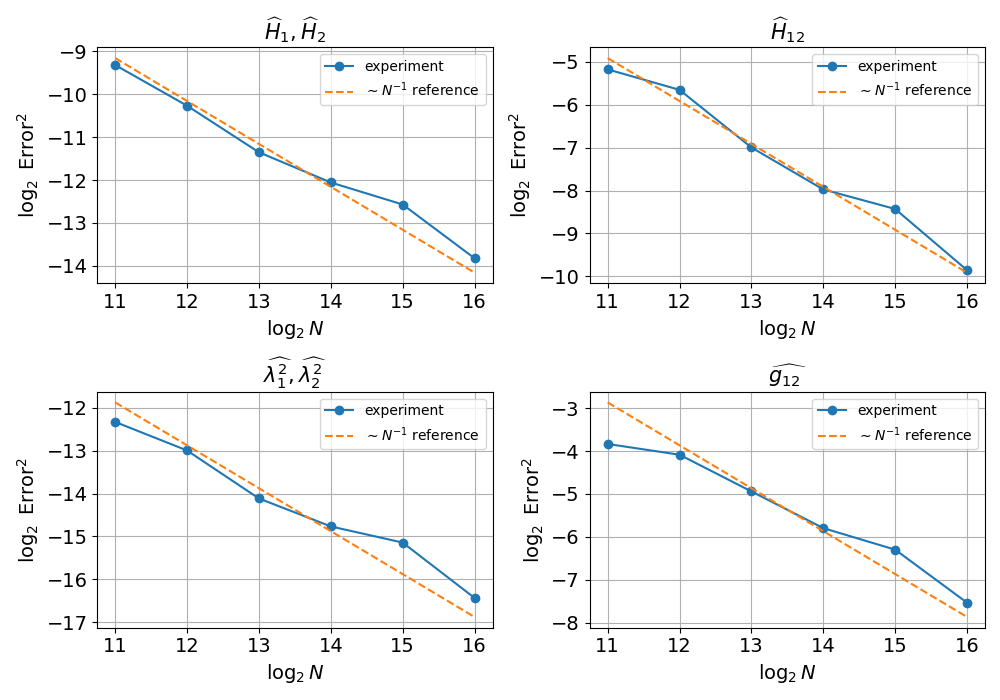}
    \end{minipage}
    \caption{The log standard deviation of the calibrated model parameters for $H_1=H_2=0.02$, $H_{1,2}=0.15$, $T=2^{14}$ and $g_{1,2}=0.5$ as a function of log length. Here, $\lambda_1^2=\lambda_2^2=0.05$. In \textbf{blue} and \textbf{orange} the respective linear fits.}
    \label{fig:logdeltaHxi}
\end{figure}

One further concern that one may be interested in is to quantify the error of the co-Hurst exponent and co-intermittency correlation estimations. In Figure \ref{fig:logdeltaHxi}, we represent those two quantities calculated using a sample of size $200$ as a function of $N$.

In Figure \ref{fig:logdeltaHxi}, we can see a good linear fit of the blue lines (log error of estimated parameters) with the orange line. The linear fit is a fortiori very pronounced which suggests that for arbitrary $(i,j)\in \llbracket 1,d \rrbracket ^2$ errors have a dependency to $N$ of the form $\frac{a}{N^{\frac{1}{2}}}$ where $a\in \R$, which is very similar to the one dimensional case investigated in Figure 6 in \cite{wu2022rough} and also showcased in the upper right and lower left graphs of Figure \ref{fig:logdeltaHxi}.
\section{Application to finance: the mLog S-fBM as a multivariate stochastic volatility model}
\label{sec:empirics}
This section is devoted to application of mLog S-fBM model to empirical stock data. First, 
we analyze the behaviour of the log volatility of an index that is built from the individual components of a mLog S-fBM model. Under mild conditions, we notably show that Hurst exponent associated with this log-volatility is predominantly determined by the  co-Hurst and co-intermittency coefficients. In a second part, we calibrate the mLog S-fBM on S\&P 500 daily market data over 3 decades and discuss the estimated values of co-Hurst as well as co-intermittency matrices. 

\subsection{The scaling of the log-volatility associated with an index of mLog S-fBM components}
\label{sec:linkwithLogSfbmNFM}

The co-Hurst and co-intermittency matrices are deeply involved when it comes to evaluate the Hurst exponent of log-volatility of the index constructed from a basket of individual stocks that are considered to be the components of  mLog S-fBM process. Given a tuple $\left(\alpha_i \right)_{i\in \llbracket 1,d \rrbracket}\in \R_{+}^d$, let us consider an index $X$ obtained as a linear combination of single stocks $X^{i}$:
\begin{eqnarray}
\label{eq:index}
X_t:=\sum_{i=1}^d \alpha_i X^{i}_t
\end{eqnarray}
where $\left(X^{i} \right)_{i\in \llbracket 1,d \rrbracket}$ is a mLog S-fBM  process characterized by a co-Hurst matrix $\bm{H}$ and co-intermittency matrix $\bm{\xi}$. As in Eq. \eqref{def:Lambda_vec}, $\Lambda$ denotes 
the vector of intermittency coefficients of each component. For any time scale $\Delta >0$, we can define $M_\Delta(t)$, the index realized volatility measure over the interval $I = [t,t+\Delta]$  as the quadratic variation over $I$ of the process defined in Eq.~\eqref{eq:index}:
\begin{eqnarray}
\label{eq:MRMaggregindex}
      M_\Delta(t):=\sum_{i=1}^d \alpha_i^2 M_{i}([t,t+\Delta]) \; .
\end{eqnarray}
For any $\tau>0$, let $V\left(\tau,\Delta \right)$ denote 
the variance of $\ln\left( M_{\Delta}(\tau )\right) - \ln\left( M_{\Delta}(0)\right)$ which represents the log volatility increment of the index process $X$ at scale $\Delta$.
If one defines  $\mathcal{V}(\tau,\Delta)$ as the small intermittency approximation 
counterpart of $V(\tau,\Delta)$ defined as: 
\begin{eqnarray}
\label{eq:variancesumsmallintermiteq}
V\left(\tau,\Delta,\Lambda \right) & = & \mathcal{V}\left(\tau,\Delta,\Lambda\right)+ o\left(\|\Lambda \| ^2\right) , 
\end{eqnarray}
then, one can establish the following approximation result:
\begin{prop}
\label{theo:variancesumsmallintermit}
For any $0 < (\tau,\Delta) < T$, the small intermittency approximation of the variance of the index $X$ log-volatility increments can be expressed as follows:
\begin{eqnarray}
\label{eq:varianceindexSI}
    \mathcal{V}\left(\tau,\Delta \right)=\sum_{1\leq i,j \leq d}  \alpha_i^2\alpha_j^2 \xi_{i,j}\left(\frac{\tau}{T}\right)^{2H_{i,j}}\Tilde{\phi}_{i,j}\left(\frac{\Delta}{\tau}\right).
\end{eqnarray}
where $\Tilde{\phi}_{i,j}$ is given in Eq.~\eqref{eq:phiijtilde}.
\end{prop}
The proof of this results can be found in Appendix \ref{app:proofoftheovariancesumsmallintermit}.

Therefore, if one assumes that $H_{i,j} = H$ while $H_i = H'$, $\xi_{i,j} = \lambda_{i}^2 = \lambda^2$, $\alpha_i = d^{-1}$, in the small intermittency approximation, considering the variance of index log-volatility increments can be decomposed as:
\begin{comment}
     \begin{eqnarray}
\label{eq:varianceindexSIdecomposition}
    \mathcal{V}\left(\tau,\Delta\right) \propto \left(\frac{\tau}{T}\right)^{2H} +  d^{-1} \left(\frac{\tau}{T}\right)^{2H'}  
\end{eqnarray}
\end{comment}

\begin{eqnarray}
\label{eq:varianceindexSIdecomposition}
    \mathcal{V}\left(\tau,\Delta,\Lambda\right)= \frac{\lambda^2\left(d^2-d\right)}{d^4} \left(\frac{\tau}{T}\right)^{2H}\Tilde{\phi}\left(\frac{\Delta}{\tau}\right)+\frac{\lambda^2}{d^3}\left(\frac{\tau}{T}\right)^{2H^{'}}\Tilde{\phi}^{'}\left(\frac{\Delta}{\tau}\right).
\end{eqnarray}
where:
\begin{eqnarray}
\begin{cases}
    \Tilde{\phi}\left(z\right) = 
    \frac{f\left(z,2H \right)}{2H(2H - 1)} -\left(\frac{\Delta}{T}\right)^{2H}\frac{1}{H(1+H)(4H^2-1)}-\frac{2H^{'}-2H}{(2H - 1)(2H^{'} - 1)}\left(f\left(z,1\right)-\frac{\Delta}{6T}\right)\\
    
    \Tilde{\phi}^{'}\left(z\right) = 
    \frac{f\left(z,2H^{'}  \right)}{2H^{'}(2H^{'} - 1)} -\left(\frac{\Delta}{T}\right)^{2H^{'}}\frac{1}{H^{'}(1+H^{'})(4H^{{'}^2}-1)}
\end{cases}
\end{eqnarray}
The ratio of the first term over the second term simplifies to $(d-1)
\left(\frac{\tau}{T}\right)^{2(H-H')}
\frac{
\tilde{\phi}\!\left(\frac{\Delta}{\tau}\right)
}{
\tilde{\phi}'\!\left(\frac{\Delta}{\tau}\right)
}$. One can notice that:
\begin{equation}
 (d-1)
\left(\frac{\tau}{T}\right)^{2(H-H')}
\frac{
\tilde{\phi}\!\left(\frac{\Delta}{\tau}\right)
}{
\tilde{\phi}'\!\left(\frac{\Delta}{\tau}\right)
}  \geq  (d-1) \left(\frac{\tau}{T}\right)^{2(H-H')}   \frac{1}{r_{H^{'},H}\left(\frac{\Delta}{\tau}\right)}
\end{equation}

where 
\begin{equation}
\label{eq:def_r}
   r_{H^{'},H}(z) = \frac{ \frac{f\left(z,2H^{'}  \right)}{2H^{'}(2H^{'} - 1)} -\left(\frac{\Delta}{T}\right)^{2H^{'}}\frac{1}{H^{'}(1+H^{'})(4H{{'}^2}-1)}}{\frac{f\left(z,2H \right)}{2H(2H - 1)} -\left(\frac{\Delta}{T}\right)^{2H}\frac{1}{H(1+H)(4H^2-1)}} \; .
\end{equation}
Conducting the same computations as in the setting of Section 4.3 in \cite{LogSfbmNFM}, one has in the limit $H^{'}\rightarrow 0$:
\begin{equation}
 (d-1)
\left(\frac{\tau}{T}\right)^{2(H-H')}
\frac{
\tilde{\phi}\!\left(\frac{\Delta}{\tau}\right)
}{
\tilde{\phi}'\!\left(\frac{\Delta}{\tau}\right)
}  \geq  (d-1) \left(\frac{\tau}{T}\right)^{2H}   C_H\left(3-2 \ln\left(\frac{\Delta}{\tau}\right)\right)
\end{equation}
with $C_H = \frac{H(1-2H)(1+2H)(1+H)}{2\left(2^{2H}-1\right)}$. In the configuration where $H\approx 0.15$ (see Section \ref{sec:empirics}) and $\frac{\Delta}{\tau}=\frac{1}{5}$ for instance, the lower bound of the ratio is $\approx 2d$.

This is another way to rationalize the findings of \cite{wu2022rough}: when $d$ is large, despite single stock volatilities are characterized by a multifractal behavior with $H' \approx 0$, the  scaling behavior of the index volatility is dominated by the value of the co-Hurst exponents $H \approx 0.15$ (see the next Section \ref{sec:empirics}). It is noteworthy that a very similar approach has already been proposed by Zarhali \textit{ et al.} \cite{LogSfbmNFM}. In this work, the authors considered a ``Nested Factor Model'' (N-SfFM) to account for nonlinear correlations in equity markets. In this framework, returns are modeled through a standard factor structure, whereas the log-volatilities of factors and residuals admit an additional factor decomposition. A leading volatility mode jointly drives market factors, sector factors and idiosyncratic residuals. 
The equation (36) in \cite{LogSfbmNFM} is the strict equivalent of our previous Eq. \eqref{eq:varianceindexSIdecomposition}. In this equation, the first term corresponds to the contribution of the "factor" where we can find the interaction of all single stock pairs through the co-intermittency and co-Hurst upper diagonals whereas in the second term, we find only the contribution of the single stocks which reminds of the "residual" part.

\subsection{Empirical study using the S\&P500 market data}
\label{sec:empirics}
As mentioned earlier, the mLog S-fBM model can be seen from a financial modeling perspective as a multivariate stochastic volatility model of financial assets. In fact, a basket of assets dynamic can be described through the multivariate log price or return dynamic of Eq.~\eqref{eq:mlogsfbm} where the cross covariation of the assets is encoded through the multivariate log volatility process being a mS-fBM process.

In the following, we are interested in applying the estimation method developed in Section \ref{sec:estimation} in the calibration of the mLog S-fBM model to stock market data. The application to one dimensional Log S-fBM model to describe empirical data has been consider in  \cite{wu2022rough}. By analyzing a large panel of individual stocks from YahooFinance databases, the authors found values of Hurst parameters distributed around $H=0$ while the intermittency parameters spread around $\lambda^2 = 0.06$. This latter observation means that the small intermittency approximation used in \cite{wu2022rough} as well as in Section \ref{sec:multidimlogsfbm} is relevant in practice. For the purpose of this study, we collected historical daily Open, High, Low and Close quote time series for 105 assets of the S\&P 500 stocks at random from Yahoo Finance, covering the period from January 1, 1990, to January 1, 2022 and run our estimation procedure. Using these data, we are able to compute the Garman-Klass (GK) daily volatility estimator, defined as:
\begin{equation}
    \hat{\sigma}_{GK}^2 = \frac{1}{2} \ln\left(\frac{High}{Low}\right)^2 - \left(2\log(2)-1\right) \ln\left(\frac{Close}{Open}\right)^2,
\end{equation}
where $High$, $Low$, $Open$ and $Close$ denote the High, Low, Open and Close prices for a particular trading day. The Garman-Klass estimator will be used as a daily log-volatility proxy in the calibration procedure explained in the previous section.

\begin{figure}[H]
    \centering
    \includegraphics[width=18.5cm,height=6cm]{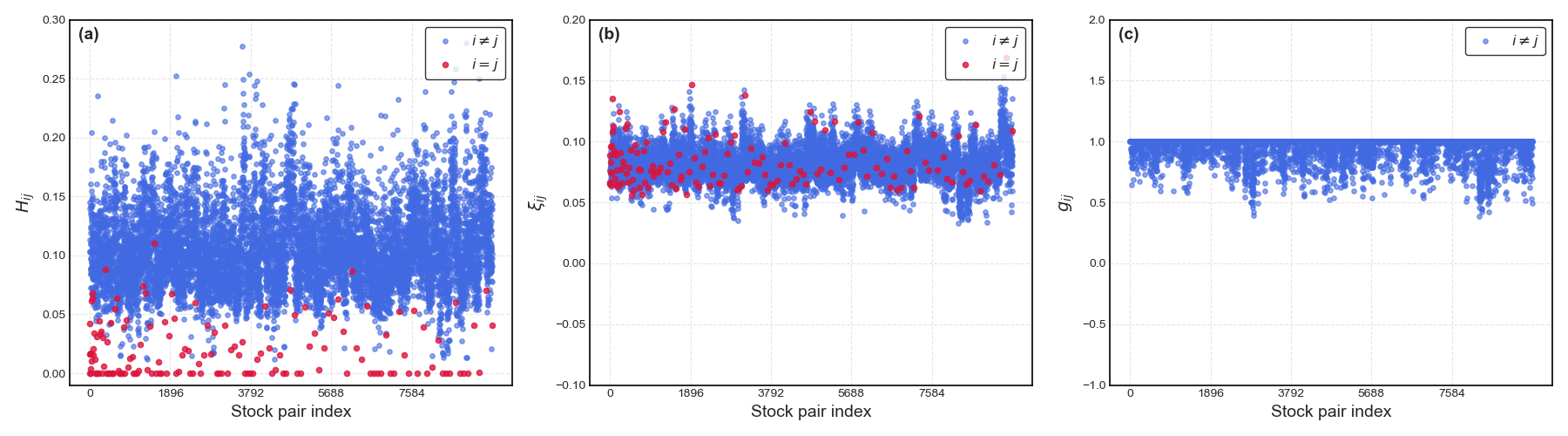}
   
    \caption{Scatter plots for all the selected assets of the S\&P500 of: the Hursts and co-Hurst exponents \textbf{(a)}, the intermittency and co-intermittency  coefficients \textbf{(b)} and the co-intermittency correlation coefficients \textbf{(c)}.} 
    \label{fig:Hassets}
\end{figure}

\begin{figure}[H]
	\centering
	\includegraphics[width=1.0\textwidth]{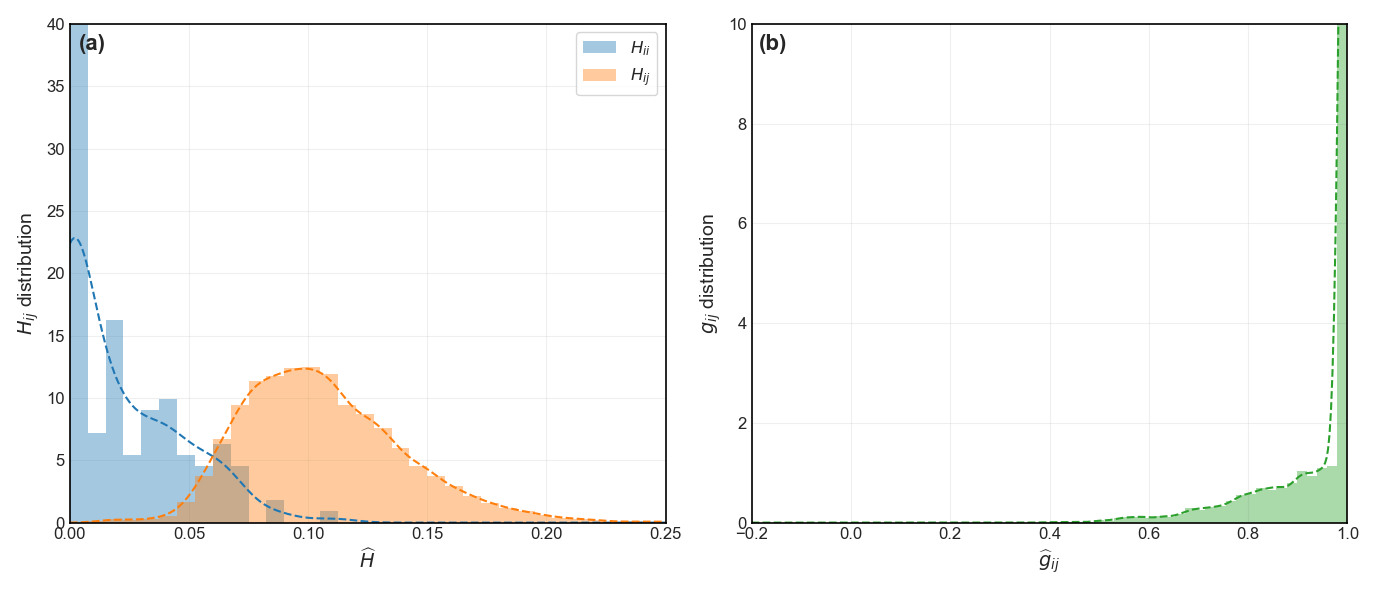}
	\caption{Histograms of empirical values of ${\widehat H}_{i,j}$ and ${\widehat g}_{i,j}$ from Fig. \ref{fig:Hassets}}
	\label{fig:emp_histo}
    %\vspace*{-1cm}
\end{figure}

All our empirical results are reported in Figures \ref{fig:Hassets} and \ref{fig:emp_histo}.
In Fig. \ref{fig:Hassets}\textbf{(a)}, are reported, as a scatter plot,
all the estimated Hurst exponent for all assets as well as the estimated co-Hurst exponents for all asset pairs. We clearly see that the single stocks of the S\&P500 index have globally a Hurst exponent close to $0$, which is coherent with reported empirical results of \cite{wu2022rough}. Furthermore, the co-Hurst exponent are spread around $\simeq 0.12$. These results can also be observed in Fig. \ref{fig:emp_histo}\textbf{(a)} where are displayed the histogram of stock log-volatily Hurst exponent (blue) which is peaked at $H = 0$ and the histogram of pair of stock log-volatility co-Hurst exponents (orange) which is centered around $H=0.12$. This value is close to the Hurst exponent of the S\&P500 log-volatility as reported in many works such as \cite{gatheral2018volatility} and \cite{LogSfbmNFM}. The results of Section
\ref{sec:linkwithLogSfbmNFM}, notably Eq. \eqref{eq:varianceindexSIdecomposition} allows us to interpret this latter results: under reasonable hypothesis, the apparent Hurst exponent of the index log-volatility corresponds to the values of the co-Hurst exponent governing the behavior of stock log-volatility correlations across scales.

The scatter plot of Fig. \ref{fig:Hassets}\textbf{(b)} reveals that the
intermittency coefficients of single stocks are close to $\lambda^2 = 0.06$
in agreement with the estimations reported in \cite{wu2022rough}. One can also observe that the co-intermittency coefficients of all asset pairs are of the same order of magnitude as intermittency coefficients. In Fig \ref{fig:Hassets}\textbf{(b)}, where are displayed the estimated values ${\hat g}_{i,j}$ of all pairs of stocks, one observes that all values are very close to $1$. This is confirmed by the shape histogram of ${\hat g}_{ij}$ in Fig. 
\ref{fig:emp_histo}\textbf{(b)} which is strongly peaked at $g_{ij} = 1$.
These observation indicate that, at very large scale, there could be a single source of volatility randomness while as the scale decreases, idiosyncratic volatility fluctuations become more and more important for each asset. In the light of Eq.~\eqref{eq:covmultidimsfbm_dG}, our empirical results suggest that the correlation coefficients between two asset log volatilities scales in terms of the co-Hurst exponent in the following way: asset log volatilities have a correlation coefficient of order $\simeq 1$ in high scales and get decorrelated in small scales with a decorrelation speed being to the power $\simeq 0.11$ in this context. Consequently, a simpler correlation structure of the driving gaussian noise vector $\left(dG_i\right)_{i\in \llbracket 1,d \rrbracket}$ (see Eq.~\eqref{eq:corrmultidimsfbm_dG} and Definition \ref{def:multidimsfbm}) being inline with the empirical results would be:
\begin{equation}
\label{eq:corrmultidimsfbm_dGempiricalmodel}
    \forall (i,j)\in \llbracket 1,d \rrbracket^2,\tab \rho_{G}^{i,j}\left(h\right):= \begin{cases}
        \left(\frac{h}{T}\right)^{2H_{i,j}}, h\leq T\\
        1, h>T
    \end{cases},
\end{equation}
where for any $(i,j)\in \llbracket 1,d \rrbracket^2$, $H_{i,j}\simeq 0.11$ .

\section{Conclusion and prospects}

In this work, we have introduced the mLog S-fBM as a natural and flexible extension of the Log S-fBM model proposed by Wu \textit{et al.}~\cite{wu2022rough}. The key novelty of the mLog S-fBM framework lies in its ability to jointly model multiple log volatility processes whose dependence structure is governed by two components: a co-intermittency matrix controlling nonlinear dependence and a co-Hurst matrix governing joint roughness across assets. This construction preserves the marginal properties of the classical Log S-fBM, allowing each component to recover both standard rough and multifractal volatility dynamics, while simultaneously introducing a rich and interpretable cross dependence structure. As such, the proposed model provides a unified setting for describing multivariate rough and multifractal volatility phenomena.\\

A central theoretical result of this study is the ability of the mLog S-fBM to encompass both the rough volatility regime and the multifractal regime within a single modeling framework. In particular, we have shown that, as the entries of the co-Hurst matrix tend to zero, the associated mLog S-fBM random measure converges weakly to a multidimensional multifractal random measure (mMRM). This property mirrors the behavior previously established in the one dimensional setting by Wu \textit{et al.}~\cite{wu2022rough} and confirms the internal consistency of the multidimensional generalization. From a modeling perspective, this convergence result highlights the versatility of the mLog S-fBM, which can smoothly interpolate between rough and multifractal dynamics depending on the co-Hurst regime, thereby offering a broader descriptive power than models restricted to a single scaling paradigm.

The analysis of the cross covariance structure further illustrates the relevance of the model for capturing empirically observed  dependence patterns. In particular, the derived expressions reveal a scale dependent behavior of both the correlation coefficients and the increments of log volatilities under the small intermittency regime. This behavior is reminiscent of the scaling properties of the \(p\)-th order moments of fractional Brownian motion reported by Gatheral \textit{et al.}, with the important distinction that, in the present framework, the scale dependence is governed by powers of the co-Hurst exponents rather than by the classical Hurst exponent. As a consequence, the model predicts that any pair of log volatility marginals \(i\) and \(j\) becomes asymptotically decorrelated at small time scales while remaining correlated at larger scales, with a decorrelation speed controlled by the corresponding co-Hurst entry \(H_{i,j}\). This multiscale correlation structure provides a natural explanation for the heterogeneous dependence patterns observed across financial assets and time horizons. This point will be specifically analyzed at an empirical level in a forthcoming study.

From a statistical standpoint, the multidimensional small intermittency approximation plays a crucial role in making the model tractable for estimation. By exploiting this approximation, we have developed a generalized method of moments (GMM) procedure that allows for the joint estimation of the co-intermittency coefficients, the co-intermittency correlation matrix and the co-Hurst exponents. Extensive numerical experiments conducted on synthetic data with fixed parameters demonstrate that the proposed estimation strategy is both stable and accurate, successfully recovering the underlying model parameters within reasonable accuracy. These results validate the practical feasibility of the mLog S-fBM framework and provide evidence that the theoretical structure of the model can be effectively leveraged in empirical applications.

The application of the model to U.S.\ equity market data further highlights its empirical relevance. In particular, the estimated upper diagonal entries of the co-Hurst matrix associated with S\&P500 assets are found to be of order \( \simeq 0.12 \), which is remarkably close to the Hurst exponent typically reported for the S\&P~500 index itself. In addition, the estimated nonlinear co-intermittency correlation matrix exhibits entries of order \( \simeq 1 \), indicating a strong level of non linear cross dependence in the volatility dynamics. These findings suggest that the mLog S-fBM captures salient features of multivariate financial volatility, including both roughness and nonlinear dependence, in a parsimonious and interpretable manner.

\vfill
\textbf{Acknowledgment:} This work has received support from the French government, managed by the National Research Agency (ANR), under the "France 2030" program with reference "ANR-23-IACL-0008.

\newpage
%\bibliography{biblio}
%\bibliographystyle{unsrt} 
%\bibliographystyle{elsarticle-num} 

%\bibliography{bibliography} 
%\nocite{*}

\nocite{*}
\bibliographystyle{plain}
\bibliography{biblio}

@article{allez2011individual,
  title={Individual and collective stock dynamics: intra-day seasonalities},
  author={Allez, Romain and Bouchaud, Jean-Philippe},
  journal={New Journal of Physics},
  volume={13},
  number={2},
  pages={025010},
  year={2011},
  publisher={IOP Publishing}
}

@article{cizeau2001correlation,
  title={Correlation structure of extreme stock returns},
  author={Cizeau, Pierre and Potters, Marc and Bouchaud, Jean-Philippe},
  year={2001},
  publisher={Taylor \& Francis}
}

@article{lillo2000symmetry,
  title={Symmetry alteration of ensemble return distribution in crash and rally days of financial markets},
  author={Lillo, Fabrizio and Mantegna, Rosario N},
  journal={The European Physical Journal B-Condensed Matter and Complex Systems},
  volume={15},
  pages={603--606},
  year={2000},
  publisher={Springer}
}

@article{prokhorov1956convergence,
  author = {Prokhorov, Yuri V.},
  title = {Convergence of Random Processes and Limit Theorems in Probability Theory},
  journal = {Theory of Probability \& Its Applications},
  year = {1956},
  volume = {1},
  number = {2},
  pages = {157--214}
}

@article{bibinger2025,
    title={Modeling and Forecasting Realized Volatility with Multivariate Fractional Brownian Motion}, 
    author={Markus Bibinger and Jun Yu and Chen Zhang},
    year={2025},
    eprint={2504.15985},
    archivePrefix={arXiv}
}

@book{billingsley1999convergence,
  author = {Billingsley, Patrick},
  title = {Convergence of Probability Measures},
  publisher = {Wiley},
  year = {1999},
  edition = {2nd}
}

@article{aldous1978stopping,
  author = {Aldous, David},
  title = {Stopping Times and Tightness},
  journal = {Annals of Probability},
  year = {1978},
  volume = {6},
  number = {2},
  pages = {335--340}
}

@article{mitoma1983tightness,
  author = {Mitoma, Itaru},
  title = {Tightness of Probabilities on \( C([0,1]; \mathcal{S}') \) and \( D([0,1]; \mathcal{S}') \)},
  journal = {Annals of Probability},
  year = {1983},
  volume = {11},
  number = {4},
  pages = {989--999}
}

@article{bennedsen2022decoupling,
  title={Decoupling the short- and long-term behavior of stochastic volatility},
  author={Bennedsen, Mikkel and Lunde, Asger and Pakkanen, Mikko S.},
  journal={Journal of Financial Econometrics},
  volume={20},
  number={5},
  pages={961--1006},
  year={2022},
  publisher={Oxford University Press}
}

@article{bayer2016pricing,
  title={Pricing under rough volatility},
  author={Bayer, Christian and Friz, Peter and Gatheral, Jim},
  journal={Quantitative Finance},
  volume={16},
  number={6},
  pages={887--904},
  year={2016},
  publisher={Taylor \& Francis}
}

@article{fukasawa2021volatility,
  title={Volatility has to be rough},
  author={Fukasawa, Masaaki},
  journal={Quantitative finance},
  volume={21},
  number={1},
  pages={1--8},
  year={2021},
  publisher={Taylor \& Francis}
}

@article{ElEuchRosenbaum2019,
  author  = {El Euch, Omar and Rosenbaum, Mathieu},
  title   = {The Characteristic Function of Rough Heston Models},
  journal = {Mathematical Finance},
  volume  = {29},
  number  = {1},
  pages   = {3--38},
  year    = {2019},
  doi     = {10.1111/mafi.12173}
}

@article{livieri2018rough,
  title={Rough volatility: evidence from option prices},
  author={Livieri, Giulia and Mouti, Saad and Pallavicini, Andrea and Rosenbaum, Mathieu},
  journal={IISE transactions},
  volume={50},
  number={9},
  pages={767--776},
  year={2018},
  publisher={Taylor \& Francis}
}

@article{bolko2020roughness,
  title={Roughness in spot variance? A GMM approach for estimation of fractional log-normal stochastic volatility models using realized measures},
  author={Bolko, Anine Eg and Christensen, Kim and Veliyev, Bezirgen and Pakkanen, Mikko},
  journal={A GMM Approach for Estimation of Fractional Log-Normal Stochastic Volatility Models Using Realized Measures (October 9, 2020)},
  year={2020}
}

@article{fukasawa2019volatility,
  title={Is volatility rough?},
  author={Fukasawa, Masaaki and Takabatake, Tetsuya and Westphal, Rebecca},
  journal={arXiv preprint arXiv:1905.04852},
  year={2019}
}

@article{ghashghaie1996turbulent,
  title={Turbulent cascades in foreign exchange markets},
  author={Ghashghaie, Shoaleh and Breymann, Wolfgang and Peinke, Joachim and Talkner, Peter and Dodge, Yadollah},
  journal={Nature},
  volume={381},
  number={6585},
  pages={767--770},
  year={1996},
  publisher={Nature Publishing Group UK London}
}

@article{mandelbrot1997multifractal,
  title={A multifractal model of asset returns},
  author={Mandelbrot, Benoit B and Fisher, Adlai J and Calvet, Laurent E},
  year={1997},
  publisher={Cowles Foundation discussion paper}
}

@article{forde2022riemann,
  title={The Riemann--Liouville field and its GMC as H→ 0, and skew flattening for the rough Bergomi model},
  author={Forde, Martin and Fukasawa, Masaaki and Gerhold, Stefan and Smith, Benjamin},
  journal={Statistics \& Probability Letters},
  volume={181},
  pages={109265},
  year={2022},
  publisher={Elsevier}
}

@article{forde2020rough,
  title={The Rough Bergomi model as H→ 0--skew flattening/blow up and non-Gaussian rough volatility},
  author={Forde, Martin and Fukasawa, Masaaki and Gerhold, Stefan and Smith, Benjamin},
  journal={preprint},
  year={2020}
}

@article{fyodorov2016fractional,
  title={Fractional Brownian motion with Hurst index H=0 and the Gaussian unitary ensemble},
  author={Fyodorov, YV and Khoruzhenko, BA and Simm, NJ3531684},
  year={2016}
}

@article{hager2022multiplicative,
  title={The multiplicative chaos of H= 0 fractional Brownian fields},
  author={Hager, Paul and Neuman, Eyal},
  journal={The Annals of Applied Probability},
  volume={32},
  number={3},
  pages={2139--2179},
  year={2022},
  publisher={Institute of Mathematical Statistics}
}

@article{bacry2003log,
  title={Log-infinitely divisible multifractal processes},
  author={Bacry, Emmanuel and Muzy, Jean Fran{\c{c}}ois},
  journal={Communications in Mathematical Physics},
  volume={236},
  pages={449--475},
  year={2003},
  publisher={Springer}
}

@article{muzy2002multifractal,
  title={Multifractal stationary random measures and multifractal random walks with log infinitely divisible scaling laws},
  author={Muzy, Jean-Fran{\c{c}}ois and Bacry, Emmanuel},
  journal={Physical review E},
  volume={66},
  number={5},
  pages={056121},
  year={2002},
  publisher={APS}
}

@article{muzy2000modelling,
  title={Modelling fluctuations of financial time series: from cascade process to stochastic volatility model},
  author={Muzy, Jean-Fran{\c{c}}ois and Delour, Jean and Bacry, Emmanuel},
  journal={The European Physical Journal B},
  volume={17},
  number={4},
  pages={537--548},
  year={2000},
  publisher={Springer}
}

@article{bacry2001multifractal,
  title={Multifractal random walk},
  author={Bacry, Emmanuel and Delour, Jean and Muzy, Jean-Fran{\c{c}}ois},
  journal={Physical Review E},
  volume={64},
  number={2},
  pages={026103},
  year={2001},
  publisher={APS}
}

@article{kahane1985chaos,
  author={Kahane, Jean-Pierre},
  title={Sur le chaos multiplicatif},
  journal={Annales scientifiques de l’École Normale Supérieure},
  volume={18},
  pages={171--246},
  year={1985}
}

@article{RhodesVargas2014,
author = {R{\'e}mi Rhodes and Vincent Vargas},
title = {{Gaussian multiplicative chaos and applications: A review}},
volume = {11},
journal = {Probability Surveys},
number = {none},
publisher = {Institute of Mathematical Statistics and Bernoulli Society},
pages = {315 -- 392},
year = {2014},
doi = {10.1214/13-PS218}
}

@article{Dietrich1997,
  author    = {C. R. Dietrich and G. N. Newsam},
  title     = {Fast and Exact Simulation of Stationary Gaussian Processes through Circulant Embedding of the Covariance Matrix},
  journal   = {SIAM Journal on Scientific Computing},
  volume    = {18},
  number    = {4},
  pages     = {1088--1107},
  year      = {1997}
}

@article{HelgasonPipirasAbry2011,
  author    = {Hannes Helgason and Vladas Pipiras and Patrice Abry},
  title     = {Fast and exact synthesis of stationary multivariate Gaussian time series using circulant embedding},
  journal   = {Signal Processing},
  volume    = {91},
  number    = {5},
  pages     = {1123--1133},
  year      = {2011},
  doi       = {10.1016/j.sigpro.2010.10.014}
}

@article{Perrin2003,
  author    = {O. Perrin and F. Harba and A. Berzin-Joseph and C. G\'erb\'e and A. Sellan},
  title     = {Fast and exact synthesis of fractional Brownian motion through the fast Fourier transform},
  journal   = {EURASIP Journal on Advances in Signal Processing},
  volume    = {2003},
  number    = {12},
  pages     = {1--11},
  year      = {2003}
}

@article{MuzyBaileBacry2013_PRE,
  title={Random cascade model in the limit of infinite integral scale as the exponential of a nonstationary 1/f noise: Application to volatility fluctuations in stock markets},
  author={Muzy, Jean-François and Baïle, Romain and Bacry, Emmanuel},
  journal={Physical Review E},
  volume={87},
  number={5},
  pages={052806},
  year={2013},
  publisher={APS}
}

@article{gatheral2018volatility,
  title={Volatility is rough},
  author={Gatheral, Jim and Jaisson, Thibault and Rosenbaum, Mathieu},
  journal={Quantitative finance},
  volume={18},
  number={6},
  pages={933--949},
  year={2018},
  publisher={Taylor \& Francis}
}

@article{wu2022rough,
  title={From rough to multifractal volatility: The log S-fBM model},
  author={Wu, Peng and Muzy, Jean-Fran{\c{c}}ois and Bacry, Emmanuel},
  journal={Physica A: Statistical Mechanics and its Applications},
  volume={604},
  pages={127919},
  year={2022},
  publisher={Elsevier}
}

@article{chicheportiche2015nested,
  title={A nested factor model for non-linear dependencies in stock returns},
  author={Chicheportiche, R{\'e}my and Bouchaud, Jean-Philippe},
  journal={Quantitative Finance},
  volume={15},
  number={11},
  pages={1789--1804},
  year={2015},
  publisher={Taylor \& Francis}
}

@article{abijaber2019affine,
  title={Affine Volterra processes},
  author={Abi Jaber, Eduardo and Larsson, Martin and Pulido, Sergio},
  journal={The Annals of Applied Probability},
  volume={29},
  number={5},
  pages={3155--3200},
  year={2019},
  publisher={Institute of Mathematical Statistics}
}

@article{bacry2001modelling,
  title={Modelling financial time series using multifractal random walks},
  author={Bacry, Emmanuel and Delour, Jean and Muzy, Jean-Fran{\c{c}}ois},
  journal={Physica A: statistical mechanics and its applications},
  volume={299},
  number={1-2},
  pages={84--92},
  year={2001},
  publisher={Elsevier}
}

@article{comte1998long,
  title={Long memory in continuous-time stochastic volatility models},
  author={Comte, Fabienne and Renault, Eric},
  journal={Mathematical finance},
  volume={8},
  number={4},
  pages={291--323},
  year={1998},
  publisher={Wiley Online Library}
}

@article{bacry2008log,
  title={Log-Normal continuous cascades: aggregation properties and estimation. Application to financial time-series},
  author={Bacry, E and Kozhemyak, A and Muzy, J-F},
  journal={Quantitative finance},
  volume={13},
  number={5},
  pages={795--818},
  year={2012}
}

@article{mouti2023rough,
  title={Rough volatility: evidence from range volatility estimators},
  author={Mouti, Saad},
  journal={arXiv preprint arXiv:2312.01426},
  year={2023}
}

@article{garmanestimation,
  title={On the Estimation of Security Price Volatility from Historical Data 2},
  author={Garman, Mark B and Klass, Michael J}
}

@inproceedings{bachelier1900theorie,
  title={Th{\'e}orie de la sp{\'e}culation},
  author={Bachelier, Louis},
  booktitle={Annales scientifiques de l'{\'E}cole normale sup{\'e}rieure},
  volume={17},
  pages={21--86},
  year={1900}
}

@inproceedings{foundationmodernprob,
  title={Foundation of modern probability},
  author={Kallenberg, Olav},
  booktitle={Annales scientifiques de l'{\'E}cole normale sup{\'e}rieure},
  year={2021},
 publisher={Springer}
}

@article{LogSfbmNFM,
  title={A Nested Factor Model for Equity Markets: Reconciling
Multifractal Stock Returns and Rough Index Volatilities},
  author={Zarhali, Othmane and Aubrun, Cecilia and  Bacry,Emmanuel and  Bouchaud,Jean-Philippe and Muzy, Jean-François },
  journal={arXiv preprint arXiv:2505.02678},
  year={2025}
}

@article{fbmmandelbrotvaness,
  title={Fractional Brownian motions, fractional noises and applications},
  author={Mandelbrot, Benoît and Van Ness, John W.},
  journal={SIAM Review, 10(4), 422-437},
  year={1968}
}

@article{mandelbrot1963variation,
  title={The variation of certain speculative prices},
  author={Mandelbrot, Benoit},
  journal={The Journal of Business},
  volume={36},
  number={4},
  pages={394--419},
  year={1963},
  publisher={University of Chicago Press}
}

@book{mandelbrot1997fractals,
  title={Fractals and scaling in finance: Discontinuity, concentration, risk},
  author={Mandelbrot, Benoit B},
  year={1997},
  publisher={Springer}
}

@article{rogers1997arbitrage,
  title={Arbitrage with fractional Brownian motion},
  author={Rogers, L Chris G},
  journal={Mathematical Finance},
  volume={7},
  number={1},
  pages={95--105},
  year={1997},
  publisher={Wiley Online Library}
}

@article{Gatheral2010,
  author  = {Gatheral, Jim},
  title   = {No-Dynamic-Arbitrage and Market Impact},
  journal = {Quantitative Finance},
  volume  = {10},
  number  = {7},
  pages   = {749--759},
  year    = {2010},
  doi     = {10.1080/14697680903373692}
}

@article{leland1994self,
  title={On the self-similar nature of Ethernet traffic (extended version)},
  author={Leland, Will E and Taqqu, Murad S and Willinger, Walter and Wilson, Daniel V},
  journal={IEEE/ACM Transactions on Networking},
  volume={2},
  number={1},
  pages={1--15},
  year={1994},
  publisher={IEEE}
}

@article{metzler2000random,
  title={The random walk's guide to anomalous diffusion: a fractional dynamics approach},
  author={Metzler, Ralf and Klafter, Joseph},
  journal={Physics Reports},
  volume={339},
  number={1},
  pages={1--77},
  year={2000},
  publisher={Elsevier}
}

@article{peng1992long,
  title={Long-range correlations in nucleotide sequences},
  author={Peng, C-K and Buldyrev, Sergey V and Goldberger, Ary L and Havlin, Shlomo and Sciortino, Francesco and Simons, Michael and Stanley, H Eugene},
  journal={Nature},
  volume={356},
  number={6365},
  pages={168--170},
  year={1992},
  publisher={Nature Publishing Group}
}

@article{arneodoDNA,
  title = {Characterizing Long-Range Correlations in DNA Sequences from Wavelet Analysis},
  author = {Arneodo, A. and Bacry, E. and Graves, P. V. and Muzy, J. F.},
  journal = {Phys. Rev. Lett.},
  volume = {74},
  issue = {16},
  pages = {3293--3296},
  numpages = {0},
  year = {1995},
  month = {Apr},
  publisher = {American Physical Society}
}

@inproceedings{musgrave1989synthesis,
  title={The synthesis and rendering of eroded fractal terrains},
  author={Musgrave, F Kenton and Kolb, Craig E and Mace, Robert S},
  booktitle={Computer Graphics (SIGGRAPH '89 Proceedings)},
  volume={23},
  number={3},
  pages={41--50},
  year={1989},
  organization={ACM}
}

@article{lillo2004long,
  title={The long memory of the efficient market},
  author={Lillo, Fabrizio and Farmer, J Doyne},
  journal={Studies in Nonlinear Dynamics and Econometrics},
  volume={8},
  number={3},
  year={2004},
  publisher={De Gruyter}
}

@article{bouchaud2004fluctuations,
  title={Fluctuations and response in financial markets: the subtle nature of 'random' price changes},
  author={Bouchaud, Jean-Philippe and Gefen, Yuval and Potters, Marc and Wyart, Matthieu},
  journal={Quantitative Finance},
  volume={4},
  number={2},
  pages={176--190},
  year={2004},
  publisher={Taylor \& Francis}
}

@inproceedings{Oksendal,
  title={Stochastic Calculus for Fractional Brownian Motion and Applications, Springer},
  author={Biagini, Francesca  and Hu, Yaozhong and Øksendal,Bernt and Zhang,Tusheng   },
  year={2008}
}

@article{mfbm,
  title={Basic properties of the multivariate fractional brownian
motion},
  author={Amblard, Pierre-Olivier  and Coeurjolly, Jean-François and Lavancier,Frédéric  and Philippe,Anne   },
  year={2013}
}

@article{mMRW,
author = {J-F. Muzy and D. Sornette and J. Delour and A. Arneodo},
title = {Multifractal returns and hierarchical portfolio theory},
journal = {Quantitative Finance},
volume = {1},
number = {1},
pages = {131--148},
year = {2001},
publisher = {Taylor \& Francis},
doi = {10.1080/713665541}
}

@article{mMRW1,
  title={A multivariate multifractal model for return fluctuations},
  author={Bacry, Emmanuel  and Delour, Jean and Muzy,Jean-François },
  year={2013}
}

@article{Schoenberg1938,
  author = {Schoenberg, Isaac J.},
  title = {Metric Spaces and Completely Monotone Functions},
  journal = {Annals of Mathematics},
  series = {Second Series},
  volume = {39},
  number = {4},
  pages = {811--841},
  year = {1938}
}

@article{woodchan,
  title={Simulation of stationary Gaussian vector
Fields},
  author={Chan, G  and Wood, A.T.A  },
journal={Statistics and Computing},
  year={1999}
}

@article{Dugo2026,
title = {The multivariate fractional Ornstein–Uhlenbeck process},
journal = {Stochastic Processes and their Applications},
volume = {192},
pages = {104814},
year = {2026},
issn = {0304-4149},
doi = {https://doi.org/10.1016/j.spa.2025.104814},
author = {Ranieri Dugo and Giacomo Giorgio and Paolo Pigato}
}

@article{MDE,
  title={A minimum distance estimator for long-memory processes},
  author={A. Tieslau, Margie  and Schmidt, Peter and T. Baillie, Richard },
  journal = {Journal of Econometrics},
  volume={71},
  pages={249--261},
  year={1996}
}

\newpage
\begin{appendices}
\section*{Appendix}
\label{app:Appendix}

\section{Proof of Lemma \ref{lemma:covdGi}}
\label{app:proofexistencedGi}
We start by recalling the following theorem:
\begin{theo}[Schoenberg,\cite{Schoenberg1938}]
Let \(M=(M_{ij})_{1\le i,j\le d}\in \R^{d\times d}\) be a symmetric matrix.
\item \(M\) is conditionally negative definite (see \hyperref[eq:negdef-coHurst]{\(\bm{\mathcal{H}_{\mathrm{1}}}\)}) if an only if
for every \(t\ge 0\), the matrix
\[
\big(e^{-t M_{ij}}\big)_{1\le i,j\le d}
\]
is positive definite.
\end{theo}
We consider an arbitrary $x:=\left(x_i\right)_{1\leq i \leq d} \in \R^d$.\\
For $h>T$, one has:
 \begin{eqnarray}
        \sum_{1\leq i,j \leq d}x_ix_j\xi_{i,j} \frac{h^{2\bH_{i,j}-2}}{T^{2\bH_{i,j}}}=\frac{1}{h^2}\sum_{1\leq i,j \leq d} \left[x_i\left(\frac{h}{T}\right)^{H_i}\right]\left[x_j\left(\frac{h}{T}\right)^{H_j}\right]\xi_{i,j} \geq 0
    \end{eqnarray}
    by positive definiteness of $\bm{\xi}$.\\
For $0<h\leq T$, set
\[
t := -2\log(h/T)\ge 0.
\]
By Schoenberg’s theorem, since \(\bm{H}\) is conditionally negative definite, the matrix $\bm{N}:=\big(e^{-t H_{ij}}\big)_{1\le i,j\le d}$
is positive definite. Moreover, as \(\bm{\xi}\) is positive definite,  $\bm{\xi} \circ \bm{N}$ is positive definite ( \(\circ\) denotes the entrywise Hadamard product) using the Schur product theorem which leads to the claimed result.
\tab\tab\tab\tab\tab\tab\tab\tab\tab\tab\tab\tab\tab\tab\tab\tab\tab\tab\tab\tab\tab\tab\tab\tab\tab\tab\tab\tab\tab\tab\tab\tab\tab\tab $\blacksquare$

\section{The mS-fBM cross covariances}
\label{app:multidimautocovs}
This section is devoted to the proof of the mS-fBM cross covariance formula of Eq.~\eqref{eq:covmultidimsfbm_dG} and Eq.~\eqref{eq:covOmega}.
\subsection{Proof of Proposition \ref{prop:autocovariancemultidimsfbm}}
\label{app:proofautocovariancemultidimsfbm}
Given $(t,s)\in \R_{+}^2$ such that $\tau:=\left|t-s\right|\leq T$ and $(i,j)\in \llbracket 1,d \rrbracket^2$, thanks to Eq.~\eqref{eq:setindexcov} we have:

\begin{eqnarray}
    \mathrm{cov}\left(\omega_{i}(t),\omega_{j}(s)\right) = \int_{C_T(t)\cap C_T(s)} \mathrm{cov}\left({\rm d}G_{i}(t,h),{\rm d}G_{j}(t,h)\right)
    \nonumber
\end{eqnarray}

Without loss of generality, suppose that $t \geq s$. In that case, from Eq.~\eqref{eq:covmultidimsfbm_dG} we have:
\begin{eqnarray}
    \int_{C_T(t)\cap C_T(s)} \mathrm{cov}\left({\rm d}G_{i}(t,h),{\rm d}G_{j}(t,h)\right)=\frac{\xi_{i,j}}{T^{2H_{i,j}}}\int_{\tau}^T\int_{t-\frac{h}{2}}^{s+\frac{h}{2}}h^{2H_{i,j}-2}{\rm d}t^{'}{\rm d}h+\frac{\xi_{i,j}}{T^{H_i+H_j}} \int_T^{+\infty}\int_{t-\frac{T}{2}}^{s+\frac{T}{2}}h^{H_i+H_j-2}{\rm d}t^{'}{\rm d}h\nonumber
\end{eqnarray}
Which leads, thanks to $\tau = t-s$, to:
\begin{eqnarray}
     \int_{C_T(t)\cap C_T(s)} \mathrm{cov}\left({\rm d}G_{i}(t,h),{\rm d}G_{j}(t,h)\right)=\frac{\xi_{i,j}}{T^{2H_{i,j}}}\int_{\tau}^T\left(h-\tau \right)h^{2H_{i,j}-2}{\rm d}h+\frac{\xi_{i,j}}{T^{H_i+H_j}}  \int_T^{+\infty}\left(T-\tau \right)h^{H_i+H_j-2}{\rm d}h\nonumber
\end{eqnarray}
After some simple algebra this directly gives expression \eqref{eq:autocovmultidimsfbm}.\\
\tab\tab\tab\tab\tab\tab\tab\tab\tab\tab\tab\tab\tab\tab\tab\tab\tab\tab\tab\tab\tab\tab\tab\tab\tab\tab\tab\tab\tab\tab\tab\tab\tab\tab $\blacksquare$
\subsection{Proof of Proposition \ref{prop:autocovarianceOmegamultidim}}
We suppose that $t>s$ as they play symmetric roles. we suppose as well that $\tau>\Delta$.\\
By construction, $((\Omega_{i,\Delta}(t))_t,{i\in \llbracket 1,d \rrbracket})$ is a centered gaussian field.
By definition, one has:
\begin{eqnarray}
    \mathrm{cov}\left(\Omega_{i,\Delta}(t),\Omega_{j,\Delta}(s) \right)= \frac{1}{\lambda_i\lambda_j}\mathrm{cov}\left(\int_t^{t+\Delta}\omega_{i}(u)du,\int_s^{s+\Delta}\omega_{j}(v)dv\right) \nonumber
\end{eqnarray}

By direct computation, we have:
\begin{eqnarray}
    \int_t^{t+\Delta}\int_s^{s+\Delta}|u-v|^{2H_{i,j}}dudv = \frac{|\tau+\Delta|^{2H_{i,j}+2}-
        \tau^{2H_{i,j}+2}}{(1+2H_{i,j})(2H_{i,j}+2)}\nonumber \\
        +\frac{1}{(1+2H_{i,j})}\left(\int_t^{s+\Delta}|u-s-\Delta|^{2H_{i,j}}(u-s-\Delta) du+\int_{s+\Delta}^{t+\Delta}|u-s-\Delta|^{2H_{i,j}}(u-s-\Delta) du\right) \nonumber
\end{eqnarray}
By integrating the term, one has:
\begin{eqnarray}
    \int_t^{t+\Delta}\int_s^{s+\Delta}|u-v|^{2H_{i,j}}dudv = \frac{|\tau+\Delta|^{2H_{i,j}+2}-\tau^{2H_{i,j}+2}+|\tau-\Delta|^{2H_{i,j}+2}-\tau^{2H_{i,j}+2}}{(1+2H_{i,j})(2H_{i,j}+2)}\nonumber
\end{eqnarray}

Meaning that if $\tau>\Delta$ (similar computations are performed otherwise):
\begin{eqnarray}
    \int_t^{t+\Delta}\int_s^{s+\Delta}|u-v|^{2H_{i,j}}dudv = \frac{|\tau+\Delta|^{2H_{i,j}+2}-2\tau^{2H_{i,j}+2}+|\tau-\Delta|^{2H_{i,j}+1}(\tau-\Delta)}{(1+2H_{i,j})(2H_{i,j}+2)}\nonumber
\end{eqnarray}
Consequently, 
\begin{eqnarray}
    \int_t^{t+\Delta}\int_s^{s+\Delta}|u-v|^{2H_{i,j}}dudv = \frac{|\tau+\Delta|^{2H_{i,j}+2}-2\tau^{2H_{i,j}+2}+|\tau-\Delta|^{2H_{i,j}+2}}{(1+2H_{i,j})(2H_{i,j}+2)}\nonumber
\end{eqnarray}
Similarly,
\begin{eqnarray}
    \int_t^{t+\Delta}\int_s^{s+\Delta}|u-v|dudv = \frac{|\tau+\Delta|^{3}-2\tau^{3}+|\tau-\Delta|^{3}}{6}\nonumber
\end{eqnarray}
Putting the terms together, one has:

\begin{eqnarray}
\mathrm{cov}\left( \Omega_{i,\Delta}(t), \Omega_{j,\Delta}(s) \right) =
 \Delta^2 g_{i,j} \Bigg[
\left( \frac{1}{2H_{i,j}} - \frac{1}{H_i + H_j - 1} \right)  - f\left(\tau,2H_{i,j}  \right)
     \left( \frac{1}{2H_{i,j}} - \frac{1}{2H_{i,j} - 1} \right) \nonumber \\
 +  f\left(\tau,1\right)\left( \frac{1}{H_i + H_j - 1} 
     - \frac{1}{2H_{i,j} - 1} \right)
\Bigg] \nonumber
\end{eqnarray}
Leading to the claimed result.\\
As the autocovariance function depends only on the shift variable $\tau$, so we conclude that the gaussian field $((\Omega_{i,\Delta}(t))_t,{i\in \llbracket 1,d \rrbracket})$ is stationary.\\
\tab\tab\tab\tab\tab\tab\tab\tab\tab\tab\tab\tab\tab\tab\tab\tab\tab\tab\tab\tab\tab\tab\tab\tab\tab\tab\tab\tab\tab\tab\tab\tab\tab\tab $\blacksquare$

\subsection{The cross covariance of the multidimensional multifractal random measure}
\label{app:proofautocovMRM}
This appendix is devoted to developments related to the autocovariance function of the variance measure of Eq.~\eqref{eq:mrmoverint} evoked in Section \ref{sec:multidimlogsfbmmrm}.\\
We denote $\gamma$ the lower incomplete gamma function:
\begin{eqnarray}
   \forall(s,x)\in \left(\R^{*}_{+}\right)^2,\tab  \gamma(s, x) = \int_0^x t^{s-1} e^{-t} \, dt . \nonumber
\end{eqnarray} 
For any $(i,j)\in \llbracket 1,d \llbracket^2$, we introduce the following:
\begin{eqnarray}
    \begin{cases}
    K_{i,j}=\frac{\xi_{i,j}}{2TH_{i,j}(1-2H_{i,j})}\\
    L_{i,j}=\frac{\xi_{i,j}}{T}.\frac{2H_{i,j}-2 \bH_{i,j}}{(2H_{i,j}-1)(1-2 \bH_{i,j})}
\\
M_{i,j}=e^{\xi_{i,j}\frac{1+2 H_{i,j}-2\bH_{i,j}}{2H_{i,j}(1-2 \bH_{i,j})} }
    \end{cases}
\end{eqnarray}
We start with this preliminary corollary, whose proof relies on a change of variable and straightforward algebra and will be useful in the following results.
\begin{Corollary}
    \label{corr:gammafunctionidentity}
For any $a<b$ in $\R_+$ and $p\in\N$ and $L>0$, the following holds:
\begin{eqnarray}
    \int_a^b u^p e^{-L u} \, du = \frac{1}{L^{p+1}} \left[ \gamma\left(p+1, L b\right) - \gamma\left(p+1, L a\right) \right]
\end{eqnarray}
\end{Corollary}
The first result concerns a series expansion of the cross covariance function.
\begin{prop}
\label{prop:autocovMRM}
    For any $(t,\tau)\in \R_{+}^2$, $(i\neq j)\in \llbracket 1,d \rrbracket^2$ and $\Delta>0$ such that $\Delta<\tau$, we have:
  \begin{eqnarray}
  \label{eq:crossautocovmrmexpand}
        \E\left(M_{i,\Delta}(t)M_{j,\Delta}(t+\tau)\right) = M_{i,j}\sum_{n=0}^{+\infty}\frac{(-K_{i,j})^n}{L_{i,j}^{2Hn+1}n!}F_n\left(H_{i,j},L_{i,j},\Delta,\tau \right)
    \end{eqnarray}
Where 
\begin{multline*}
     F_n\left(H,L,\Delta,\tau \right)=
(\Delta + \tau) \cdot \frac{1}{L_{i,j}^{2H_{i,j} n + 1}} 
\left( \gamma(2H_{i,j} n + 1, L_{i,j}(\Delta + \tau)) 
- \gamma(2H_{i,j} n + 1, L_{i,j} \tau) \right) \\
- \frac{1}{L_{i,j}^{2H_{i,j} n + 2}} 
\left( \gamma(2H_{i,j} n + 2, L_{i,j}(\Delta + \tau)) 
- \gamma(2H_{i,j} n + 2, L_{i,j} \tau) \right) \\
+ \frac{1}{L_{i,j}^{2H_{i,j} n + 2}} 
\left( \gamma(2H_{i,j} n + 2, L_{i,j} \tau) 
- \gamma(2H_{i,j} n + 2, L_{i,j} (\tau - \Delta)) \right) \\
+ (\Delta - \tau) \cdot \frac{1}{L_{i,j}^{2H_{i,j} n + 1}} 
\left( \gamma(2H_{i,j} n + 1, L_{i,j} \tau) 
- \gamma(2H_{i,j} n + 1, L_{i,j} (\tau - \Delta)) \right)
\end{multline*}

\end{prop}
\textit{\textbf{Proof}}\\
    We fix  $(t,\tau)\in \R_{+}^2$ and $(i\neq j)\in \llbracket 1,d \rrbracket^2$.\\
   
 Using the fact that  $\left(\omega_{i}(.) \right)_{i\in \llbracket 1,d \rrbracket}$ is a stationary gaussian field, we get:
    \begin{eqnarray}
        \E\left(M_{i,\Delta}(t)M_{j,\Delta}(t+\tau)\right)=\int_0^{\Delta}\int_{\tau}^{\tau+\Delta}e^{\xi_{i,j} \Bigg[
\frac{1+2 H_{i,j}-2\bH_{i,j}}{2H_{i,j}(1-2 \bH_{i,j})} 
- \left(\frac{|x-y|}{T}\right)^{2H_{i,j}} \frac{1}{2H_{i,     j}(1-2H_{i,j})}  - \frac{|x-y|}{T}  \left(\frac{2H_{i,j}-2 \bH_{i,j}}{(2H_{i,j}-1)(1-2 \bH_{i,j})}
\right)
\Bigg]}{\rm d}x{\rm d}y \nonumber
    \end{eqnarray}
    After some algebra, one has:
    \begin{eqnarray}
        \E\left(M_{i,\Delta}(t)M_{j,\Delta}(t+\tau)\right)=e^{\xi_{i,j}\frac{1+2 H_{i,j}-2\bH_{i,j}}{2H_{i,j}(1-2 \bH_{i,j})} }\int_0^{\Delta}\int_{\tau}^{\tau+\Delta}e^{-K_{i,j}|x-y|^{2H_{i,j}}-L_{i,j}|x-y|}{\rm d}x{\rm d}y \nonumber
    \end{eqnarray}

In the sequel, we denote:
\begin{eqnarray}
    f_{i,j}(z):=e^{-K_{i,j}|z|^{2H_{i,j}}-L_{i,j}|z|}
\end{eqnarray}

Given that $f_{i,j}$ is a symmetric function, we have using Eq (83) in \cite{wu2022rough}:

\begin{eqnarray}
\int_0^{\Delta}\int_{\tau}^{\tau+\Delta} f_{i,j}(x - y)\, {\rm d}x\, {\rm d}y 
&=&  
\underbrace{\int_0^{\Delta} (\Delta - z) f_{i,j}(z + \tau)\, {\rm d}z}_{\text{(I)}}
+ 
\underbrace{\int_0^{\Delta} (\Delta - z) f_{i,j}(z - \tau)\, {\rm d}z}_{\text{(II)}} \nonumber
\end{eqnarray}
Let's deal with each term separately.\\

By substituting the exponential series expansion into the integral, one has:
\begin{align*}
\text{(I)}
&= \int_0^{\Delta} (\Delta - z) e^{-L_{i,j}(z + \tau)} \sum_{n=0}^\infty \frac{(-K_{i,j})^n}{n!} (z + \tau)^{2H_{i,j} n} dz \\
&= e^{-L_{i,j} \tau} \sum_{n=0}^\infty \frac{(-K_{i,j})^n}{n!} \int_0^{\Delta} (\Delta - z)(z + \tau)^{2H_{i,j} n} e^{-L_{i,j} z} dz
\end{align*}
After some algebra, one obtains:
\begin{align*}
\int_0^{\Delta}(\Delta - z)(z + \tau)^{2H_{i,j} n} e^{-L_{i,j} z} dz
&= \int_\tau^{\Delta+\tau} (\Delta + \tau - u) u^{2H_{i,j} n} e^{-L_{i,j}(u - \tau)} du \\
&= e^{L_{i,j} \tau} \int_\tau^{\Delta+\tau} (\Delta + \tau - u) u^{2H_{i,j} n} e^{-L_{i,j} u} du
\end{align*}

By expanding, the integral becomes:
\begin{align*}
\int_\tau^{\Delta+\tau} (\Delta + \tau - u) u^{2H_{i,j} n} e^{-L_{i,j} u} du
= (\Delta + \tau) \int_\tau^{\Delta+\tau} u^{2H_{i,j} n} e^{-L_{i,j} u} du
- \int_\tau^{\Delta+\tau} u^{2H_{i,j} n + 1} e^{-L_{i,j} u} du
\end{align*}

Using corollary \ref{corr:gammafunctionidentity}, one has:
\begin{align*}
\int_\tau^{\Delta+\tau} u^{2H_{i,j} n} e^{-L_{i,j} u} du &= \frac{1}{L_{i,j}^{2H_{i,j} n + 1}} \left[ \gamma(2H_{i,j} n + 1, L_{i,j}(\Delta + \tau)) - \gamma(2H_{i,j} n + 1, L_{i,j} \tau) \right] \\
\int_\tau^{\Delta+\tau} u^{2H_{i,j} n + 1} e^{-L_{i,j} u} du &= \frac{1}{L_{i,j}^{2H_{i,j} n + 2}} \left[ \gamma(2H_{i,j} n + 2, L_{i,j}(\Delta + \tau)) - \gamma(2H_{i,j} n + 2, L_{i,j} \tau) \right]
\end{align*}

Leading to:
\begin{align*}
\text{(I)}
= \sum_{n=0}^\infty \frac{(-K_{i,j})^n}{n!} \Bigg[
& (\Delta + \tau) \cdot \frac{1}{L_{i,j}^{2H_{i,j} n + 1}} \left( \gamma(2H_{i,j} n + 1, L_{i,j}(\Delta + \tau)) - \gamma(2H_{i,j} n + 1, L_{i,j} \tau) \right) \\
& - \frac{1}{L_{i,j}^{2H_{i,j} n + 2}} \left( \gamma(2H_{i,j} n + 2, L_{i,j}(\Delta + \tau)) - \gamma(2H_{i,j} n + 2, L_{i,j} \tau) \right)
\Bigg]
\end{align*}

Conducting similar computations, one also has:
\begin{align*}
\text{(II)}
&= \int_0^{\Delta} (\Delta - z) e^{-L_{i,j}|z - \tau|} \sum_{n=0}^\infty \frac{(-K_{i,j})^n}{n!} (z - \tau)^{2H_{i,j} n} dz = \sum_{n=0}^\infty \frac{(-K_{i,j})^n}{n!}\int_0^{\Delta} (z - \tau)^{2H_{i,j} n}(\Delta - z) e^{-L_{i,j}|z - \tau|}dz
\end{align*}

The integral, after a change of variable becomes:
\begin{align*}
\int_0^{\Delta} (\tau - z)^{2H_{i,j} n} (\Delta - z) e^{-L_{i,j}(\tau - z)} dz= \int_{\tau - \Delta}^{\tau} u^{2H_{i,j} n + 1} e^{-L_{i,j} u} du
+ (\Delta - \tau) \int_{\tau - \Delta}^{\tau} u^{2H_{i,j} n} e^{-L_{i,j} u} du
\end{align*}

Thanks to the corollary \ref{corr:gammafunctionidentity}, we have:
\begin{align*}
\int_{\tau - \Delta}^{\tau} u^{2H_{i,j} n} e^{-L_{i,j} u} du
&= \frac{1}{L_{i,j}^{2H_{i,j} n + 1}} \left[ \gamma(2H_{i,j} n + 1, L_{i,j} \tau) - \gamma(2H_{i,j} n + 1, L_{i,j} (\tau - \Delta)) \right] \\
\int_{\tau - \Delta}^{\tau} u^{2H_{i,j} n + 1} e^{-L_{i,j} u} du
&= \frac{1}{L_{i,j}^{2H_{i,j} n + 2}} \left[ \gamma(2H_{i,j} n + 2, L_{i,j} \tau) - \gamma(2H_{i,j} n + 2, L_{i,j} (\tau - \Delta)) \right]
\end{align*}

Which means that:
\begin{align*}
\int_0^{\Delta} (z - \tau)^{2H_{i,j} n} (\Delta - z) e^{-L_{i,j}|z - \tau|} dz
= &\; \frac{1}{L_{i,j}^{2H_{i,j} n + 2}} \left[ \gamma(2H_{i,j} n + 2, L_{i,j} \tau) - \gamma(2H_{i,j} n + 2, L_{i,j} (\tau - \Delta)) \right] \\
&+ (\Delta - \tau) \cdot \frac{1}{L_{i,j}^{2H_{i,j} n + 1}} \left[ \gamma(2H_{i,j} n + 1, L_{i,j} \tau) - \gamma(2H_{i,j} n + 1, L_{i,j} (\tau - \Delta)) \right]
\end{align*}
Consequently,
\begin{multline*}
\text{(II)} =
\sum_{n=0}^\infty \frac{(-K_{i,j})^n}{n!}
\Bigg[
\frac{1}{L_{i,j}^{2H_{i,j} n + 2}} 
\left( \gamma(2H_{i,j} n + 2, L_{i,j} \tau) 
- \gamma(2H_{i,j} n + 2, L_{i,j} (\tau - \Delta)) \right) \\
+ (\Delta - \tau) \cdot \frac{1}{L_{i,j}^{2H_{i,j} n + 1}} 
\left( \gamma(2H_{i,j} n + 1, L_{i,j} \tau) 
- \gamma(2H_{i,j} n + 1, L_{i,j} (\tau - \Delta)) \right)
\Bigg]
\end{multline*}

Which leads to the claimed result.
    \\
\tab\tab\tab\tab\tab\tab\tab\tab\tab\tab\tab\tab\tab\tab\tab\tab\tab\tab\tab\tab\tab\tab\tab\tab\tab\tab\tab\tab\tab\tab\tab\tab\tab\tab $\blacksquare$
One way to avoid doing a series expansion is to rely on the small intermittency approximation, which is the aim of the following result.
\begin{prop}
  For any $(t,\tau)\in \R_{+}^2$, $\Delta>\tau$ and $(i,j)\in \llbracket 1,d \rrbracket^2$, The following approximation holds:
    \begin{multline}
    \label{eq:crossautocovmrmSIA}
\E\left(M_{i,\Delta}(t)M_{j,\Delta}(t+\tau)\right) = M_{i,j} \Bigg\{ 
F_{i,j}(\tau + \Delta) + F_{i,j}(\tau - \Delta) - 2F_{i,j}(\tau)  \\
- L_{i,j} \Bigg[-\frac{1}{2H_{i,j} \, K_{i,j}^{\frac{n+13}{2H_{i,j}}}} \bigg[
\gamma\left( \frac{3}{2H_{i,j}},\, K_{i,j} \left(\tau+\Delta\right)^{2H_{i,j}} \right) 
- \gamma\left( \frac{3}{2H_{i,j}},\, K_{i,j} \tau^{2H_{i,j}} \right)
\bigg]+\\
\left(\Delta+2\tau\right)\frac{1}{2H_{i,j} \, K_{i,j}^{\frac{2}{2H_{i,j}}}} \bigg[
\gamma\left( \frac{2}{2H_{i,j}},\, K_{i,j} \left(\tau+\Delta\right)^{2H_{i,j}} \right) 
- \gamma\left( \frac{2}{2H_{i,j}},\, K_{i,j} \tau^{2H_{i,j}} \right)
\bigg]-\\
\frac{\left(\Delta+\tau\right)\tau}{2H_{i,j} \, K_{i,j}^{\frac{1}{2H_{i,j}}}} \bigg[
\gamma\left( \frac{1}{2H_{i,j}},\, K_{i,j} \left(\tau+\Delta\right)^{2H_{i,j}} \right) 
- \gamma\left( \frac{1}{2H_{i,j}},\, K_{i,j} \tau^{2H_{i,j}} \right)
\bigg]\\
-\frac{1}{2H_{i,j} \, K_{i,j}^{\frac{3}{2H_{i,j}}}} \bigg[
\gamma\left( \frac{3}{2H_{i,j}},\, K_{i,j} \tau^{2H_{i,j}} \right) 
- \gamma\left( \frac{3}{2H_{i,j}},\, K_{i,j} \left(\tau-\Delta\right)^{2H_{i,j}} \right)
\bigg]+\\
\frac{2\tau-\Delta}{2H_{i,j} \, K_{i,j}^{\frac{21}{2H_{i,j}}}} \bigg[
\gamma\left( \frac{2}{2H_{i,j}},\, K_{i,j} \tau^{2H_{i,j}} \right) 
- \gamma\left( \frac{2}{2H_{i,j}},\, K_{i,j} \left(\tau-\Delta\right)^{2H_{i,j}} \right)
\bigg]-\\
\frac{\left(\Delta-\tau\right)\tau}{2H_{i,j} \, K_{i,j}^{\frac{1}{2H_{i,j}}}} \bigg[
\gamma\left( \frac{1}{2H_{i,j}},\, K_{i,j} \tau^{2H_{i,j}} \right) 
- \gamma\left( \frac{1}{2H_{i,j}},\, K_{i,j} \left(\tau-\Delta\right)^{2H_{i,j}} \right)
\bigg]
\Bigg]
\Bigg\} \\
+ \frac{L_{i,j}}{\xi_{i,j}}\, o\left(\|\Lambda_{i,j}\|^2\right)
\end{multline}
where:
\begin{eqnarray}
    F_{i,j}(x) = \frac{1}{2H_{i,j}} \left(
\frac{x}{K_{i,j}^{1/(2H_{i,j})}}\, \gamma\left(\frac{1}{2H_{i,j}}, K_{i,j} x^{2H_{i,j}}\right)
- \frac{1}{K_{i,j}^{1/H_{i,j}}}\, \gamma\left(\frac{1}{H_{i,j}}, K_{i,j} x^{2H_{i,j}}\right)
\right).\nonumber
\end{eqnarray}

\end{prop}
\textit{\textbf{Proof}}\\
Using the fact that  $\left(\omega_{i}(.) \right)_{i\in \llbracket 1,d \rrbracket}$ is a stationary gaussian field, we get:
    \begin{eqnarray}
        \E\left(M_{i,\Delta}(t)M_{j,\Delta}(t+\tau)\right)=\int_0^{\Delta}\int_{\tau}^{\tau+\Delta}e^{\xi_{i,j} \Bigg[
\frac{1+2 H_{i,j}-2\bH_{i,j}}{2H_{i,j}(1-2 \bH_{i,j})} 
- \left(\frac{|x-y|}{T}\right)^{2H_{i,j}} \frac{1}{2H_{i,     j}(1-2H_{i,j})}  - \frac{|x-y|}{T}  \left(\frac{2H_{i,j}-2 \bH_{i,j}}{(2H_{i,j}-1)(1-2 \bH_{i,j})}
\right)
\Bigg]}{\rm d}x{\rm d}y \nonumber
    \end{eqnarray}
    After some algebra, one has:
    \begin{eqnarray}
        \E\left(M_{i,\Delta}(t)M_{j,\Delta}(t+\tau)\right)=M_{i,j}\int_0^{\Delta} \int_{\tau}^{\tau+\Delta}e^{-K_{i,j}|x-y|^{2H_{i,j}}-L_{i,j}|x-y|}{\rm d}x{\rm d}y \nonumber
    \end{eqnarray}
Using a first order approximation, one has:
\begin{multline*}
\E\left(M_{i,\Delta}(t)M_{j,\Delta}(t+\tau)\right)
= M_{i,j} \bigg\{ 
\underbrace{\int_0^{\Delta} \int_{\tau}^{\tau+\Delta}e^{-K_{i,j} |x - y|^{2H_{i,j}}} \,{\rm d}x\,{\rm d}y}_{\text{(I)}} \\
- L_{i,j} \underbrace{\int_0^{\Delta} \int_{\tau}^{\tau+\Delta} |x - y|\, e^{-K_{i,j} |x - y|^{2H_{i,j}}} \,{\rm d}x\,{\rm d}y}_{\text{(II)}}
\bigg\}
+ \frac{L_{i,j}}{\xi_{i,j}}o\left(\|\Lambda_{i,j}\|^2\right)
\end{multline*}
Let's treat each term separately.\\
$(x,y)\mapsto e^{-K_{i,j} |x - y|^{2H_{i,j}}} $ is a symmetric function. Using Eq (83) in \cite{wu2022rough} we have:
\begin{eqnarray}
     \text{(I)}=\int_0^{\Delta} (\Delta - z)\left(e^{-K_{i,j} |z+\tau|^{2H_{i,j}}} + e^{-K_{i,j} |z-\tau|^{2H_{i,j}}}\right)\, dz \nonumber
\end{eqnarray}
After some algebra, one has:
\begin{eqnarray}
   \text{(I)}=\int_{\tau}^{\tau + \Delta} (\Delta + \tau - z)\, e^{-K_{i,j} |z|^{2H}}\, dz 
+  \int_{\tau - \Delta}^{\tau} (\Delta - \tau + z)\, e^{-K_{i,j} |z|^{2H}}\, dz 
\nonumber
\end{eqnarray}
Which can be expressed as:
\begin{eqnarray}
    \text{(I)}= F_{i,j}(\tau + \Delta) + F_{i,j}(\tau - \Delta) - 2F_{i,j}(\tau) \nonumber
\end{eqnarray}

where $F_{i,j}(x) =  x \int_0^x e^{-K_{i,j} |z|^{2H_{i,j}}}\, dz - \int_0^x z\, e^{-K_{i,j} |z|^{2H_{i,j}}}\, dz 
$.\\
By direct computations, one has:
\begin{eqnarray}
    F_{i,j}(x) = \frac{1}{2H_{i,j}} \left(
\frac{x}{K_{i,j}^{1/(2H_{i,j})}}\, \gamma\left(\frac{1}{2H_{i,j}}, K_{i,j} x^{2H_{i,j}}\right)
- \frac{1}{K_{i,j}^{1/H_{i,j}}}\, \gamma\left(\frac{1}{H_{i,j}}, K_{i,j} x^{2H_{i,j}}\right)
\right).\nonumber
\end{eqnarray}

On the other hand, as the function $(x,y)\mapsto |x - y|\, e^{-K_{i,j} |x - y|^{2H_{i,j}}}$ is again symmetric, one has:
\begin{eqnarray}
    \text{(II)}=\int_0^{\Delta} (\Delta - z)z\left(e^{-K_{i,j} |z+\tau|^{2H_{i,j}}} + e^{-K_{i,j} |z-\tau|^{2H_{i,j}}}\right)\, dz \nonumber
\end{eqnarray}
Besides,

\begin{eqnarray*}
 \text{(II)}
=
\int_{\tau}^{\tau+\Delta} (\Delta + \tau-z)(z-\tau)e^{-K_{i,j} z^{2H_{i,j}}}\, dz +\int_{\tau-\Delta}^{\tau} (\Delta - \tau+z)(\tau-z) e^{-K_{i,j} z^{2H_{i,j}}}\, dz 
\end{eqnarray*}
After some algebra, one has for any $a<b$ in  $\R_+^*$ and $n\in \N$:
\begin{eqnarray*}
\int_a^b z^n e^{-K_{i,j} z^{2H_{i,j}}} \, dz = 
\frac{1}{2H_{i,j} \, K_{i,j}^{\frac{n+1}{2H_{i,j}}}} \bigg[
\gamma\left( \frac{n+1}{2H_{i,j}},\, K_{i,j} b^{2H_{i,j}} \right) 
- \gamma\left( \frac{n+1}{2H_{i,j}},\, K_{i,j} a^{2H_{i,j}} \right)
\bigg]
\end{eqnarray*}
Substituting this into the previous equation gives:
\begin{multline*}
\text{(II)} =-\frac{1}{2H_{i,j} \, K_{i,j}^{\frac{3}{2H_{i,j}}}} \bigg[
\gamma\left( \frac{3}{2H_{i,j}},\, K_{i,j} \left(\tau+\Delta\right)^{2H_{i,j}} \right) 
- \gamma\left( \frac{3}{2H_{i,j}},\, K_{i,j} \tau^{2H_{i,j}} \right)
\bigg]+\\
\left(\Delta+2\tau\right)\frac{1}{2H_{i,j} \, K_{i,j}^{\frac{2}{2H_{i,j}}}} \bigg[
\gamma\left( \frac{2}{2H_{i,j}},\, K_{i,j} \left(\tau+\Delta\right)^{2H_{i,j}} \right) 
- \gamma\left( \frac{2}{2H_{i,j}},\, K_{i,j} \tau^{2H_{i,j}} \right)
\bigg]-\\
\frac{\left(\Delta+\tau\right)\tau}{2H_{i,j} \, K_{i,j}^{\frac{1}{2H_{i,j}}}} \bigg[
\gamma\left( \frac{1}{2H_{i,j}},\, K_{i,j} \left(\tau+\Delta\right)^{2H_{i,j}} \right) 
- \gamma\left( \frac{1}{2H_{i,j}},\, K_{i,j} \tau^{2H_{i,j}} \right)
\bigg]\\
-\frac{1}{2H_{i,j} \, K_{i,j}^{\frac{3}{2H_{i,j}}}} \bigg[
\gamma\left( \frac{3}{2H_{i,j}},\, K_{i,j} \tau^{2H_{i,j}} \right) 
- \gamma\left( \frac{3}{2H_{i,j}},\, K_{i,j} \left(\tau-\Delta\right)^{2H_{i,j}} \right)
\bigg]+\\
\frac{2\tau-\Delta}{2H_{i,j} \, K_{i,j}^{\frac{21}{2H_{i,j}}}} \bigg[
\gamma\left( \frac{2}{2H_{i,j}},\, K_{i,j} \tau^{2H_{i,j}} \right) 
- \gamma\left( \frac{2}{2H_{i,j}},\, K_{i,j} \left(\tau-\Delta\right)^{2H_{i,j}} \right)
\bigg]-\\
\frac{\left(\Delta-\tau\right)\tau}{2H_{i,j} \, K_{i,j}^{\frac{1}{2H_{i,j}}}} \bigg[
\gamma\left( \frac{1}{2H_{i,j}},\, K_{i,j} \tau^{2H_{i,j}} \right) 
- \gamma\left( \frac{1}{2H_{i,j}},\, K_{i,j} \left(\tau-\Delta\right)^{2H_{i,j}} \right)
\bigg]
\end{multline*}
Substituting each term gives the claimed result.\\
\tab\tab\tab\tab\tab\tab\tab\tab\tab\tab\tab\tab\tab\tab\tab\tab\tab\tab\tab\tab\tab\tab\tab\tab\tab\tab\tab\tab\tab\tab\tab\tab\tab\tab $\blacksquare$
It's worth mentioning that in the small intermitency approximation of Eq.~\eqref{eq:crossautocovmrmSIA}, if $i=j$, $L_{i,j}=0$ meaning that:
\begin{eqnarray}
    \label{eq:crossautocovmrmSIAii}
\forall i\in \llbracket 1,d \rrbracket,\tab \E\left(M_{i,\Delta}(t)M_{i,\Delta}(t+\tau)\right) = M_{i,i} \left[ 
F_{i,i}(\tau + \Delta) + F_{i,i}(\tau - \Delta) - 2F_{i,i}(\tau)  \right]
\end{eqnarray}
which corresponds to the one dimensional autocovariance function's formula of Eq (21) in \cite{wu2022rough}.

\section{Proof of Proposition \ref{prop:mutidimensionalmultifractalMRW}}
\label{app:proofmutidimensionalmultifractalMRW}

    Given $l>0$, $(i,j)\in \llbracket 1,d \rrbracket^2$ and $\alpha \in ]0,1]$, there exists, thanks to Eq.~\eqref{eq:mlogmrm}, a gaussian random variable $ \Omega^{i,j}_{\alpha}$ independent of $\tilde{\omega}^{i}_{\ell,T}$ and $\tilde{\omega}^{j}_{\ell,T}$ such that:
    \begin{eqnarray}
    \begin{cases}
        \tilde{\omega}^{i}_{\alpha l,T}(\alpha t) \overset{\text{law}}{=} \Omega^{i,j}_{\alpha} + \tilde{\omega}^{j}_{\ell,T}(t)\\
        \tilde{\omega}^{j}_{\alpha l,T}(\alpha t) \overset{\text{law}}{=} \Omega^{i,j}_{\alpha} + \tilde{\omega}^{i}_{\ell,T}(t)
    \end{cases}
         \nonumber
    \end{eqnarray}
    of mean $\frac{\xi_{i,j}}{2}\ln\left(\alpha\right)$ variance $-\xi_{i,j}\ln\left(\alpha\right)$ (this comes from the normalization condition of Eq.~\eqref{eq:normalizationlogmrm}).\\
    This means that:
    \begin{eqnarray}
    \begin{cases}
        \tilde{M}^{i}_{\alpha \ell,T}(\alpha t) \overset{\text{law}}{=} \alpha e^{\Omega^{i,j}_{\alpha}} \tilde{M}^{j}_{\ell,T}( t)\\
        \tilde{M}^{j}_{\alpha \ell,T}(\alpha t) \overset{\text{law}}{=} \alpha e^{\Omega^{i,j}_{\alpha}} \tilde{M}^{i}_{\ell,T}( t)
    \end{cases}
        \nonumber
    \end{eqnarray}
By the weak limit (see Definition 5 and Theorem 6 in \cite{bacry2003log}), one has:
\begin{eqnarray}
    \E\left[ \widetilde{M}^i_{T}(t)^p\widetilde{M}^j_{T}(t)^q\right]=\left(\frac{t}{T}\right)^{p+q}\E\left[e^{p\Omega^{i,j}_{\frac{t}{T}}+q\Omega^{i,j}_{\frac{t}{T}}}\right] \E\left[\widetilde{M}^{i}_{T}(T)^p\widetilde{M}^{j}_{T}(T)^q\right] \nonumber
\end{eqnarray}
By direct computations:
\begin{eqnarray} 
\E\left[e^{p\Omega^{i,j}_{\frac{t}{T}}+q\Omega^{i,j}_{\frac{t}{T}}}\right] = \left(\frac{t}{T}\right)^{ -\frac{\xi_{i,j}}{2}(p+q)(p+q-1) }. \nonumber 
\end{eqnarray}
Consequently, this can be rewritten as:
\begin{eqnarray}
    \E\left[ \widetilde{M}^i_{T}(t)^p\widetilde{M}^j_{T}(t)^q\right]=K^{i,j}_{p,q}\left(\frac{t}{T}\right)^{\zeta_{i,j}(p,q)} \nonumber
\end{eqnarray}
where:
\begin{eqnarray}
    \begin{cases}
        K^{i,j}_{p,q}=\E\left[\widetilde{M}^{i}_{T}(T)^p\widetilde{M}^{j}_{T}(T)^q\right] \\
        \zeta_{i,j}(p,q)=p+q-\phi_{i,j}(p,q)\\
        \phi_{i,j}(p,q)=-\frac{\xi_{i,j}}{2}(p+q)(p+q-1)
    \end{cases}\nonumber
\end{eqnarray}
leading to the multidimensional scale invariance property.\\
\tab\tab\tab\tab\tab\tab\tab\tab\tab\tab\tab\tab\tab\tab\tab\tab\tab\tab\tab\tab\tab\tab\tab\tab\tab\tab\tab\tab\tab\tab\tab\tab\tab\tab $\blacksquare$

\section{Proofs of the weak convergence results }
\label{app:linkMMRW}

 In the sequel, we consider for any $\bm{H}\in \mathcal{S}_d$, $\|\bm{H} \|_{\infty}:=\underset{1\leq i,j \leq d}\sup\left|H_{i,j} \right|$.\\
 We denote:
 
\begin{eqnarray}
    \begin{cases}
        I:=\{i\in \llbracket 1,d \rrbracket, H_{i,i}=0\}\\
        J:=\{i\in \llbracket 1,d \rrbracket, H_{i,i}\neq 0\}
    \end{cases}\nonumber
\end{eqnarray}

\begin{comment}
     \subsection{\texorpdfstring{Tightness $\left(\bm{M}_{\bm{H}_n,T}(.)\right)_{n\in \N}$ for $\left(\bm{H}_n\right)_{n\in \N} \in \mathcal{S}_d^{\N}$}{Tightness (M_Hn,T(.)_n in N)}}
\end{comment}
\subsection{Tightness and weak convergence of random measures}

\label{app:proofproplinkMMRW}
We start by an introductory definition of tightness of stochastic processes valued in $\R^d$ in the space of compactly supported continuous paths.\\
\begin{mydeff}
    Let $ f : D \subset \mathbb{R} \to \mathbb{R}^d $ be a function. The \textbf{modulus of continuity} of $ f $ is the function $ \omega_f : [0, \infty) \to [0, \infty) $ defined by:
\begin{eqnarray}
    \omega_f(\delta) = \sup \left\{ \|f(x) - f(y)\| : x, y \in D, \, \left|x - y\right| \leq \delta \right\},
\end{eqnarray}

where $ \|.\|$ is the $\mathcal{L}^2$ canonical norm of $\R^d$.
\end{mydeff}
Similarly, one can define the modulus of continuity of a continuous stochastic process  $X$ valued in $\R^d$ as follows:
\begin{eqnarray}
    \omega_X(\delta) = \sup \left\{ \|X_t - X_s\| : x, y \in D, \, \left|t- s\right| \leq \delta \right\},
\end{eqnarray}
This leads to defined the notion of tightness of a vector valued sequence of stochastic processes.
\begin{mydeff}(Tightness)\\
    A $\R^d$ valued sequence of continuous stochastic processes $\left(X^n\right)_{n\in \N}$ is tight if:
    \begin{eqnarray}
    \forall \epsilon > 0 \quad \lim_{\delta \to 0} \sup_{n \in \mathbb{N}} \, \mathbb{P}\left( \omega_{X^n}(\delta) \geq \epsilon \right) = 0
\end{eqnarray}
\end{mydeff}
The tightness is usually used to establish weak convergence of sequences of random processes. A sequence is tight if its distributions concentrate increasingly within compact sets. Prohorov's theorem provides the equivalence between tightness and relative compactness in Polish spaces (see \cite{prokhorov1956convergence}). In Skorokhod space $ D[0,1] $, Billingsley gives sufficient conditions for tightness via the control of finite-dimensional distributions and a modulus of continuity condition \cite{billingsley1999convergence}. This framework can be adapted to the multidimensional setting, starting with the multidimensional version of Kolmogorov–Chentsov Theorem announced in \cite{foundationmodernprob} (see Theorem 23.7).
\begin{theo}
\label{theo:tightnessKolmogorov}
Let $B>0$.\\
    Let $(X^{n})_{n \in \mathbb{N}}$ be a $\R^d$ valued sequence of stochastic processes in $C([0, B])$, satisfying the following two conditions:
\begin{enumerate}
    \item The sequence of random variables $(X^{n}(0))_{n \in \mathbb{N}}$ is tight in $\mathbb{R}^d$.
    \item There exist constants $a \geq 1$, $b > 1$ and $C > 0$ such that, for all $n \in \mathbb{N}$ and all $s, t \in [0, B]$, we have:
    \begin{eqnarray}
         \mathbb{E}\left[\, \left|  X^{n}_s - X^{n}_t \right| ^a \,\right] \leq C |s - t|^b.
    \end{eqnarray}
   
\end{enumerate}
Then, the sequence $(X^{n})_{n \in \mathbb{N}}$ is tight.
\end{theo}

The second condition is also known as the Kolmogorov criterion. Provided the latter results, the convergence in distribution (weak convergence) is supported by the argument of Theorem 14.5 in \cite{foundationmodernprob}.

\subsection{Proof of Proposition \ref{prop:linkmLogSfBMMMRW}}
\label{app:prooflinkmLogSfBMMMRW}

To prove the tightness, it is sufficient to verify that the $\R^d$ valued process $\left(\bm{M}_{\bm{H}}(t)\right)_{t\geq 0}$ satisfies the previous assumptions of Theorem \ref{theo:tightnessKolmogorov}. The first point of Theorem \ref{theo:tightnessKolmogorov} is straightforward. Concerning the Kolmogorov criterion, we suppose that $t>s$ as they play symmetric roles. One has:
\begin{eqnarray}
    \E\left( \|\bm{M}_{\bm{H}}(t)-\bm{M}_{\bm{H}}(s)\|^2\right)=\underset{ i\in J}\sum\E \left[ \left(M_{i} \left([0, t]\right) - M_{i} \left([0, s]\right)\right)^2 \right]  \nonumber
\end{eqnarray}
Then, after some algebra
\begin{eqnarray}
    \E \left[ \left(M_{i} \left([0, t]\right) - M_{i} \left([0, s]\right)\right)^2 \right]=e^{ \frac{\lambda_i^2}{2H_{i}\left(1 - 2H_{i}\right)}}\int_s^t\int_s^te^{- \frac{\lambda_i^2}{2H_{i}\left(1 - 2H_{i}\right)}  \left(\frac{\left|u-v\right|}{T}\right)^{2H_{i}}}dudv \nonumber
\end{eqnarray}
One can see straightforwardly that:
\begin{eqnarray}
    \E \left[ \left(M_{i} \left([0, t]\right) - M_{i} \left([0, s]\right)\right)^2 \right]\leq e^{ \frac{\lambda_i^2}{2H_{i}\left(1 - 2H_{i}\right)}}\left|t-s\right|^2 \nonumber
\end{eqnarray}
which ends the proof.

Now, we proceed to the convergence of the finite dimensional distribution.\\
We begin with the following result.
\begin{prop}
\label{prop:linkMMRW} 
For any fixed $t\geq 0$ and $A$ a symmetric matrix in $\R^{d \times d}$, the following holds:
\begin{eqnarray}
\label{eq:limitmmrw}
    A\left(\bm{M}_{\bm{H}}(t)-\widetilde{\bm{M}}(t)\right)\xrightarrow[\|\bm{H} \|_{\infty}\rightarrow 0]{\mathcal{L}^2}0
\end{eqnarray}

\end{prop}
\textbf{\textit{Proof}}\\
First, we recall the geometric construction of the gaussian field $\left(\widetilde{\omega}^i\right)_{i\in \llbracket 1,d \rrbracket}$.\\
 For fixed $(t,h)\in C_{\ell,T}(.)$, we consider $(dG^0_{i}(t,h))_{i\in \llbracket 1,d \rrbracket} $ a centered gaussian vector with covariance measure:
     \begin{equation}
        \forall(i,j)\in \llbracket 1,d \rrbracket^2, \tab \mathrm{cov}\left({\rm d}G^0_i(t,h),{\rm d}G^0_j(t,h)\right) = \xi_{i,j}h^{-2} {\rm d}t {\rm d}h
        \label{eq:covmultidimsfbm_dGmrw}
    \end{equation}

Then we define a gaussian field $\left(\omega^i_{\ell,T}(.),i\in \llbracket 1,d \rrbracket \right)$ as follows:
\begin{eqnarray}
   \omega^i_{\ell,T} (t) = \mu^i_{\ell,T} + \int_{C_{\ell,T}(t)} \, dG^0_i(t^{'},h)
\end{eqnarray}
where similarly to the S-fBM process, $\mu^i_{\ell,T} $ is the mean of the $i^{th}$ marginal and defined such that $ \E\left(e^{\omega^i_{\ell,T} (t)} \right)=1$. Conduting similar computations as in Appendix \ref{app:proofautocovariancemultidimsfbm}, one ends up with the cross covariance function of Eq.~\eqref{eq:mlogmrm}.\\

By definition, one has:
\begin{flalign*}
    &\E \left[ \left(M_{i} \left([0, t]\right) - \widetilde{M}^i_{\ell} \left([0, t]\right)\right)  
    \left(M_{j} \left([0, t]\right) - \widetilde{M}^j_{\ell,T} \left([0, t]\right)\right) \right] & \\ 
    &= \int_0^t \int_0^t \left( 
    e^{\mathrm{cov}\left(\omega_{i} (u),\omega_{j} (v)\right) }
    + e^{\mathrm{cov}\left(\omega^i_{\ell,T} (u),\omega^j_{\ell,T} (v)\right) } \right.
    \left. - e^{\mathrm{cov}\left(\omega_{i} (u),\omega^j_{\ell,T} (v)\right)}
    - e^{\mathrm{cov}\left(\omega_{j} (u),\omega^i_{\ell,T} (v)\right)}
    \right) du \, dv.
\end{flalign*}
By similar computations as in Appendix \ref{app:multidimautocovs}, one has:

\begin{eqnarray}
\mathrm{cov}(\omega_{i}(t), \omega^j_{\ell,T}(t+\tau)) =
    \begin{cases}
    \xi_{i,j} \Bigg[\frac{1+2 H_{i,j}-2\bH_{i,j}}{2H_{i,j}(1-2 \bH_{i,j})} 
- \left(\frac{\tau}{T}\right)^{2H_{i,j}} \frac{1}{2H_{i,j}(1-2H_{i,j})}  - \frac{\tau}{T}  \left(\frac{2H_{i,j}-2 \bH_{i,j}}{(2H_{i,j}-1)(1-2 \bH_{i,j})}
\right)
\Bigg], if \tab  l\leq \tau \leq T
    \\
        \xi_{i,j} \Bigg[ 
\frac{2\bar{H}_{i,j} - 2H_{i,j} - 1}{2H_{i,j}(2\bar{H}_{i,j} - 1)} 
+ \frac{2H_{i,j} - 2\bar{H}_{i,j}}{(2\bar{H}_{i,j} - 1)(2H_{i,j} - 1)} \cdot \frac{\tau}{T} \\
- \frac{1}{2H_{i,j}} \left(\frac{l}{T}\right)^{2H_{i,j}} 
+ \frac{1}{2H_{i,j} - 1} \cdot \frac{\tau}{T} \cdot \left(\frac{l}{T}\right)^{2H_{i,j} - 1}
\Bigg], if \tab  \tau<l
    \end{cases}
\end{eqnarray}

Following the same computations as in Appendix A.3 in \cite{wu2022rough}, we have:
\begin{flalign*}
    &\E \left[ \left(M_{i} \left([0, t]\right) - \widetilde{M}^i_{\ell} \left([0, t]\right)\right)  
    \left(M_{j} \left([0, t]\right) - \widetilde{M}^j_{\ell,T} \left([0, t]\right)\right) \right] = I_1+I_2
\end{flalign*}
where:
\begin{eqnarray}
    \begin{cases}
        I_1 = \int_{l}^{t} (t - z) \left( 
    e^{C_{i,j}^{\omega}(z)}
    + \left(\frac{T}{z}\right)^{\xi_{i,j}} 
    - 2 e^{C_{i,j}^{\omega}(z)}
\right) dz\\
I_2 = \int_{0}^{l} (t - z) \left( 
     e^{C_{i,j}^{\omega}(z)}
    + e^{{\xi_{i,j}} \left(\frac{T}{\ell} + 1 - \frac{z}{\ell} \right)} 
    - 2e^{C_{i,j}^{l}(z)}
\right) dz.
    \end{cases}\nonumber
\end{eqnarray}
and:
\begin{eqnarray}
   C_{i,j}^{l}(z)=\xi_{i,j} \Bigg[ 
\frac{2\bar{H}_{i,j} - 2H_{i,j} - 1}{2H_{i,j}(2\bar{H}_{i,j} - 1)} 
+ \frac{2H_{i,j} - 2\bar{H}_{i,j}}{(2\bar{H}_{i,j} - 1)(2H_{i,j} - 1)} \cdot \frac{z}{T} 
- \frac{1}{2H_{i,j}} \left(\frac{l}{T}\right)^{2H_{i,j}} 
+\nonumber \\ \frac{1}{2H_{i,j} - 1} \cdot \frac{z}{T} \cdot \left(\frac{l}{T}\right)^{2H_{i,j} - 1}
\Bigg] \nonumber
\end{eqnarray}
One can conclude for any arbitrary $(i,j)\in I^2$, as in Appendix A.3 of \cite{wu2022rough}, using the dominated convergence theorem that:
\begin{eqnarray}
    \E \left[ \left(M_{i} \left([0, t]\right) - \widetilde{M}^i_{T} \left([0, t]\right)\right)  
    \left(M_{j} \left([0, t]\right) - \widetilde{M}^j_{T} \left([0, t]\right)\right) \right]\xrightarrow[H_{i,j}\rightarrow 0]{}0 .\nonumber
\end{eqnarray}

On the other hand, one has straightforwardly that:
\begin{eqnarray}
    \E\left( \left\|A\left(\bm{M}_{\bm{H}}(t)-\widetilde{\bm{M}}(t)\right)\right\|^2\right)=\underset{1\leq i,j\leq d}\sum q_{i,j}\E \left[ \left(M_{i} \left([0, t]\right) - \widetilde{M}^i_{T} \left([0, t]\right)\right)  
    \left(M_{j} \left([0, t]\right) - \widetilde{M}^j_{T} \left([0, t]\right)\right) \right]\nonumber\\ \xrightarrow[ \|\bm{H} \|_{\infty}\rightarrow 0]{}0 \nonumber
\end{eqnarray}
where $\|.\|$ is the $\mathcal{L}^2$ norm in $\R^d$ and $\left(q_{i,j}\right)_{1\leq i,j\leq d}=Q=A^TA$.\\
\\
\tab\tab\tab\tab\tab\tab\tab\tab\tab\tab\tab\tab\tab\tab\tab\tab\tab\tab\tab\tab\tab\tab\tab\tab\tab\tab\tab\tab\tab\tab\tab\tab\tab\tab $\blacksquare$

If $A=I_d$ in Eq.~\eqref{eq:limitmmrw}, one has:
\begin{eqnarray}
\label{eq:limitmmrw}
    \bm{M}_{\bm{H}}(t)\xrightarrow[\|\bm{H} \|_{\infty}\rightarrow 0]{\mathcal{L}^2}\widetilde{\bm{M}}(t)
\end{eqnarray}
which means similarly that the finite dimensional distribution of $\left(\bm{M}_{\bm{H}}(t_1),\bm{M}_{\bm{H}}(t_2),...,\bm{M}_{\bm{H}}(t_n) \right)$ converge to the distribution of $\left(\widetilde{\bm{M}}(t_1),\widetilde{\bm{M}}(t_2),...,\widetilde{\bm{M}}(t_n) \right)$ for an arbitrary tuple $\left(t_1,t_2,...,t_n\right)\in \R_+^n$. Combined with the tightness criterion demonstrated earlier, this leads to convergence in distribution of the process $\bm{M}_{\bm{H}}$ towards $\widetilde{\bm{M}}$ when $\|\bm{H} \|_{\infty}\rightarrow 0$, supported by the argument of Theorem 14.5 in \cite{foundationmodernprob}.\\

One can notice that the result still holds when instead of considering $\|\bm{H} \|_{\infty}\rightarrow 0$, the co-Hurst exponents (upper diagonal entries of $\bm{H}$) are finite and $\| \bm{H} \|^{\operatorname{diag}}_\infty\rightarrow 0$ where $\|\bm{H} \|^{\operatorname{diag}}_\infty = \underset{1 \leq i \leq d}\sup \left|H_{i,i}\right|$.

\subsection{Proof of Proposition \ref{prop:mixedmlogsfbmMRM}}
\label{app:prooflemmaexistencemixedmlogsfmbmrm}
 By definition, one has for any $t\geq 0$:
\begin{eqnarray}
    \E\left( \left\|\left(\bm{M}_{\bm{H},\ell}(t)-\bm{M}_{\bm{H}}(t)\right)\right\|^2\right)=\underset{j\in I}\sum \E \left[\left( \widetilde{M}^j_{\ell} \left([0, t]\right) - \widetilde{M}^j \left([0, t]\right)\right)^2
    \right]\nonumber
\end{eqnarray}
Given the weak convergence result of the one dimensional case of Eq.~\eqref{eq:weakcvgmrm1D}, the latter quantity vanishes as $\ell \rightarrow 0$. This ensures the finite dimensional distribution of $\left(\bm{M}_{\bm{H},\ell}(t_1),\bm{M}_{\bm{H},\ell}(t_2),...,\bm{M}_{\bm{H},\ell}(t_n) \right)$ to the distribution of $\left(\bm{M}_{\bm{H}}(t_1),\bm{M}_{\bm{H}}(t_2),...,\bm{M}_{\bm{H}}(t_n) \right)$ for arbitrary $(t_1,t_2,...,t_n)\in \R_+^n$.

By definition, one has:
\begin{eqnarray}
    \E \left[\left( \widetilde{M}^j_{\ell} \left([0, t]\right) - \widetilde{M}^j_{\ell} \left([0, s]\right)\right)^2
    \right]=\int_s^{t}\int_s^{t} \left(e^{-\lambda_j^2 \ln({\left|u-v\right|}/{T})} \mathbf{1}_{\{\ell < |u-v| < T\}}+ e^{ -\lambda_j^2 (\ln({\ell}/{T}) - 1 + \left|u-v\right|/\ell)}\mathbf{1}_{\{\ell < |u-v| < T\}}\right)  dudv \nonumber
\end{eqnarray}
One can upper bound the latter quantity:
\begin{eqnarray}
    \E \left[\left( \widetilde{M}^j_{\ell} \left([0, t]\right) - \widetilde{M}^j_{\ell} \left([0, s]\right)\right)^2
    \right] \leq 2e^{-\lambda_j^2 \ln({\ell}/{T})}\left|t-s \right|^2  \nonumber
\end{eqnarray}
Thanks to Theorem \ref{theo:tightnessKolmogorov}, the tightness criterion is verified. Consequently, the claimed result holds.

\subsection{Proof of Theorem \ref{theo:weakcontinuitylinkmLogSfBMMMRW}}
\label{app:proofweakcontinuitylinkmLogSfBMMMRW}
The convergence in distribution of the finite dimensional counterparts rely on continuity arguments. For that, we refer to the computations of the proof of Proposition \ref{prop:linkmLogSfBMMMRW} (Appendix \ref{app:prooflinkmLogSfBMMMRW}) as a more involving case. The tightness criterion is satisfied and demonstrated in a similar way as in Appendix \ref{app:prooflinkmLogSfBMMMRW} leading to the weak convergence claimed in the theorem.

\tab\tab\tab\tab\tab\tab\tab\tab\tab\tab\tab\tab\tab\tab\tab\tab\tab\tab\tab\tab\tab\tab\tab\tab\tab\tab\tab\tab\tab\tab\tab\tab\tab\tab $\blacksquare$

\section{Proof of theorem \ref{thm:smallintermlogmrmtheo}}
\label{app:thm1proof}
At this stage, we recall of the so called Wick's theorem which is an important tool in small intermittency approximations. In the sequel, we denote $\mathcal{P}\left( \mathcal{I}_m\right)$ the set of all pairings of $\mathcal{I}_m$ (all distinct ways of partitioning $\mathcal{I}_m$ into pairs) where $ \mathcal{I}_m$ is a subset of $\mathcal{I}_d:=\llbracket 1,d \rrbracket$.\\
\begin{theo}
    (Wick)\\
    For a given $n\in \N$, if $\left(X_i\right)_{0\leq i\leq n}$ a centered gaussian vector, then:
    \begin{equation}
    \E\left(\prod_{i=1}^{n}X_i\right)=
    \begin{cases*}
    \underset{P\in\mathcal{P}\left( \mathcal{I}_n\right)} \sum \underset{\{i,j\}\in P} \prod\E\left(X_iX_j\right)$  if n even$\\ 
      0  $  if n odd$
    \end{cases*}
    \end{equation}.
    \label{thm:wick}
\end{theo}
We will proceed progressively by exposing the necessary intermediary results.\\
In the sequel, we consider arbitrary subintervals $\left(I_i\right)_{i\in \llbracket 1,d \rrbracket}\subset K^d$ where the interval $K\subset \R_{+}$ with $|K|\leq T$
\begin{prop} (\textbf{Step 1/4})\\
For any subintervals $\left(I_i\right)_{i\in \llbracket 1,d \rrbracket} \subset K$, the following holds:
   \begin{eqnarray}
        E\left(\prod_{i=1}^{d}\frac{\Omega_{i}(I_i)}{|I_i|}\right)=
    \begin{cases*}
     \smashoperator[r]{\sum_{\{i_k,j_k\}_{k\in \left\llbracket 1,\frac{d}{2}\right\rrbracket}\in \mathcal{P}\left( \mathcal{I}_d\right)}}\hspace{1.2cm}\smashoperator[r]{\prod_{k=1}^{\frac{d}{2}}}\int_{I_{i_k}}\frac{du_{i_k}}{|I_{i_k}|}\int_{I_{j_k}}\frac{du_{j_k}}{|I_{j_k}|}\frac{g_{i_k,j_k}}{2H_{i_k,j_k}(1-2H_{i_k,j_k})}\left(1-\left(\frac{\left|u_{i_k}-u_{j_k} \right|}{T}\right)^{2H_{i_k,j_k}}\right) ,\\
     \hspace{10cm} if \hspace{1mm} d \hspace{1mm} even\\ 
      0 ,if \hspace{1mm} d  \hspace{1mm}odd
    \end{cases*}.\nonumber
    \end{eqnarray}
\end{prop}
\textbf{\textit{Proof}}\\
    Notice that $\left(\frac{\Omega_{i}(I_i)}{|I_{i}|},i\in \llbracket 1,d \rrbracket \right)$ is a centered gaussian vector with the following covariance structure:
    $$\forall(i,j)\in\llbracket 1,d \rrbracket^2,\mathrm{cov}\left(\frac{\Omega_{i}(I_i)}{|I_{i}|},\frac{\Omega_{j}(I_j)}{|I_{j}|} \right)=\int_{I_{i}}\int_{I_{j}}\frac{du}{|I_{i}|}\frac{dv}{|I_{j}|}\frac{g_{i,j}}{2H_{i,j}(1-2H_{i,j})}\left(1-\left(\frac{\left|u-v \right|}{T}\right)^{2H_{i,j}}\right).$$
    We have thanks to Wick's theorem:
    \begin{eqnarray}
        \E\left(\prod_{i=1}^{d}\frac{\Omega_{i}(I_i)}{|I_i|}\right)=
    \begin{cases*}
    \underset{\substack{\mathcal{P}\in \mathcal{P}\left( \mathcal{I}_d\right)}}
     \sum\underset{\substack{\{i,j\}\in \mathcal{P}}}\prod\int_{I_{i}}\frac{du_{i}}{|I_{i}|}\int_{I_{j}}\frac{du_{j}}{|I_{j}|}\frac{g_{i,j}}{2H_{i,j}(1-2H_{i,j})}\left(1-\left(\frac{\left|u_i-u_j \right|}{T}\right)^{2H_{i,j}}\right),  if\hspace{1mm} d\hspace{1mm}  even\\ 
      0 ,  if\hspace{1mm}  d \hspace{1mm} odd
    \end{cases*}.
    \nonumber
    \end{eqnarray}
    By partitioning the sum into pairs $\{i_k,j_k\}_{k\in \llbracket 1,\frac{d}{2} \rrbracket} $, can be rewritten as:
    \begin{eqnarray}
        E\left(\prod_{i=1}^{d}\frac{\Omega_{i}(I_i)}{|I_i|}\right)=
    \begin{cases*}
   \underset{\{i_k,j_k\}_{k\in \left\llbracket 1,\frac{d}{2}\right\rrbracket}\in \mathcal{P}\left( \mathcal{I}_d\right)}
     \sum\prod_{k=1}^{\frac{d}{2}}\int_{I_{i_k}}\frac{du_{i_k}}{|I_{i_k}|}\int_{I_{j_k}}\frac{du_{j_k}}{|I_{j_k}|}\frac{g_{i_k,j_k}}{2H_{i_k,j_k}(1-2H_{i_k,j_k})}\left(1-\left(\frac{\left|u_{i_k}-u_{j_k} \right|}{T}\right)^{2H_{i_k,j_k}}\right) ,\\
     \hspace{10cm} if \hspace{1mm} d \hspace{1mm} even\\ 
      0 ,if \hspace{1mm} d  \hspace{1mm}odd
    \end{cases*},\nonumber
    \end{eqnarray}
    which gives the claimed result.
    \\
\tab\tab\tab\tab\tab\tab\tab\tab\tab\tab\tab\tab\tab\tab\tab\tab\tab\tab\tab\tab\tab\tab\tab\tab\tab\tab\tab\tab\tab\tab\tab\tab\tab\tab $\blacksquare$
Using the same previous arguments, we notice that:
        \begin{eqnarray}
        \E\left(\prod_{i=1}^{d}\frac{\lambda_i\Omega_{i}(I_i)}{|I_i|}\right)=\begin{cases*}
   \underset{\{i_k,j_k\}_{k\in \left\llbracket 1,\frac{d}{2}\right\rrbracket}\in \mathcal{P}\left( \mathcal{I}_d\right)}
     \sum\prod_{k=1}^{\frac{d}{2}}\int_{I_{i_k}}\frac{du_{i_k}}{|I_{i_k}|}\int_{I_{j_k}}\frac{du_{j_k}}{|I_{j_k}|}\frac{g_{i_k,j_k}\lambda_{i_k}\lambda_{j_k}}{2H_{i_k,j_k}(1-2H_{i_k,j_k})}\left(1-\left(\frac{\left|u_{i_k}-u_{j_k} \right|}{T}\right)^{2H_{i_k,j_k}}\right) ,\\
     \hspace{10cm} if \hspace{1mm} d \hspace{1mm} even\\ 
      0 ,if \hspace{1mm} d  \hspace{1mm}odd
    \end{cases*}.\nonumber
        \label{eq:momentOmegaforproofstep1}
    \end{eqnarray}
For the sake of simplicity, we consider the function for any $(i,j)\in \llbracket 1,d \rrbracket^2$:
\begin{eqnarray}
    \rho_{i,j}\left(t\right):=\frac{1}{2H_{i,j}\left(1-2H_{i,j}\right)}\left(1-\left(\frac{\left|t\right|}{T}\right)^{2H_{i,j}}\right) \tab , |t|\leq T.\nonumber
\end{eqnarray}
\begin{prop}(\textbf{Step 2/4})\\
    For any subintervals $\left(I_i\right)_{i\in \llbracket 1,d \rrbracket} \subset[0,T]^d$, the following claim holds:
        \begin{eqnarray}
            \E\left(\prod_{i=1}^{d}\left(\frac{M_{i}(I_i)}{|I_{i}|}-1 \right) \right)=\sum_{m=0}^d\left(-1\right)^{d-m}\sum_{\substack{\mathcal{I}_m=\left\{i_k,k\in \llbracket 1,m \rrbracket\right\} \\ \mathcal{I}_m \subset \mathcal{I}_d}}\left(\prod_{k=1}^{m}\int_{I_{i_k}}\frac{du_{i_k}}{|I_{i_k}|} \right) e^{\sum_{1\leq s\leq l\leq m}\lambda_{i_s}\lambda_{i_{l}}g_{i_s,i_{l}}\rho_{i_s,i_{l}}\left(u_{i_s}-u_{i_{l}} \right)}.\nonumber\\
            \label{eq:step2}
        \end{eqnarray}
\end{prop}

\textbf{\textit{Proof}}\\
For simplicity we denote $M_i:=\frac{M_{i}(I_i)}{|I_{i}|},i\in \llbracket 1,d \rrbracket$.\\
    We have:
    $$\prod_{i=1}^{d}\left(M_i-1 \right)=\sum_{\substack{\mathcal{I}_m=\{i_k,k\in \llbracket 1,m \rrbracket\} \\ \mathcal{I}_m \subset \mathcal{I}_d}}\left(-1\right)^{d-|\mathcal{I}_m|} \prod_{j\in \mathcal{I}_m}M_{j},$$
    which leads to:\\
    $$\E\left( \prod_{i=1}^{d}\left(M_i-1 \right) \right)=\sum_{m=0}^d\left(-1\right)^{d-m} \sum_{\substack{\mathcal{I}_m=\{i_k,k\in \llbracket 1,m \rrbracket\} \\ \mathcal{I}_m \subset \mathcal{I}_d}} \E\left( \prod_{k=1}^{m}M_{i_k} \right).$$

    By straightforward computations, we have:
     $$\E\left( \prod_{k=1}^{m}M_{i_k} \right)=\left(\prod_{k=1}^{m}\int_{I_{i_k}}\frac{du_{i_k}}{|I_{i_k}|} \right)\E\left( e^{\sum_{k=1}^m \omega_{i_k}(u_{i_k}) }\right).$$
     We know that $\left(\omega_{i}(.),i\in \llbracket 1,d \rrbracket\right)$ is gaussian field whose marginals are gaussians with mean $\left(\mu_{i}\right)_{i\in \llbracket 1,d \rrbracket}$ and variances $\left(\frac{\nu_i^2}{2}\right)_{i\in \llbracket 1,d \rrbracket}$ such that:
$$\forall i\in \llbracket 1,d \rrbracket,\tab \mu_{i}=-\frac{\nu_i^2}{4}$$

     and its covariance structure can be rewritten this way:

     $$\forall (i,j)\in\llbracket 1,d \rrbracket^2,\forall (t,s)\in\R_{+}^2,\tab \mathrm{cov}\left( \omega_{i}(t),\omega_{j}(s)\right)=\lambda_i\lambda_jg_{i,j}\rho_{i,j}\left(t-s \right) .$$
     Thus we have for any $k\in \llbracket 1,m \rrbracket $:
      $$\E\left( e^{\sum_{k=1}^m \omega_{i_k}(u_{i_k}) }\right) = e^{\sum_{1\leq s\leq l\leq m}\lambda_{i_s}\lambda_{i_{l}}g_{i_s,i_{l}}\rho_{i_s,i_{l}}\left(u_{i_s}-u_{i_{l}} \right)}.$$
      which leads to the claimed result.\\
\tab\tab\tab\tab\tab\tab\tab\tab\tab\tab\tab\tab\tab\tab\tab\tab\tab\tab\tab\tab\tab\tab\tab\tab\tab\tab\tab\tab\tab\tab\tab\tab\tab\tab $\blacksquare$
Now, we are ready for the proof.\\\\
\textbf{Step 3/4}\\
For $m\in\llbracket 1,d \rrbracket$ and $\left(i_k\right)_{k\in \llbracket 1,m \rrbracket}:=\mathcal{I}_m$, we consider the following continuously differentiable function for any fixed $\left(u_i\right)_{i\in \llbracket 1,m \rrbracket}\in \R^m$:\\
    \[\psi(\lambda_{i_1},...,\lambda_{i_m}):=e^{\sum_{1\leq s\leq l\leq m}\lambda_{i_s}\lambda_{i_{l}}g_{i_s,i_{l}}\rho_{i_s,i_{l}}\left(u_{i_s}-u_{i_{l}}\right)}.\]
    We have for any $1\leq s< l\leq m$:\\
    \begin{eqnarray}
        \begin{cases*}
    \frac{\partial \psi(\lambda_{i_1},...,\lambda_{i_m})}{\partial\lambda_{i_s}}|_{\bm{\lambda}=0_{\R^d}}=0\\
      \frac{\partial^2 \psi(\lambda_{i_1},...,\lambda_{i_m})}{\partial\lambda_{i_s}\partial \lambda_{i_{l}}}|_{\bm{\lambda}=0_{\R^d}}=g_{i_s,i_{l}}\rho_{i_s,i_{l}}\left(u_{i_s}-u_{i_{l}} \right)
    \end{cases*}.
    \end{eqnarray}
    We can derive by induction a closed form formula for the following partial derivatives:
   \begin{equation}
\label{eq:partialderivativepsi}
\left. \frac{\partial^{l} \psi(\lambda_{i_1}, \ldots, \lambda_{i_m})}{\partial \lambda_{i_1} \cdots \partial \lambda_{i_l}} \right|_{\boldsymbol{\lambda} = 0_{\mathbb{R}^d}} =
\begin{cases}
\displaystyle\prod_{\substack{\{k, k'\} \in P_l}} g_{i_k, i_{k'}} \rho_{i_k, i_{k'}}(u_{i_k} - u_{i_{k'}}) \quad \text{if } l \text{ is even} \\[1.2ex]
0 \quad \hspace{4.5cm}\text{if } l \text{ is odd}
\end{cases}
\end{equation}
where $P_l$ is the partition of pairs of successive integers of $\llbracket 1,l \rrbracket $ and all $i_k$'s are distinct.\\\\
We temporary denote :
\begin{eqnarray}
\lambda_{i,m} = \begin{pmatrix}
\lambda_{i_1}\\
\lambda_{i_2} \\
\vdots \\
\lambda_{i_m}
\end{pmatrix} \in \R_{+}^m .\nonumber
\end{eqnarray}
A Taylor expansion around $0_{\R^d}$ of order $m$ on $\left(\lambda_{i_1},...,\lambda_{i_m}\right) \mapsto\psi\left(\lambda_{i_1},...,\lambda_{i_m}\right)$ gives the following:\\\\
    \[\psi\left(\lambda_{i_1},...,\lambda_{i_m}\right)=\psi(0_{\R^m})+\left<\nabla_{\bm{\lambda}_{i,m}=0_{\R^m}}\psi,\bm{\lambda}_d\right>+\sum_{\substack{1<|\beta|\leq m \\|\beta|=\sum_{i=1}^m\alpha_i }}\frac{1}{|\beta|!}\left(\prod_{k=1}^{m}\lambda_{i_k}^{\alpha_k} \right)\frac{\partial^{|\beta|} \psi(\lambda_{i_1},...,\lambda_{i_m})}{\partial\lambda_{i_1}^{\alpha_1}...\partial \lambda_{i_{m}}^{\alpha_m}}|_{\bm{\lambda}_{i,m}=0_{\R^m}}+ o\left(\|\lambda_{i,m} \|^m\right),\]
    where $\beta=\left(\alpha_i\right)_{i\in \llbracket 1,d \rrbracket}$ and $|\beta|!:=\prod_{i=1}^m\alpha_i!$.\\\\
    If any of $\alpha_i$'s is odd (not all equal to 1), the partial derivative above will vanish. Thus, it's non null only if on the the two following conditions hold: all $\alpha_i$'s are even, in which case, it's equal to :
    \begin{eqnarray}
        \prod_{k=1}^m2^{\frac{\alpha_k}{2}}\left(g_{i_k,i_k}\rho_{i_k,i_k}(0)\right)^{\alpha_k}=\prod_{k=1}^m2^{\frac{\alpha_k}{2}}\left(\rho_{i_k,i_k}(0)\right)^{\alpha_k} \nonumber
    \end{eqnarray}
      
      If they are all equal to 1, in which case, we can apply Eq.~\eqref{eq:partialderivativepsi}.\\
Consequently, one has:
\begin{eqnarray}
    \psi\left(\lambda_{i_1},...,\lambda_{i_m}\right)=1+\underset{\substack{s \in \llbracket1, m\rrbracket \\ s \hspace{0.1cm}even }}\sum\underset{\substack{\left\{k,k^{'}\right\} \in P_s}}\prod\lambda_{i_k}\lambda_{i_{k^{'}}} g_{i_k,i_{k^{'}}}\rho_{i_k,i_{k^{'}}}\left(\left|u_{i_k}-u_{i_{k^{'}}} \right|\right)+\underset{\substack{1<|\beta|< m \\|\beta|=\sum_{i=1}^m\alpha_i\\ \alpha_i's \hspace{0.1cm} even }}\sum\frac{2^{\frac{|\beta|}{2}}}{|\beta|!}\prod_{k=1}^m\left(\lambda_{i_k}\rho_{i_k,i_k}(0)\right)^{\alpha_k}\nonumber \\+o\left(\|\lambda_{i,m} \|^m\right).
    \nonumber
\end{eqnarray}

      We denote:
    \begin{eqnarray}
        \bm{S}_{-m}(d):=\sum_{\substack{\mathcal{I}_m=\left(i_k\right)_{k\in \llbracket 1,m \rrbracket} \\ \mathcal{I}_m \subset \mathcal{I}_d}}\sum_{\substack{1<|\beta|< m \\|\beta|=\sum_{i=1}^m\alpha_i\\ \alpha_i's \hspace{0.1cm} even }}\frac{1}{|\beta|!}\left(\prod_{k=1}^{m}\lambda_{i_k}^{\alpha_k} \right)\left(\prod_{k=1}^m2^{\frac{\alpha_k}{2}}\left(\rho_{i_k,i_k}(0)\right)^{\alpha_k}\right). \nonumber
    \end{eqnarray}
    After some algebra, it can be rewritten as follows:

    \begin{eqnarray}
        \bm{S}_{-m}(d)=\sum_{\substack{\mathcal{I}_m=\left(i_k\right)_{k\in \llbracket 1,m \rrbracket} \\ \mathcal{I}_m \subset \mathcal{I}_d}}\sum_{\substack{1<|\beta|< m \\|\beta|=\sum_{i=1}^m\alpha_i\\ \alpha_i's \hspace{0.1cm} even }}\frac{2^{\frac{|\beta|}{2}}}{|\beta|!}\prod_{k=1}^m\left(\lambda_{i_k}\rho_{i_k,i_k}(0)\right)^{\alpha_k} .\nonumber
    \end{eqnarray}

    Consequently, using Eq.~\eqref{eq:step2} leads to:    
    
   \begin{multline*}
       \E\left(\prod_{i=1}^{d}\left(M_i-1 \right) \right)=\sum_{\substack{m=0 }}^d\left(-1\right)^{d-m} \sum_{\substack{\mathcal{I}_m=\left(i_k\right)_{k\in \llbracket 1,m \rrbracket} \\ \mathcal{I}_m \subset \mathcal{I}_d}}\left(\prod_{k=1}^{m}\int_{I_{i_k}}\frac{du_{i_k}}{|I_{i_k}|} \right)\\\left[
       \underset{\substack{s \in \llbracket1, m\rrbracket \\ s \hspace{0.1cm}even }}\sum\underset{\substack{\left\{k,k^{'}\right\} \in P_s}}\prod\lambda_{i_k}\lambda_{i_{k^{'}}}g_{i_k,i_{k^{'}}}\rho_{i_k,i_{k^{'}}}\left(\left|u_{i_k}-u_{i_{k^{'}}} \right|\right)\right]+\bm{S}_{-m}(d)+o\left(\|\Lambda \|^d\right).
   \end{multline*}
   Written differently, one has:

   \begin{multline*}
       \E\left(\prod_{i=1}^{d}\left(M_i-1 \right) \right)=\sum_{\substack{m=0 }}^d\left(-1\right)^{d-m} \sum_{\substack{\mathcal{I}_m=\left(i_k\right)_{k\in \llbracket 1,m \rrbracket} \\ \mathcal{I}_m \subset \mathcal{I}_d}}\\\left[
       \underset{\substack{s \in \llbracket1, m\rrbracket \\ s \hspace{0.1cm}even }}\sum\underset{\substack{\left\{k,k^{'}\right\} \in P_s}}\prod\lambda_{i_k}\lambda_{i_{k^{'}}}\E\left(\frac{\Omega_{i_k}(I_{i_k})}{|I_{i_k}|} \frac{\Omega_{i_{k^{'}}}(I_{i_{k^{'}}})}{|I_{i_{k^{'}}}|}     \right)  \right]+\bm{S}_{-m}(d)+o\left(\|\Lambda \|^d\right).
   \end{multline*}
Consequently, one gets:
 \begin{multline}
 \label{eq:SIintermediaryresult}
       \E\left(\prod_{i=1}^{d}\left(M_i-1 \right) \right)=\sum_{\substack{m=0 }}^d\left(-1\right)^{d-m} \sum_{\substack{\mathcal{I}_m=\left(i_k\right)_{k\in \llbracket 1,m \rrbracket} \\ \mathcal{I}_m \subset \mathcal{I}_d}}\left[
       \sum_{s=1}^{m}\underset{\substack{\left\{k,k^{'}\right\} \in P_s}}\prod\lambda_{i_k}\lambda_{i_{k^{'}}}\E\left(\frac{\Omega_{i_k}(I_{i_k})}{|I_{i_k}|} \frac{\Omega_{i_{k^{'}}}(I_{i_{k^{'}}})}{|I_{i_{k^{'}}}|}     \right)  \right]+\\ \bm{S}_{-m}(d)+o\left(\|\Lambda \|^d\right).
   \end{multline}
where:
         \begin{eqnarray}
             \begin{cases}
                 \bm{S}_{-m}(d) := \sum_{\substack{\mathcal{I}_m = (i_k)_{k \in \llbracket 1, m \rrbracket} \\ \mathcal{I}_m \subset \mathcal{I}_d}} \sum_{\substack{|\beta| = \sum_{i=1}^m \alpha_i \\ 1 < |\beta| < m \\ \alpha_i \text{ even}}} \frac{2^{\frac{|\beta|}{2}}}{|\beta|!} \prod_{k=1}^m \left(\frac{\lambda_{i_k}^2}{2H_{i_k}(1-2H_{i_k})}\right)^{\alpha_k}\\
                 \mathcal{I}_d:=\llbracket 1,d \rrbracket
             \end{cases}
         \end{eqnarray}
   
 \begin{comment}
      \begin{multline*}
      \E\left(\prod_{i=1}^{d}\left(M_i-1 \right) \right)=\sum_{\substack{m=0}}^d\left(-1\right)^{d-m} \left[\bm{S}_{-m}(d)+
       \underset{\substack{s \in \llbracket1, m\rrbracket \\ s \hspace{0.1cm}even }}\sum\sum_{\substack{\mathcal{I}_s=\left(i_k\right)_{k\in \llbracket 1,j \rrbracket} \\ \mathcal{I}_s \subset \mathcal{P}\left(\mathcal{I}_d\right)}}\underset{\substack{\left\{i_{k^{'}},i_{k^{''}}\right\} \in I_s}}\prod\int_{I_{i_{k^{'}}}}\frac{du_{i_{k^{'}}}}{|I_{i_{k^{'}}}|}\int_{I_{i_{k^{''}}}}\frac{du_{i_{k^{''}}}}{|I_{i_{k^{''}}}|} \lambda_{i_{k^{'}}}\lambda_{i_{k^{''}}}\\
        g_{i_{k^{'}},i_{k^{''}}}\rho_{i_{k^{'}},i_{k^{''}}}\left(\left|u_{i_{k^{'}}}-u_{i_{k^{''}}} \right|\right)\right]+o\left(\|\Lambda \|^d\right).
   \end{multline*}
    Finally, we end up thanks to Eq.~\eqref{eq:momentOmegaforproofstep1} with the claimed result:
   \begin{eqnarray}
   \label{eq:SIintermediaryresult}
     \E\left(\prod_{i=1}^{d}\left(M_i-1 \right) \right)=\sum_{\substack{m=0 }}^d\left(-1\right)^{d-m} \left[\bm{S}_{-m}(d)+
        \sum_{s=1}^{m}\E\left(\prod_{i=1}^{s}\frac{\lambda_i\Omega_{i}(I_i)}{|I_i|}\right)\right]+o\left(\|\Lambda \|^d\right) .
   \end{eqnarray}
 \end{comment}  
\tab\tab\tab\tab\tab\tab\tab\tab\tab\tab\tab\tab\tab\tab\tab\tab\tab\tab\tab\tab\tab\tab\tab\tab\tab\tab\tab\tab\tab\tab\tab\tab\tab\tab $\blacksquare$
\\\\
\textbf{Step 4/4}\\
 We denote:
    \begin{itemize}
        \item $M:=\left(\frac{M_{i}(I_i)}{|I_{i}|}\right)_{i\in\llbracket 1,d \rrbracket}$
        \item $B_{\epsilon}:=\{x\in\R^d, \|x-\mathbf{1}\|_{\infty}\leq\epsilon \}$
    \end{itemize}
 We have using the union bound that:
    \begin{eqnarray}
        \Prob\left(M\notin B_{\epsilon}\right)\leq \sum_{i=1}^d\Prob\left(\left|\frac{M_{i}(I_i)}{|I_{i}|}-1\right|>\epsilon \right). \nonumber
    \end{eqnarray}

    Using Bienaymé-Tchebychev inequality, we have $\forall m\in \N$:
    \begin{eqnarray}
        \Prob\left(M\notin B_{\epsilon}\right)\leq \sum_{i=1}^d\frac{\E\left[\left|\frac{M_{i}(I_i)}{|I_{i}|}-1 \right|^{2m+1}\right]}{\epsilon^{2m+1}} .\nonumber
    \end{eqnarray}
    Given that each marginal of the  mLog S-fBM model is a Log S-fBM model, thus, using Proposition 12 in \cite{bacry2008log} (that leads to Proposition 4 in \cite{wu2022rough} in the Log S-fBM case) we have the small intermittency approximation: 
\begin{eqnarray}
    \forall (i,n) \in \llbracket 1,d \rrbracket \times \N, \forall (I_1,...,I_n)\subset K^n, \E\left(\prod_{k=1}^n\left(\frac{M_{i}(I_k)}{|I_k|}-1\right) \right)=\E\left(\prod_{k=1}^n\frac{\Omega_{i}\left(I_k\right)}{|I_k|} \right)+o\left(\lambda_i^n\right). \nonumber
\end{eqnarray}
In particular for any $i\in \llbracket 1,d \rrbracket$ and $k\in \llbracket 1,n \rrbracket$:
\begin{eqnarray}
    \forall m\in \N, \tab \E\left(\left(\frac{M_{i}(I_k)}{|I_k|}-1\right)^{2m+1}\right)=\lambda_i^{2m+1}\E\left(\left(\frac{\Omega_{i}\left(I_k\right)}{|I_k|} \right)^{2m+1}\right)+o\left(\lambda_i^{2m+1}\right). \nonumber
\end{eqnarray}
Given that $\frac{\Omega_{i}\left(I_k\right)}{|I_k|} $ is a gaussian random variable, we have thanks to Wick's theorem (Theorem \ref{thm:wick}) that:
\begin{eqnarray}
    \E\left(\left(\frac{\Omega_{i}\left(I_k\right)}{|I_k|} \right)^{2m+1}\right) =0.\nonumber
\end{eqnarray}
Meaning that:
\begin{eqnarray}
    \forall m\in \N, \tab \E\left(\left(\frac{M_{i}(I_i)}{|I_i|}-1\right)^{2m+1}\right)=o\left(\lambda_i^{2m+1}\right). \nonumber
\end{eqnarray}
Consequently, we obtain:
\begin{eqnarray}
    \Prob\left(M\notin B_{\epsilon}\right)\leq \sum_{i=1}^do\left(\lambda_i^{2m+1}\right), \nonumber
\end{eqnarray}
    leading to:
    \begin{eqnarray}
        \Prob\left(M\notin B_{\epsilon}\right)\leq o\left(\|\Lambda\|^{2m+1}\right). \nonumber
    \end{eqnarray}
The latter equality is similar to Lemma 5 in \cite{bacry2008log}.
As a result, we obtain by the same computations as in Proposition 13 in \cite{bacry2008log}:
\begin{eqnarray}
            \E\left(\prod_{i=1}^{d} \ln\left(\frac{M_{i}(I_i)}{|I_i|}\right) \right)=\E\left(\prod_{i=1}^{d}\left(\frac{M_{i}(I_i)}{|I_i|}-1 \right) \right)+o\left(\|\Lambda\|^{d}\right)
            \nonumber
\end{eqnarray}
Using Eq.~\eqref{eq:SIintermediaryresult}, the claimed result is obtained.\\
\tab\tab\tab\tab\tab\tab\tab\tab\tab\tab\tab\tab\tab\tab\tab\tab\tab\tab\tab\tab\tab\tab\tab\tab\tab\tab\tab\tab\tab\tab\tab\tab\tab\tab $\blacksquare$
\begin{remark}
\label{rmk:SIAdim2simplification}
Given an interval $K\subset \R_{+}$ with $|K|\leq T$ and $(i,j)\in \llbracket 1,d \rrbracket^2$, one has for any arbitrary subintervals $(I,J) \in K^2$ as $\bm{S}_{-1}(2)=\bm{S}_{-2}(2)=0$. This leads to Eq.~\eqref{eq:smallintermitformulamomentdim2}.
\end{remark}

\section{Simulation procedure of two mLog S-fBM marginals}
\label{app:simulationmarginalsmlogsfbm}

The starting point of the estimation procedure is to simulate a mS-fBM process with the respective co-Hurst exponents and co-intermittency parameters. To simulate over a period of scale $\Delta$ sample paths of $\left(\ln\left(\frac{M_{i,\Delta}(.)}{\Delta}\right),i\in \llbracket 1,d \rrbracket \right)$, one needs to simulate trajectories say of length $L$ of the mS-fBM process that are sufficiently long in order to be able to subsample $\Delta$ observations necessary to construct trajectories of size $N$ of  $\left(\ln\left(\frac{M_{i,\Delta}(.)}{\Delta}\right),i\in \llbracket 1,d \rrbracket \right)$, in that case, $L=N\times \Delta$. For that purpose, we leverage the spectral-domain construction that generates a pair of coupled mS-fBM distinct marginals. This procedure is a standard extension of the circulant-embedding method for Gaussian processes studied in \cite{Dietrich1997, Perrin2003,HelgasonPipirasAbry2011}.\\
Nonetheless, one can see according to Eq.~\eqref{eq:covOmeganormalized} that in the small intermittency regime, for any $t\geq 0$, the process $\frac{\lambda\Omega_{H,T,\Delta}(t)}{\Delta}$ is a first order approximation of $\omega_{H,T}(t)-\mu_H$ when $\Delta$ is arbitrary small. Besides, 
given Eq.~\eqref{eq:smallintermitformulamomentdim2qv}, one can see that $\frac{\lambda\Omega_{H,T,\Delta}(t)}{\Delta}$ is also an approximation of $\ln\left(\frac{\left<X\right>_{[t,t+\Delta]}}{\Delta}\right)$ in small intermittency.
This infers that one may consider that the latter processes are equivalent in the small intermittency regime.

\section{Proof of Theorem \ref{theo:variancesumsmallintermit}}
\label{app:proofoftheovariancesumsmallintermit}
Let $K\subset \R_+$ be an interval such that $|K|\leq T$ and consider 
any sub-intervals $(I,J)\subset K^2$. We introduce the following notations:

\begin{eqnarray}
    \begin{cases}
         C^{\ln}(I,J):=\E\left(\ln\left(\frac{M\left(I\right)}{|I|}\right)\ln\left(\frac{M\left(J\right)}{|J|}\right)\right)\\
          C_{i,j}(I,J):=\E\left(\frac{\Omega_{i}(I)}{|I|}\frac{\Omega_{j}(J)}{|J|}\right), (i,j)\in \llbracket 1,d \rrbracket^2
    \end{cases}
    \nonumber
    \end{eqnarray}

Beforehand, we need this intermediary result.
\begin{theo}
\label{thm:smallinterautocovinterval}
For any sub intervals $I$ and $J$ of $K$, the following approximations holds:

\begin{eqnarray}
         C^{\ln}(I,J)-C^{\ln}(I,I)=\sum_{1\leq i,j \leq d} \alpha_i^2\alpha_j^2 \lambda_{i}\lambda_{j}\left[C_{i,j}\left(I,J\right)-C_{i,j}\left(I,I\right)\right] +o\left(\|\Lambda\|^2\right),
\end{eqnarray}
where $||.||$ is the $\mathcal{L}^2$ norm in $\R^d$ and the tuples $\left(\alpha_i\right)_{i\in \llbracket 1,d \rrbracket}$ is defined in the subsection \ref{sec:multidimlogsfbm}.
\end{theo}

\textbf{\textit{Proof}}\\
%\begin{enumerate}[label=\alph*)]  

Given two arbitrary sub-intervals $I$ and $J$ of $K$, we have
\begin{eqnarray}
    M\left(I\right)M\left(J\right)=\sum_{1\leq i,j \leq d} \alpha_i^2 \alpha_j^2 M_{i}\left(I\right)M_{j}\left(J\right). \nonumber
\end{eqnarray}

For any $(i,j)\in \llbracket 1,d \rrbracket^2$, we have:

\begin{eqnarray}
\E \left( \frac{M_{i}(I)}{|I|}\frac{M_{j}(J)}{|J|} \right) = \int_{I \times J}\frac{dxdy}{|I||J|} \E \left( e^{\omega_{i}(x)+\omega{H_j,T}(y)}\right)  . \nonumber
\end{eqnarray}

We have for any $(x,y)\in I\times I$:
\begin{eqnarray}
    \E\left(e^{\omega_{i}(x)+\omega{H_j,T}(y)} \right) = e^{\mu_i + \mu_j + \frac{1}{2} \sum_{(m,n) \in \{i,j\}} \lambda_m \lambda_n \rho_{m,n}(|x-y|)}, \nonumber
\end{eqnarray}

where for any $(m,n)\in \llbracket 1,d \rrbracket^2$:
\begin{eqnarray}
    \rho_{m,n}\left(\tau\right)=\frac{g_{m,n}}{2H_{m,n}(1-2H_{m,n})}\left(1-\left(\frac{\tau}{T}\right)^{2H_{m,n}}\right).\nonumber
\end{eqnarray}

Which gives after some algebra:
\begin{eqnarray}
    \E\left(e^{\omega_{H_{i},T}(x)+\omega_{H_{j},T}(y)}\right) = e^{\lambda_{i }\lambda_{j}\rho_{i,j}(|x-y|)}. \nonumber
\end{eqnarray}
By denoting the functional:
  $$\psi:\left(\lambda_{i},\lambda_{j}\right) \mapsto e^{\lambda_{i }\lambda_{j}\rho_{i ,j}(|x-y|)},  $$
we perform a first order Taylor expansion:
 \[\psi\left(\lambda_{i},\lambda_{j}\right)=\psi(0_{\R^2})+\left<\nabla_{\bm{\lambda}_d=0_{\R^2}}\psi,\bm{\lambda}\right>+\frac{1}{2}\left<Hess\left(\psi\right)|_{\bm{\lambda}=0_{\R^2}}\bm{\lambda},\bm{\lambda}\right>+ o\left(\|\Lambda_{i,j}\|^2\right),\]
Where 
$$ \Lambda_{i,j}= \begin{pmatrix}
\lambda_i \\
\lambda_j 
\end{pmatrix}
$$
We get:

 \[\psi\left(\lambda_{i},\lambda_{j}\right)=1+\lambda_{i}\lambda_{j}\rho_{i,j}(|x-y|)+ o\left(\|\Lambda_{i,j} \|^2\right).\]
Thus, we conclude that:

\begin{eqnarray}
    \E\left(\frac{M_{i}\left(I\right)}{|I|}\frac{M_{j}\left(J\right)}{|J|}\right)=
    1+\lambda_{i}\lambda_{j}\E\left(\frac{\omega_{i}(I)}{|I|}\frac{\omega_{j}(J)}{|J|}\right)+ o\left(\|\Lambda_{i,j} \|^2\right) .
    \nonumber
\end{eqnarray}

As a result, we obtain:
\begin{eqnarray}
          \E\left(\frac{M\left(I\right)}{|I|}\frac{M\left(J\right)}{|J|}\right)=\sum_{1\leq i,j \leq d} \alpha_i^2\alpha_j^2 \left(1+\lambda_{i}\lambda_{j}\E\left(\frac{\omega_{i}(I)}{|I|}\frac{\Omega_{j}(J)}{|J|}\right) \right)+o\left(\|\Lambda \|^2\right) ,\nonumber
     \end{eqnarray}

which can be rewritten as:
\begin{eqnarray}
          \E\left(\frac{M\left(I\right)}{|I|}\frac{M\left(J\right)}{|J|}\right)=\sum_{1\leq i,j \leq d} \alpha_i^2\alpha_j^2 \lambda_{i}\lambda_{j}\E\left(\frac{\omega_{i}(I)}{|I|}\frac{\Omega_{j}(J)}{|J|}\right) +\left(\sum_{i=1}^d \alpha_i^2\right)^2 +o\left(\|\Lambda\|^2\right).\nonumber
\end{eqnarray}
Thus, we obtain:
\begin{eqnarray}
         \E\left(\left(\frac{M\left(I\right)}{|I|}-1\right)\left(\frac{M\left(J\right)}{|J|}-1\right)\right)=\sum_{1\leq i,j \leq d} \alpha_i^2\alpha_j^2 \lambda_{i}\lambda_{j}\E\left(\frac{\Omega_{H_iT}(I)}{|I|}\frac{\Omega_{j}(J)}{|J|}\right) +\left(\sum_{i=1}^d \alpha_i^2\right)^2 -2\sum_{i=1}^d\alpha_i^2 \nonumber \\
         +1 +o\left(\|\Lambda\|^2\right).\nonumber
\end{eqnarray}
Using the Proposition 13 in \cite{bacry2008log}, we have:
\begin{eqnarray}
         \E\left(\ln\left(\frac{M\left(I\right)}{|I|}\right)\ln\left(\frac{M\left(J\right)}{|J|}\right)\right)=\sum_{1\leq i,j \leq d} \alpha_i^2\alpha_j^2 \lambda_{i}\lambda_{j}\E\left(\frac{\omega_{i}(I)}{|I|}\frac{\Omega_{j}(J)}{|J|}\right) +\left(\sum_{i=1}^d \alpha_i^2\right)^2 -2\sum_{i=1}^d\alpha_i^2 \nonumber \\
         +1 +o\left(\|\Lambda\|^2\right).\nonumber
\end{eqnarray}
In particular, we have:
\begin{eqnarray}
         \E\left(\ln\left(\frac{M\left(I\right)}{|I|}\right)\ln\left(\frac{M\left(I\right)}{|I|}\right)\right)=\sum_{1\leq i,j \leq d} \alpha_i^2\alpha_j^2 \lambda_{i}\lambda_{j}\E\left(\frac{\omega_{i}(I)}{|I|}\frac{\Omega_{j}(I)}{|I|}\right) +\left(\sum_{i=1}^d \alpha_i^2\right)^2 -2\sum_{i=1}^d\alpha_i^2 \nonumber \\
         +1 +o\left(\|\Lambda\|^2\right),\nonumber
\end{eqnarray}
which leads to:
\begin{eqnarray}
         C^{\ln}(I,J)-C^{\ln}(I,I)=\sum_{1\leq i,j \leq d} \alpha_i^2\alpha_j^2 \lambda_{i}\lambda_{j}\left[C_{i,j}\left(I,J\right)-C_{i,j}\left(I,I\right)\right] +o\left(\|\Lambda\|^2\right).\nonumber
\end{eqnarray}

%\end{enumerate}
\tab\tab\tab\tab\tab\tab\tab\tab\tab\tab\tab\tab\tab\tab\tab\tab\tab\tab\tab\tab\tab\tab\tab\tab\tab\tab\tab\tab\tab\tab\tab\tab\tab\tab $\blacksquare$
Now, we proceed to the proof of theorem \ref{theo:variancesumsmallintermit}.\\\\
\textbf{\textit{Proof}}\\
\label{proof:variancesumsmallintermitequation}
In this proof, we denote:
\begin{eqnarray}
C^{\ln}\left(\Delta,\tau\right):=C^{\ln}([0,\Delta],[\tau,\tau+\Delta])
\end{eqnarray}

and for any $(i,j)\in \llbracket 1,d \rrbracket^2$:
\begin{eqnarray}
    C_{i,j}\left(\Delta,\tau\right):=C_{i,j}([0,\Delta],[\tau,\tau+\Delta]).
\end{eqnarray}

 This part is analogous to the previous one. We have:
 \begin{eqnarray}
        V\left(\tau,\Delta,\Lambda\right)=2\textnormal{var}\left(\ln\left(\frac{M_{\Delta}(\tau)}{\Delta}\right) \right)-2C^{\ln}\left(\Delta,\tau\right). \nonumber
    \end{eqnarray}
    Consequently, we get using theorem \ref{thm:smallinterautocovinterval}:
\begin{eqnarray}
        V\left(\tau,\Delta,\Lambda\right)=\sum_{1\leq i,j \leq d} \frac{2\alpha_i^2\alpha_j^2 \lambda_{i}\lambda_{j}}{\Delta}\left[C_{i,j}\left(\Delta,0\right)-C_{i,j}\left(\Delta,\tau\right)\right] +o\left(\|\Lambda\|^2\right).\nonumber
\end{eqnarray}

After some algebra using Proposition \ref{prop:autocovarianceOmegamultidim} and Eq.~\eqref{eq:phiijtilde}, we obtain:
 \begin{eqnarray}
        V\left(\tau,\Delta,\Lambda\right)=\sum_{1\leq i,j \leq d}  \alpha_i^2\alpha_j^2 g_{i,j}\lambda_{i}\lambda_{j}\left(\frac{\tau}{T}\right)^{2H_{i,j}}\Tilde{\phi}_{i,j}\left(\frac{\Delta}{\tau}\right)+o\left(\|\Lambda\|^2\right),\nonumber
\end{eqnarray}
where for any $(i,j)\in \llbracket 1,d \rrbracket^2$, the function $\phi_{i,j}(.)$ is defined in Proposition \ref{prop:autocovarianceOmegamultidim}. \\\\ 
This leads to:
\begin{eqnarray}
    \mathcal{V}\left(\tau,\Delta,\Lambda\right)=\sum_{1\leq i,j \leq d}  \alpha_i^2\alpha_j^2 g_{i,j}\lambda_{i}\lambda_{j}\left(\frac{\tau}{T}\right)^{2H_{i,j}}\Tilde{\phi}_{i,j}\left(\frac{\Delta}{\tau}\right). \nonumber
\end{eqnarray}

\tab\tab\tab\tab\tab\tab\tab\tab\tab\tab\tab\tab\tab\tab\tab\tab\tab\tab\tab\tab\tab\tab\tab\tab\tab\tab\tab\tab\tab\tab\tab\tab\tab\tab $\blacksquare$

\section{Proof of Theorem \ref{thm:converD}}
\label{app:proofconverD}
We fix $k\in \Tau_{Q}$ and $\theta_{i,j}\in \R\times\left]0,\frac{1}{4}\right[\times\R_+$.\\
For notational simplicity, we denote here:
\begin{eqnarray}
    C_{i,j}^{\Omega}(k):=C^{\Omega}\left(k,\theta_{i,j}\right) \nonumber
\end{eqnarray}

\begin{enumerate}
    \item We have straightforwardly:
    \begin{eqnarray}
        \E\left(\frac{N}{N-k}\Hat{D}^{\Omega}_{i,j}(k) \right)=\frac{N}{N-k}\E\left(\hat{C}^{\Omega}_{i,j}(k)-\hat{C}^{\Omega}_{i,j}(0) \right)=C^{\Omega}\left(k,\theta_{i,j}\right)-C^{\Omega}\left(0,\theta_{i,j}\right)=D^{\Omega}\left(k,\theta_{i,j}\right). \nonumber
    \end{eqnarray}
     On the other hand, we have as well:
     \begin{eqnarray}
         Var\left( \frac{N}{N-k}\Hat{D}^{\Omega}_{i,j}(k)\right)=\left(\frac{N}{N-k} \right)^2\frac{1}{N^2}\sum_{l=1}^{N-k}\sum_{s=1}^{N-k}\mathrm{cov}\left(\frac{\Omega_{i,\Delta}(l\Delta)}{\Delta}Z^{k}_{j}(l),\frac{\Omega_{i,\Delta}(s\Delta)}{\Delta}Z^{k}_{j}(s) \right). \nonumber
     \end{eqnarray}
After some algebra, one has:

\begin{eqnarray}
   \Delta^2 Var\left( \frac{N}{N-k}\Hat{D}^{\Omega}_{i,j}(k)\right)=\left(\frac{N}{N-k} \right)^2\frac{1}{N^2}\sum_{l=1}^{N-k}\mathrm{cov}\left(\frac{\Omega_{i,\Delta}(l\Delta)}{\Delta}Z^{k}_{j}(l),\frac{\Omega_{i,\Delta}(l\Delta)}{\Delta}Z^{k}_{j}(l) \right)+ \nonumber\\
    \left(\frac{N}{N-k} \right)^2\frac{2}{N^2}\sum_{l=1}^{N-k}\sum_{s=l+1}^{N-k}\mathrm{cov}\left(\Omega_{i,\Delta}(l\Delta)Z^{k}_{j}(l),\Omega_{i,\Delta}(s\Delta)Z^{k}_{j}(s) \right), \nonumber
\end{eqnarray}
 where $Z^{k}_{j}(l):=\Omega_{j,\Delta}\left(\left(l+k\right)\Delta\right)-\Omega_{j,\Delta}(l\Delta)$.\\\\
 As  $\left(\left(\Omega_{i,\Delta}(t)\right)_t,i\in \llbracket 1,d \rrbracket\right)$ is a centered gaussian field, thanks to Wick's theorem, we have that:
    \begin{eqnarray}
    \mathrm{cov}\left(\Omega_{i,\Delta}(l\Delta),Z^{k}_{j}(l),\Omega_{i,\Delta}(s\Delta)Z^{k}_{j}(s) \right)=\mathrm{cov}\left(Z^{k}_{j}(l),Z^{k}_{j}(s) \right)\mathrm{cov}\left(\Omega_{i,\Delta}(l\Delta),\Omega_{i,\Delta}(s\Delta)\right)+\nonumber \\ \mathrm{cov}\left(Z^{k}_{j}(l),\Omega_{i,\Delta}(s\Delta) \right)\mathrm{cov}\left(\Omega_{i,\Delta}(l\Delta),Z^{k}_{j}(s)\right), \nonumber
\end{eqnarray}
meaning that:
 \begin{eqnarray}
    \mathrm{cov}\left(\Omega_{i,\Delta}(l\Delta)Z^{k}_{j}(l),\Omega_{i,\Delta}(s\Delta)Z^{k}_{j}(l) \right)&=&C^{\Omega}_{i,i}(l-s)\left(2C^{\Omega}_{j,j}(l-s)-C^{\Omega}_{j,j}(l+k-s)-C^{\Omega}_{j,j}(s+k-l) \right)+\nonumber \\
    &&
    \left( C^{\Omega}_{i,j}(s+k-l)-C^{\Omega}_{i,j}(l-s)\right)\left( C^{\Omega}_{i,j}(l+k-s)-C^{\Omega}_{i,j}(l-s)\right). \nonumber
\end{eqnarray}
Using Eq.~\eqref{eq:covOmega}, we obtain the following upper bound after some algebra:
\begin{eqnarray}
    \mathrm{cov}\left(\Omega_{i,\Delta}(l\Delta)Z^{k}_{j}(l),\Omega_{i,\Delta}(s\Delta)Z^{k}_{j}(l) \right)\leq CN^{4H_{i,j}}\left(1+|l-s|\right)^{4H_{i,j}-2}, \nonumber
\end{eqnarray}
where $C$ is a positive constant.\\\\
This leads to:
\begin{eqnarray}
    Var\left( \Hat{D}^{\Omega}(x,y)(k)\right)\leq \left(C N^{4H_{i,j}-1}+2CN^{4H_{i,j}-2}\sum_{l=1}^{N}\int_l^Nu^{4H_{i,j}-2}du\right), \nonumber
\end{eqnarray}
which gives the following upper bound:
\begin{eqnarray}
    Var\left( \frac{N}{N-k} \Hat{D}^{\Omega}_{i,j}(k)\right)\leq C^{'}\left(\frac{N}{N-k} \right)^2N^{4H_{i,j}-1}, \nonumber
\end{eqnarray}
where $C^{'}=C\left(1+\frac{2}{4H_{i,j}-1} \right)$ .\\\\
We conclude that $Var\left( \frac{N}{N-k} \Hat{D}^{\Omega}_{i,j}(k)\right)\underset{N \rightarrow \infty}\longrightarrow 0 $. Consequently, we obtain the claimed result thanks to Bienaymé-Tchebychev's inequality.

\item  We denote analogously to the other notations:\\
$$Z^{\ln}_{i}(l):=\ln\left(\frac{M_{i,\Delta}(l\Delta)}{\Delta}\right)-m_{i,\Delta},l\in \llbracket 1,N \rrbracket.$$
By direct computations:
\begin{eqnarray}
    \E\left(\frac{N}{N-k}\Hat{D}^{\ln}_{i,j}(k) \right)=\frac{N}{N-k}\E\left(\hat{C}^{\ln}_{i,j}(k)-\hat{C}_{i,j}^{\ln}\left(\Delta,0\right) \right) 
    =C^{\ln}\left(k,\theta_{i,j}\right)-C^{\ln}\left(0,\theta_{i,j}\right)=D^{\ln}\left(k,\theta_{i,j}\right)\nonumber
\end{eqnarray}
 Using the first order approximation in small intermittencies in  Eq.~\eqref{eq:smallintermitformulamomentdim2qv} , we have:
 \begin{eqnarray}
     D_{i,j}^{\ln}(k) = \lambda_i\lambda_jD^{\Omega}\left(k,\theta_{i,j}\right)+o\left(\| \Lambda_{i,j} \|^2\right).\nonumber
 \end{eqnarray}
 where $$ \Lambda_{i,j}= \begin{pmatrix}
\lambda_i \\
\lambda_j 
\end{pmatrix}
$$
 Thus,
\begin{eqnarray}
    \E\left(\frac{N}{N-k}\frac{\Hat{D}_{i,j}^{\ln}(k)}{\lambda_i\lambda_j} \right)=D^{\Omega}\left(k,\theta_{i,j}\right)+ o\left(\| \Lambda_{i,j} \|^2\right) .\nonumber
\end{eqnarray}
Which means that:
 \begin{eqnarray}
       \lim_{\| \Lambda_{i,j} \|^2\to 0} \E\left(\frac{N}{N-k}\frac{\Hat{D}^{\ln}_{i,j}(k) }{\lambda_i\lambda_j}\right) =D^{\Omega}\left(k,\theta_{i,j}\right).
       \nonumber
\end{eqnarray}
The mS-fBM has continuous sample paths, thus, one has using the dominated convergence theorem alongside with Eq.~\eqref{eq:aslimit}:
 \begin{eqnarray}
       \E\left(\frac{N}{N-k}\Tilde{D}_{i,j}^{\ln}(k) \right) =D^{\Omega}\left(k,\theta_{i,j}\right).
       \nonumber
\end{eqnarray}
On the other hand:
\begin{eqnarray}
    Var\left( \Hat{D}^{\ln}_{i,j}(k)\right)=\frac{1}{N^2}\sum_{l=1}^{N-k}\sum_{s=1}^{N-k}\mathrm{cov}\left(Z^{\ln}_{i}(l)Z^{k,log}_{j}(l),Z^{\ln}_{i}(s)Z^{k}_{j}(s) \right),
\end{eqnarray}
meaning that:
\begin{eqnarray}
    Var\left( \Hat{D}^{\ln}_{i,j}(k)\right)=\frac{1}{N^2}\sum_{l=1}^{N-k}\mathrm{cov}\left(Z^{\ln}_{i}(l)Z^{k,log}_{j}(l),Z^{\ln}_{i}(l)Z^{k,log}_{j}(l) \right)+\nonumber\\
    \frac{2}{N^2}\sum_{l=1}^{N-k}\sum_{s=l+1}^{N-k}\mathrm{cov}\left(Z^{\ln}_{i}(l)Z^{k,log}_{j}(l),Z^{\ln}_{i}(s)Z^{k,log}_{j}(s) \right), \nonumber
\end{eqnarray}
 where $Z^{k,log}_{j}(l):=Z^{\ln}_{i}(l+k)-Z^{\ln}_{i}(l)$.\\\\
  Using again Eq.~\eqref{eq:smallintermitformulamomentdim2qv}, we have:
    $$\mathrm{cov}\left(Z^{\ln}_{i}(l)Z^{k,log}_{j}(l),Z^{\ln}_{i}(s)Z^{k,log}_{j}(s) \right)=\left(\lambda_i\lambda_j\right)^2\mathrm{cov}\left(\Omega_{i,\Delta}(l\Delta)Z^{k}_{j}(l),\Omega_{i,\Delta}(s\Delta)Z^{k}_{j}(s) \right)+o\left(\|\Lambda_{i,j}\| ^2\right).$$
    As  $\left(\left(\Omega_{i,\Delta}(t)\right)_t,i\in \llbracket 1,d \rrbracket\right)$ is a centered gaussian field, thanks to Wick's theorem we have that:
    \begin{eqnarray}
    \mathrm{cov}\left(\Omega_{i,\Delta}(l\Delta)Z^{k}_{j}(l),\Omega_{i,\Delta}(s\Delta)Z^{k}_{j}(s) \right)= \mathrm{cov}\left(Z^{k}_{j}(l),Z^{k}_{j}(s) \right)\mathrm{cov}\left(\Omega_{i,\Delta}(l\Delta),\Omega_{i,\Delta}(s\Delta)\right)+\nonumber \\  \mathrm{cov}\left(Z^{k}_{j}(l),\Omega_{i,\Delta}(s\Delta) \right)\mathrm{cov}\left(\Omega_{i,\Delta}(l\Delta),Z^{k}_{j}(s)\right).\nonumber
\end{eqnarray}
.\\\\
After some algebra, we obtain:
 \begin{eqnarray}
    \mathrm{cov}\left(\Omega_{i,\Delta}(l\Delta),Z^{k}_{j}(l),\Omega_{i,\Delta}(s\Delta)Z^{k}_{j}(l) \right)= C^{\Omega}_{i,i}(l-s)\left(2C^{\Omega}_{j,j}(l-s)-C^{\Omega}_{j,j}(l+k-s)-C^{\Omega}_{j,j}(s+k-l) \right)+\nonumber \\
    \left( C^{\Omega}_{i,j}(s+k-l)-C^{\Omega}_{i,j}(l-s)\right)\left( C^{\Omega}_{i,j}(l+k-s)-C^{\Omega}_{i,j}(l-s)\right). \nonumber
\end{eqnarray}
Using Eq.~\eqref{eq:covOmega}, we obtain the following upper bound after some algebra:
\begin{eqnarray}
    \mathrm{cov}\left(\Omega_{i,\Delta}(l\Delta),Z^{k}_{j}(l),\Omega_{i,\Delta}(s\Delta)Z^{k}_{j}(l) \right)\leq  C N^{4H_{i,j}}\left(1+|l-s|\right)^{4H_{i,j}+2} \nonumber
\end{eqnarray}
where $C$ is a positive constant.\\\\
This gives thanks to Eq.~\eqref{eq:aslimit}:
\begin{eqnarray}
    Var\left( \Tilde{D}_{i,j}^{\ln}(k)\right)\leq  C \left(N^{4H_{i,j}-1}+2N^{4H_{i,j}-2}\sum_{l=1}^{N}\int_l^Nu^{4H_{i,j}-2}du\right) \nonumber
\end{eqnarray}
which gives the following upper bound:
\begin{eqnarray}
    Var\left(\frac{N}{N-k} \Tilde{D}_{i,j}^{\ln}(k)\right)\leq  \left(\frac{N}{N-k}\lambda_i\lambda_j\right)^2C^{'}N^{4H_{i,j}-1} \nonumber
\end{eqnarray}
where $C^{'}=C\left(1+\frac{2}{4H_{i,j}-1} \right)$ .\\\\
Thus, We conclude that $Var\left( \frac{N}{N-k}\Tilde{D}_{i,j}^{\ln}(k)\right)\xrightarrow[\substack{N \rightarrow \infty }]{} 0 $. Consequently, we obtain the claimed result thanks to Bienaymé-Tchebychev's inequality.
\end{enumerate} 

\tab\tab\tab\tab\tab\tab\tab\tab\tab\tab\tab\tab\tab\tab\tab\tab\tab\tab\tab\tab\tab\tab\tab\tab\tab\tab\tab\tab\tab\tab\tab\tab\tab\tab $\blacksquare$

\section{Proof of theorem \ref{thm:convergenceprobcalibration}}
\label{app:proofconvergenceprobcalibration}
We fix two arbitrary marginals $i$ and $j$ in $\llbracket 1,d \rrbracket$.\\

  We introduce the following first order small intermittency approximations for any $\left(x,\Lambda_{i,j}\right) \in \Xi_{i,j}\times\R_{+}^2$:
  \begin{eqnarray}
      \bm{h}_{i,j}^{\ln}\left(x,\Lambda_{i,j}\right)=\Hat{\bm{h}}_{i,j}^{\ln}\left(x,\Lambda_{i,j}\right)+o\left(\|\Lambda_{i,j} \| ^2\right)
  \end{eqnarray}

In the sequel, we consider the objective function of Eq.~\eqref{eq:calibrationproblemlogvol}:
\begin{eqnarray}
    \forall x\in \Xi_{i,j},\tab  
    \begin{cases}
        \Tau^{\ln}\left(x\right):=\bm{h}_{i,j}^{\ln}\left(x,\Lambda_{i,j}\right)^{T}W\bm{h}_{i,j}^{\ln}\left(x,\Lambda_{i,j}\right) \\
        \Tau\left(x\right):=\bm{h}_{i,j}\left(x\right)^{T}W\bm{h}_{i,j}\left(x\right) 
    \end{cases}\nonumber
\end{eqnarray}
We denote  $\nabla.$ denotes the gradient and $\nabla^2 .$ stands for the Hessian matrix.
\begin{enumerate}
    \item  We denote as well $\theta_{i,j}^{*}$ the true vector of parameters.\\
According to Eq.~\eqref{eq:calibrationproblemlogvol}, $\theta_{i,j}^N\left(\Lambda_{i,j}\right)$ satisfies the first-order condition:
\begin{eqnarray}
\nabla  \Tau^{\ln}\left(\theta_{i,j}^N\left(\Lambda_{i,j}\right)\right) = 0,
\end{eqnarray}
Using a first-order Taylor expansion, there exists $\tilde{\theta}_{i,j}\in \Xi_{i,j}$ such that:
\begin{eqnarray}
\nabla  \Tau^{\ln}\left(\theta_{i,j}^N\left(\Lambda_{i,j}\right)\right)  &=& \nabla  \Tau^{\ln}\left(\theta_{i,j}^{*}\right) + \nabla^2  \Tau^{\ln}\left(\tilde{\theta}_{i,j}\right) \left(\theta_{i,j}^N\left(\Lambda_{i,j}\right) - \theta_{i,j}^{*}\right) \nonumber 
\end{eqnarray}
This leads to the following identity:
\begin{eqnarray}
\label{eq:identitytheta}
\theta_{i,j}^N\left(\Lambda_{i,j}\right) - \theta_{i,j}^{*} &=& - \left(\nabla^2  \Tau^{\ln}\left(\tilde{\theta}_{i,j}\right) \right)^{-1} \nabla  \Tau^{\ln}\left(\theta_{i,j}^{*}\right)
\end{eqnarray}

$\Tau^{\ln}(.)$ is differentiable, one has for any $x\in \Xi_{i,j}$:
\begin{eqnarray}
\begin{cases}
    \nabla  \Tau^{\ln}\left(x\right) = 2 J^{\ln}\left(x\right)^\top W \bm{h}_{i,j}^{\ln}\left(x,\Lambda_{i,j}\right)\\
    \nabla^2  \Tau^{\ln}\left(x\right) = 2 J^{\ln}\left(x\right)^\top WJ^{\ln}\left(x\right) + 2\left(\nabla^2\bm{h}_{i,j}^{\ln}\left(x,\Lambda_{i,j}\right)\right) W \bm{h}_{i,j}^{\ln}\left(x,\Lambda_{i,j}\right)
\end{cases} \nonumber
\end{eqnarray}
where $J^{\ln}\left(.\right)$ is the Jacobian matrix of $\bm{h}_{i,j}^{\ln}$, $\nabla^2\bm{h}_{i,j}^{\ln}$ is the hessian tensor of $\bm{h}_{i,j}^{\ln}$ and the contraction operation $\left(\nabla^2\bm{h}_{i,j}^{\ln}\left(x,\Lambda_{i,j}\right)\right) W \bm{h}_{i,j}^{\ln}\left(x,\Lambda_{i,j}\right)$ is defined as:
\begin{eqnarray}
\label{Eq:contraction}
    \left(\nabla^2\bm{h}_{i,j}^{\ln}\left(x,\Lambda_{i,j}\right)\right) W \bm{h}_{i,j}^{\ln}\left(x,\Lambda_{i,j}\right)=\sum_{k=1}^Q \left(W\bm{h}_{i,j}^{\ln}\left(x,\Lambda_{i,j}\right)\right)_k \nabla^2 \left(\bm{h}_{i,j}\right)_k^{\ln}\left(x,\Lambda_{i,j}\right) 
\end{eqnarray}
($\left(\bm{h}_{i,j}\right)_k^{\ln}$ is the $k^{th}$ component of the $\bm{h}_{i,j}^{\ln}$).\\
According to the definitions of Theorem \ref{thm:converD}, one has:
\begin{eqnarray}
    \bm{h}_{i,j}^{\ln}\left(\theta_{i,j}^{*},\Lambda_{i,j}\right)=\lambda_i\lambda_j\left(\bm{\Tilde{D}}_{i,j}^{\ln}-\bm{D}^{\Omega}\left(\theta_{i,j}^{*}\right)\right)+o\left(\|\Lambda_{i,j} \| ^2\right) \nonumber
\end{eqnarray}

rewritten differently, one has:
\begin{eqnarray}
    \bm{h}_{i,j}^{\ln}\left(\theta_{i,j}^{*},\Lambda_{i,j}\right)=\lambda_i\lambda_j\left(\begin{pmatrix}
\frac{N}{N-\tau_1}\Tilde{D}_{i,j}^{\ln}\left(\tau_1\right) \\
\frac{N}{N-\tau_2}\Tilde{D}_{i,j}^{\ln}\left(\tau_2\right) \\
\vdots \\
\frac{N}{N-\tau_Q}\Tilde{D}_{i,j}^{\ln}\left(\tau_Q\right)
\end{pmatrix}-\bm{D}^{\Omega}\left(\theta_{i,j}^{*}\right)\right)-\lambda_i\lambda_j\begin{pmatrix}
\frac{\tau_1}{N-\tau_1}\Tilde{D}_{i,j}^{\ln}\left(\tau_1\right) \\
\frac{\tau_2}{N-\tau_2}\Tilde{D}_{i,j}^{\ln}\left(\tau_2\right) \\
\vdots \\
\frac{\tau_Q}{N-\tau_Q}\Tilde{D}_{i,j}^{\ln}\left(\tau_Q\right)
\end{pmatrix}
+o\left(\|\Lambda_{i,j} \| ^2\right) \nonumber
\end{eqnarray}
We identify
\begin{eqnarray}
    \Hat{\bm{h}}_{i,j}^{\ln}\left(\theta_{i,j}^{*},\Lambda_{i,j}\right)=\lambda_i\lambda_j\left(\begin{pmatrix}
\frac{N}{N-\tau_1}\Tilde{D}_{i,j}^{\ln}\left(\tau_1\right) \\
\frac{N}{N-\tau_2}\Tilde{D}_{i,j}^{\ln}\left(\tau_2\right) \\
\vdots \\
\frac{N}{N-\tau_Q}\Tilde{D}_{i,j}^{\ln}\left(\tau_Q\right)
\end{pmatrix}-\bm{D}^{\Omega}\left(\theta_{i,j}^{*}\right)\right)-\lambda_i\lambda_j\begin{pmatrix}
\frac{\tau_1}{N-\tau_1}\Tilde{D}_{i,j}^{\ln}\left(\tau_1\right) \\
\frac{\tau_2}{N-\tau_2}\Tilde{D}_{i,j}^{\ln}\left(\tau_2\right) \\
\vdots \\
\frac{\tau_Q}{N-\tau_Q}\Tilde{D}_{i,j}^{\ln}\left(\tau_Q\right)
\end{pmatrix}\nonumber
\end{eqnarray}
According to Theorem \ref{thm:converD}:
\begin{eqnarray}
    \begin{pmatrix}
\frac{N}{N-\tau_1}\Tilde{D}_{i,j}^{\ln}\left(\tau_1\right) \\
\frac{N}{N-\tau_2}\Tilde{D}_{i,j}^{\ln}\left(\tau_2\right) \\
\vdots \\
\frac{N}{N-\tau_Q}\Tilde{D}_{i,j}^{\ln}\left(\tau_Q\right)
\end{pmatrix}-\bm{D}_{i,j}^{\Omega}\left(\theta_{i,j}^{*}\right)\xrightarrow[N\rightarrow \infty]{\Prob}0_{\R^Q} \nonumber
\end{eqnarray}
Furthermore,
\begin{eqnarray}
    \begin{pmatrix}
\frac{\tau_1}{N-\tau_1}\Tilde{D}_{i,j}^{\ln}\left(\tau_1\right) \\
\frac{\tau_2}{N-\tau_2}\Tilde{D}_{i,j}^{\ln}\left(\tau_2\right) \\
\vdots \\
\frac{\tau_Q}{N-\tau_Q}\Tilde{D}_{i,j}^{\ln}\left(\tau_Q\right)
\end{pmatrix} \xrightarrow[N\rightarrow \infty]{\Prob}0_{\R^Q}.\nonumber
\end{eqnarray}

According to Theorem \ref{thm:converD} as well as Eq.~\eqref{eq:identitytheta}, it holds that: 
\begin{eqnarray}
    \Hat{\theta}_{i,j}^N\left(\Lambda_{i,j}\right) \xrightarrow[N\rightarrow \infty]{\Prob}\theta_{i,j}^{*} \nonumber
\end{eqnarray}
\item  Conducting the same computations, there exists $\tilde{\eta}_{i,j}\in \Xi_{i,j}$ such that:
\begin{eqnarray}
\label{eq:identityeta}
\eta_{i,j}^N - \theta_{i,j}^{*} &=& - \left(\nabla^2  \Tau\left(\tilde{\eta}_{i,j}\right) \right)^{-1} \nabla  \Tau\left(\theta_{i,j}^{*}\right)
\end{eqnarray}
where$\Tau$ is the objective function of Eq.~\eqref{eq:calibrationproblemOmega} whose gradient and hessian have the closed formulas:
\begin{eqnarray}
\begin{cases}
    \nabla  \Tau\left(x\right) = 2 J\left(x\right)^\top W \bm{h}_{i,j}\left(x\right)\\
    \nabla^2  \Tau\left(x\right) = 2 J\left(x\right)^\top WJ\left(x\right) + 2\left(\nabla^2\bm{h}_{i,j}\left(x\right)\right)W \bm{h}_{i,j}\left(x\right), \tab x\in  \Xi_{i,j}
\end{cases} \nonumber
\end{eqnarray}
$J$ is the Jacobian matrix of $\bm{h}_{i,j}$ and the contraction operation is defined as in Eq.~\eqref{Eq:contraction}.
\\
Alongside, one has:
\begin{eqnarray}
    \bm{h}_{i,j}\left(\theta_{i,j}^{*}\right)=\left(\begin{pmatrix}
\frac{N}{N-\tau_1}\Hat{D}^{\Omega}_{i,j}\left(\tau_1\right) \\
\frac{N}{N-\tau_2}\Hat{D}^{\Omega}_{i,j}\left(\tau_2\right) \\
\vdots \\
\frac{N}{N-\tau_Q}\Hat{D}^{\Omega}_{i,j}\left(\tau_Q\right)
\end{pmatrix}-\bm{D}^{\Omega}\left(\theta_{i,j}^{*}\right)\right)-\begin{pmatrix}
\frac{\tau_1}{N-\tau_1}\Hat{D}^{\Omega}_{i,j}\left(\tau_0\right) \\
\frac{\tau_2}{N-\tau_2}\Hat{D}^{\Omega}_{i,j}\left(\tau_1\right) \\
\vdots \\
\frac{\tau_Q}{N-\tau_Q}\Hat{D}^{\Omega}_{i,j}\left(\tau_Q\right)
\end{pmatrix}\nonumber
\end{eqnarray}
Using Theorem \ref{thm:converD}, the first term converges in probability to $0_{\R^Q}$. By definition, the second term converges to $0_{\R^Q}$ in probability. \\
Consequently, the following holds:
\begin{eqnarray}
    \eta_{i,j}^N\left(\Lambda_{i,j}\right) \xrightarrow[N\rightarrow \infty]{\Prob}\theta_{i,j}^{*} \nonumber
\end{eqnarray}

\end{enumerate}
\tab\tab\tab\tab\tab\tab\tab\tab\tab\tab\tab\tab\tab\tab\tab\tab\tab\tab\tab\tab\tab\tab\tab\tab\tab\tab\tab\tab\tab\tab\tab\tab\tab\tab $\blacksquare$

\end{appendices}
\end{document}